\documentclass[final,12pt,authoryear]{elsarticle} 

\usepackage{amssymb}
\usepackage{bm}

\usepackage{tikz}

\usetikzlibrary{arrows, positioning,arrows.meta}
\usetikzlibrary{backgrounds}
\usetikzlibrary{calc}

\usepackage{svg}
\usepackage{array} 

\usepackage{graphicx}
\usepackage{newtxtext}
\usepackage{newtxmath}
\usepackage{natbib}
\usepackage{hyperref}

\usepackage{color}

\usepackage[caption=false,justification=centering]{subfig}

\usepackage{algorithm}
\usepackage{algpseudocode}
\usepackage{placeins}
\usepackage{textgreek}



\newcommand{\RomanNumeralCaps}[1]
\linenumbers

\newcommand\Rey{\mbox{\textit{Re}}}  

\journal{Computer Physics Communications}

\begin{document}
\begin{frontmatter}	
\title{Orbital cluster-based network modelling}
	
\author[a,b]{Antonio Colanera\corref{cor1}}
\author[c]{Nan Deng}
\author[b]{Matteo Chiatto}
\author[b]{Luigi de Luca}
\author[d,e]{Bernd~R.~Noack}
			
\cortext[cor1] {Corresponding author.\\\textit{E-mail address:} antonio.colanera@polito.it}
\address[a]{Department of Mechanical and Aerospace Engineering, Politecnico di Torino, Turin, 10129, Italy}				
\address[b]{Department of Industrial Engineering, University of Naples ``Federico II'', Naples, 80125, Italy}
\address[c]{Chair of Artificial Intelligence and Aerodynamics, School of Robotics and Advanced Manufacture,
Harbin Institute of Technology, Shenzhen,
518055, People's Republic of China}

\address[d]{ College of Mechatronics and Control Engineering, Shenzhen University, Shenzhen, 518060, People's Republic of China}
\address[e]{ Guangdong Province VTOL Manufacturing Innovation Center, Shenzhen, 518060, People's Republic of China}
\begin{abstract}
		
We propose a novel reduced-order framework to describe complex multi-frequency fluid dynamics from time-resolved snapshot data. The starting point is the Cluster-based Network Model (CNM), valued for its fully automatable development and human interpretability. Our key innovation is to model the transitions from cluster to cluster much more accurately by replacing snapshot states with short-term trajectories (``orbits'') over multiple clusters, thus avoiding non-physical diffusion of the probability distributions in the dynamics reconstruction. The proposed orbital CNM (oCNM) employs functional clustering to coarse-grain the short-term trajectories. Specifically, different filtering techniques, resulting in different temporal basis expansions, demonstrate the versatility and capability of the oCNM to adapt to diverse flow phenomena. The oCNM is illustrated on the Stuart-Landau oscillator and its post-transient solution with time-varying parameters to test its ability to capture the amplitude selection mechanism and multi-frequency behaviours. Then, the oCNM is applied to the fluidic pinball across varying flow regimes at different Reynolds numbers, including the periodic, quasi-periodic, and chaotic dynamics. This orbital-focused perspective enhances the understanding of complex temporal behaviours by incorporating high-frequency behaviour into the kinematics of short-time trajectories while modelling the dynamics of the lower frequencies. In analogy to Spectral Proper Orthogonal Decomposition, which marked the transition from spatial-only modes to spatio-temporal ones, this work advances from analysing temporal local states to examining piecewise short-term trajectories or orbits. By merging advanced analytical methods, such as the functional representation of short-time trajectories with CNM, this study paves the way for new approaches to dissect the complex dynamics characterising turbulent systems.
\end{abstract}
	
\end{frontmatter}

\section{Introduction}
\label{sec:introduction}
Over the past few decades, Reduced-Order Modelling (ROM) has gained significant attention in computational science and engineering. It has become a crucial tool to obtain fast prediction in various industry sectors, including mechanical and electronic engineering, as well as in fundamental and applied sciences such as neuroscience \citep{Brunton20161}, medicine, biology, and chemistry \citep{Quarteroni2011}. These methods are also gaining importance in emerging areas that deal with complex problems \citep{Papaioannou2022}, spanning multiple physical phenomena and scales \citep{Mendez2019}. The utility of ROM also extends to flow control \citep{Brunton2015amr} and uncertainty quantification, where it offers robust solutions \citep{Cinnella2011,Edeling2014,Xiao2019}.

A popular way to obtain a ROM is to extract physically important features or modes which characterise the flow topology and project the Navier-Stokes equations (Galerkin projection) onto a subset of these modes, resulting in a system of ordinary differential equations \citep{noack2003,Mezic2005,Rowley2017,Stabile2017CAIM,LUCHTENBURG2009}. By choosing a restricted set of modes among those available, it is possible to build a reduced model to predict the flow field behaviour with a lower computational cost.

The recent integration of machine learning \citep{brunton2020arfm} into this field has unveiled novel avenues for the analysis and modelling of complex systems \citep{Rowley2009,Lee2020,Fabiani2021,Alvarez2023,Racca2023}. 
Among the notable advances is the development of Physics-Informed Neural Networks (PINNs), which have proven effective in estimating mean fields within linearised frameworks, particularly when governing equations are well established, but data availability is sparse \citep{Raissi2020,VonSaldern2022,Oezalp2023}.

Moreover, data-driven approaches have found applications in analysing recurrent and chaotic dynamical systems via Poincaré maps \citep{Guckenheimer1983,Bramburger2020}. 
It is well-known that any chaotic attractor is densely approximated by a countably infinite ensemble of Unstable Periodic Orbits (UPOs), pivotal for understanding dynamical systems \citep{Auerbach1987,Cvitanovic1989}. Key dynamical metrics, such as Lyapunov exponents and entropy, can be quantitatively linked to these UPOs \citep{Ott2002,Cvitanovic1988}. 
\citet{Bramburger2021} showed that chaotic attractors can be understood in terms of the UPOs embedded in attractors and proposed a data-driven method to stabilise the UPOs of ordinary differential equations, by using Sparse Identification of Nonlinear Dynamics (SINDy) \citep{Brunton2016,Loiseau2018}.

Among the several ROM techniques, cluster-based modelling has gained an increasing interest in recent years due to its ability to effectively reproduce the dynamics of complex systems, optimally capturing temporal statistics and the evolution of coherent structures \citep{Li1976,Burkardt2006,Iacobello2021}. 
This method is based on a data-driven coarse-graining of the state space into a few representative flow states (centroids) or snapshot bins (clusters).
An early version of cluster-based modelling uses Markov models for the cluster population \citep{Schneider2007}.
However, the modelled dynamics feature a non-physical diffusion of the probability distributions (PDF) \citep{kaiser2014jfm}.
This diffusion of PDF is mitigated in higher-order Markov models \cite{fernex2020sa} and more fully addressed in a cluster-based network model (CNM) \citep{Hou2022,li2020jfm}.
CNM offers a fully automated framework for dynamics characterisation, parameter estimation, and model-based control \citep{Wang2023,Nair2019jfm,Colanera2024b}. CNM is also particularly effective in analysing bifurcations and secondary flow structures \citep{deng2022jfm}. 
In \citet{Colanera2023}, an analogous approach in the clustering procedure is employed to focus on specific, potentially more informative subspaces or subsets of variables with a filtered distance metric based on the filtered correlation matrix \citep{sieber2016}. 

Yet, the deterministic trajectories through a cluster are only modelled to their entry into the next cluster. 
Multiple trajectory exits of a cluster are modelled as a stochastic process or diffusion. This leads to non-physical diffusion within models, creating significant difficulties in describing transient dynamics and broadband frequency behaviour \citep{kaiser2014jfm}. This work aims to overcome this limitation by analysing orbits (piecewise trajectories) instead of temporal local states. 

\begin{figure}
\centering
\begin{tikzpicture}[node distance=4.5cm, auto]

    \tikzstyle{block} = [rectangle, draw, fill=white!20, text width=3.8cm, text centered, rounded corners, minimum height=3.6cm]
    \tikzstyle{line} = [draw,  -{Latex[scale=1.15]}]
    
    \node [block, fill=cyan!20] (block1) {\scriptsize{Data collection} \\\vspace{5pt} \includegraphics[width=3.8cm]{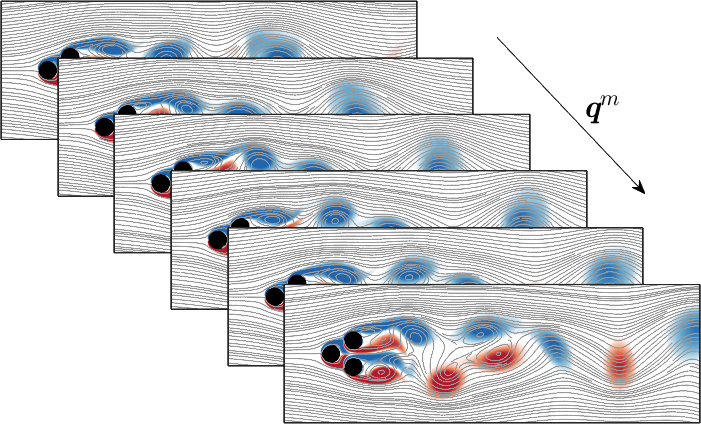}};
    \node [block, below of=block1, node distance=4.1cm,fill=cyan!20] (block3) {\scriptsize{Multivariate functional clustering \\and orbit identification} \\ \vspace{5pt} \includegraphics[trim= 0.cm 0.cm 0.cm 0cm,clip,width=3.8cm]{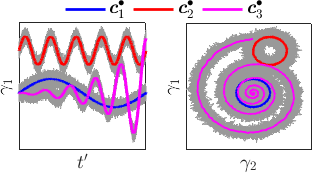}};
    \node [block, right of=block3,fill=red!20] (block4) {\scriptsize{Computation of spatio-temporal \\cluster centroids} \\ \vspace{5pt}\includegraphics[trim= 0cm 0cm 0cm 0cm,clip,width=3.5cm]{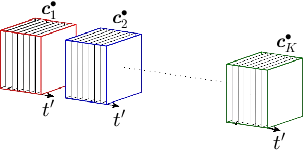}};
    
    \node [block, left of=block3,fill=yellow!20] (block5) {\scriptsize{Time evolution of orbit index} \\ \vspace{10pt}\includegraphics[trim=0cm 0.0cm 0.0cm 0.0cm,clip,width=3.8cm]{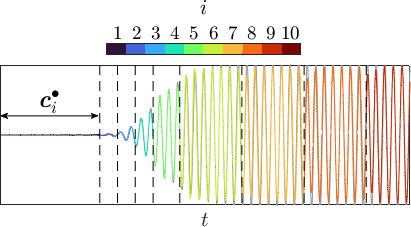}};
    
    \node [block, left of=block1, fill=yellow!20] (block6) {\scriptsize{Transitions network and probabilities} \\ \vspace{5pt}\centering\includegraphics[width=2cm]{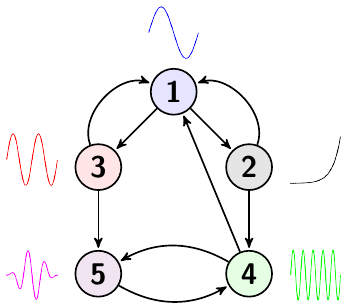} \includegraphics[width=1.6cm]{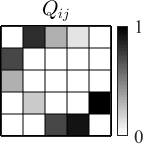} };
	\node [block, right of=block1, fill=red!20] (block8) {\scriptsize{Spatio-temporal clusters topology} \\ \vspace{5pt} \includegraphics[trim= 1.91cm 0.95cm 3.115cm 0.695cm,clip,height=1cm]{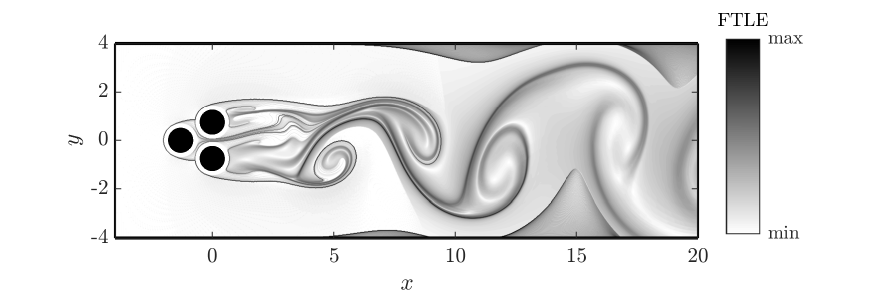}\\ \vspace{5pt}
\includegraphics[trim= 1.91cm 0.95cm 3.115cm 0.695cm,clip,height=1cm]{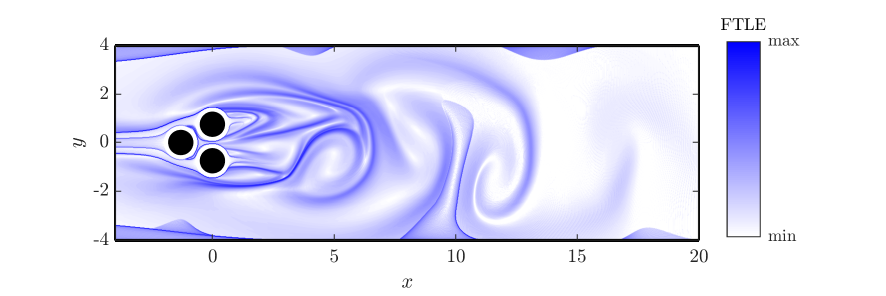}};

 \node [block, above of=block1, fill=cyan!20,minimum height=1.2cm, node distance=2.6cm] (blockdata) {Data processing};
 \node [block, above of=block6, fill=yellow!20,minimum height=1.2cm, node distance=2.6cm] (blockdata) {Dynamics};
 \node [block, above of=block8, fill=red!20,minimum height=1.2cm, node distance=2.6cm] (blockdata) {Physical\\ interpretation};
		
    \path [line] (block1) -- (block3);
    \path [line] (block3) -- (block4);
    \path [line] (block3) -- (block5);
		\path [line] (block5) -- (block6);
		\path [line] (block4) -- (block8);
\end{tikzpicture}
\caption{Schematic overview of the orbital CNM (oCNM). 
The oCNM coarse-grains the system dynamics into a set of representative orbits (piecewise trajectories) by orbital clustering.
Then, the transition dynamics between those orbits are described by a CNM.
More details on the procedure are given in \S~\ref{sec:TCNM}.
}\label{fig:oCNM}
\end{figure}
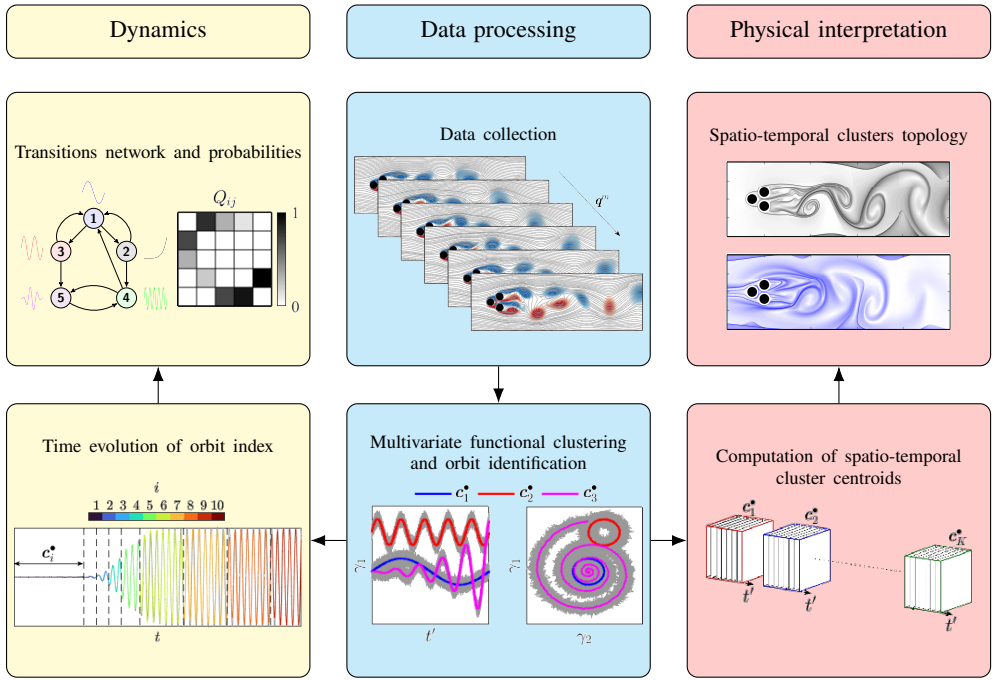

The evolution from Proper Orthogonal Decomposition (POD) to Spectral POD (SPOD) \citep{towne2018,Schmidt2019,Frame2023,Colanera2025} marked the transition from spatial-only modes to spatio-temporal ones, thereby minimising errors over a defined time window. Similarly, this work advances from analysing temporal local states to examining piecewise short-term trajectories, or namely orbits. 
This shift enables the capture of multi-frequency behaviours and amplitude modulations characteristic of complex fluid dynamics systems. 
The resulting methodology, namely orbital CNM (oCNM), is schematically shown in Figure~\ref{fig:oCNM}.

This manuscript is organised as follows:
The CNM is firstly recalled in \S~\ref{sec:CNM}. 
The oCNM based on trajectories clustering is introduced in \S~\ref{sec:TCNM}, and then is demonstrated on the Stuart-Landau oscillator and the fluidic pinball for periodic, quasi-periodic, and chaotic flow regimes in \S~\ref{sec:results}.
In the end, \S~\ref{sec:conclusion} concludes the main findings and improvements, together with several suggested future directions.

\section{Cluster-based network model for the state space}\label{sec:CNM}
Consider the state field $\bm{q}(\bm{x}, t)$ of a system within a spatial domain $\Omega$. The field data can arise from experimental measurements or computational simulations and record state variables such as velocity, pressure, etc.
The field is sampled at equidistant time steps with $\Delta t$, so $t^m = m \Delta t$ represents the time instance for the $m$-th snapshot. 
An ensemble of $M$ time-resolved snapshots serves as the foundation for the system analysis, denoted as $\bm{q}^m(\bm{x}) := \bm{q}(\bm{x}, t^m)$, where $m = 1, \ldots, M$.

For clarity, the variables used in this work are listed in Table~\ref{tab:variables}.
\begin{table}
\centering
\def~{\hphantom{0}}
  \begin{tabular}{ll}
        Variables   &   Description\\[0.5pt]
        $\bm{q}^m$    &   Time-resolved snapshots\\[0.5pt]
        $m$    &   Snapshots index\\[0.5pt]
        $M$   &   Number of snapshots\\[0.5pt]
        $t$   &   Time\\[0.5pt]
        $\Delta t$   &  Time step\\[0.5pt]
        \multicolumn{2}{l}{ -----------------  CNM ----------------- }\\[0.5pt]  
        $K$   &   Number of clusters\\[0.5pt]
        $\mathscr{C}_{i}$   &   Clusters \\[0.5pt]
        $\chi_{i}^{m}$   &   Characteristic function of the state space clustering\\[0.5pt]
        $k(m)$ &   Cluster-affiliation function to $\mathscr{C}_{k}$\\[0.5pt]
        $M_{k}$   &   Number of snapshots in cluster $\mathscr{C}_{k}$\\[0.5pt]
        $\boldsymbol{c}_{k}$   &   Centroids of phase space clusters \\[0.5pt]
        $m_{ij}$   &   Number of transitions from $\mathscr{C}_{j}$ to $\mathscr{C}_{i}$\\[0.5pt]
        $m_{j}$   &   Total number of transitions from $\mathscr{C}_{j}$\\[0.5pt]
        $Q_{ij}$   &   Cluster transition probability from $\mathscr{C}_{j}$ to $\mathscr{C}_{i}$\\[0.5pt]
        $T_{ij}$   &   Cluster transition time from $\mathscr{C}_{j}$ to $\mathscr{C}_{i}$\\[0.5pt]  
        $J$&  Intra-cluster variance\\[0.5pt]
   
        \multicolumn{2}{l}{ ----------------- Orbital CNM ----------------- }\\[0.5pt]  
        $L$&   Number of trajectory segments\\[0.5pt]       
        $\bm{q}_l$&   The $l$-th trajectory segment\\[0.5pt]
        $t'$   &   Local trajectory time\\[0.5pt]
        $t_{0l}$ &   Initial time of $l$-th segment\\[0.5pt]
        $T_l$  &   Trajectories time span\\[0.5pt]
        $L_{\mathrm{traj}}$   &   Length of  trajectory segments\\[0.5pt]
        $\bm{\mu}$  &   Mean trajectory observed\\[0.5pt]
        $\bm{\phi}_i$&   Spatio-temporal basis function\\[0.5pt]
        $w_{l}$&   Temporal weight for reconstruction\\[0.5pt]
        $J^\bullet$&   Intra-cluster functional variance\\[0.5pt]
        $V^\bullet$&   Inter-cluster functional variance\\[0.5pt]
        $\bm{c}^{\bullet}_k$ &   Orbital centroids of clusters \\[0.5pt]
        $L_{k}$   &   Number of segments in $k$-th orbital cluster\\[0.5pt]
        $d_{lj}^2$&   Distance between $l$ and $j$ functions\\[0.5pt]
        $\hat{\bm{\alpha}}_{lk}$ & The $k$-th Fourier expansion coefficient of the $l$-th segment \\[0.5pt]
        $\bm{\beta}_{lk}$& The $k$-th B-spline expansion coefficient of the $l$-th segment \\[0.5pt]
        $\bm{\gamma}_{lrk}$ & The $k$-th wavelets expansion coefficient at $r$-th level of the $l$-th segment
  \end{tabular}
\caption{Table of variables.}
\label{tab:variables}
\end{table}

\subsection{Clustering the phase space }
Cluster analysis, an unsupervised method of data organisation, aggregates similar entities into groups known as clusters, all without the need for prior data labelling or classification. 
In the context of our set of $M$ snapshots $\bm{q}^m(\bm{x})$, the method divides the data into $K$ clusters. Each cluster $\mathscr{C}_k$ is represented by a centroid $\bm{c}_k(\bm{x})$, where $k = 1, \ldots, K$. These centroids are determined via the unsupervised $k$-means++ initialization and Lloyd iteration \citep{steinhaus1956,MacQueen1967,Lloyd1982,arthur2006}, representing the characteristic flow patterns or modes by averaging the snapshots within each cluster.

The cluster-affiliation function plays a crucial role here, associating a specific velocity field $\bm{q}$ with the index of its nearest centroid:
\begin{equation}
k(\bm{q}) = \underset{i}{\arg\min}\,\|\bm{q} - \bm{c}_i\|_{\Omega},
\end{equation}
where $\|\cdot\|_{\Omega}$ denotes the Hilbert space norm within the domain $\Omega$. 
%
It is useful to establish a characteristic function $\chi_i^m$ to indicate if the $m$-th snapshot $\bm{q}^{m}$ belongs to the cluster $\mathscr{C}_i$:
\begin{equation}
\chi_i^m = 
\begin{cases}
    1, & \text{if } i = k(\bm{q}^m). \\
    0, & \text{otherwise}.
\end{cases}
\end{equation}

The effectiveness of a given set of centroids ${\bm{c}_{k=1,\ldots,K}}$, relative to a given set of snapshots ${\bm{q}^{m=1,\ldots,M}}$, can be evaluated based on the mean variance of the snapshots to their nearest centroid. This intra-cluster variance gives the cost function:
\begin{equation}\label{eq:costfun}
J(\bm{c}_{\!1}, \ldots, \bm{c}_{\!K}) = \frac{1}{M} \sum_{m=1}^M \|\bm{q}^m - \bm{c}_{k(m)}\|_{\Omega}^2,
\end{equation}
where $k(m) := k(\bm{q}^m)$. The optimal centroids ${\bm{c}_{k=1,\ldots,K}^{\star}}$ are those that minimise this intra-cluster variance:
\begin{equation}\label{eq:optCLU1}
(\bm{c}_1^{\star}, \ldots, \bm{c}_K^{\star}) = \underset{\bm{c}_{\!1}, \ldots, \bm{c}_{\!K}}{\arg\min} \,J(\bm{c}_{\!1}, \ldots, \bm{c}_{\!K}).
\end{equation}
In the following, the optimal centroids $\bm{c}_i^{\star}$ are denoted as $\bm{c}_i$ for simplicity.

To solve the optimisation problem indicated by \eqref{eq:optCLU1}, the $k$-means++ algorithm is employed. 
This algorithm begins by randomly initialising the $K$ centroids and then iteratively adjusts these centroids until $J$ falls below a predefined threshold, indicating that the clustering is sufficiently accurate. 
The $k$-means++ algorithm repeats the clustering process several times and selects the best set of centroids. 

The total number of snapshots $M_k$ within the cluster $\mathscr{C}_k$ is calculated by $M_k = \sum_{m=1}^M \chi_k^m$. Moreover, the centroids $\bm{c}_k$ that are representative of each cluster are computed as the mean of all snapshots within the respective cluster:
\begin{equation}
\bm{c}_{\!k} = \frac{1}{M_k} \sum_{\bm{q}^m \in \mathscr{C}_k} \bm{q}^m = \frac{1}{M_k} \sum_{m=1}^M \chi_k^m \bm{q}^m.
\end{equation}

\subsection{Cluster-based Network Model}
After the original snapshots are coarse-grained into $K$ clusters, each cluster can be interpreted as a distinct state or mode of the system dynamics \citep{Taira2022}. This representation reduces the data dimensionality and provides an abstract view of the complex data structure, facilitating the understanding and analysis of the underlying dynamics.

Following \citet{fernex2020sa} and \citet{li2020jfm}, these clusters or modes are used to conceptualise the system dynamics as a directed network. 
In this network, the centroids represent the nodes, each signifying a unique state of the system. The directed edges, on the other hand, denote potential finite-time transitions between these states. This network-based representation aids in uncovering the temporal flow or transition tendencies among the clusters, revealing the structure and progression of the system states.


The transitions in the cluster-based network model are characterised by the probability and time from clusters to clusters.
For a transition from $\mathscr{C}_j$ to $\mathscr{C}_i$, the direct transition probability, denoted $Q_{ij}$, and the transition time, denoted $T_{ij}$, can be deduced from the dataset. 
The transition probability is defined as:
\begin{equation}
Q_{ij} = \frac{m_{ij}}{m_j}, \quad i, j = 1, \ldots, K,
\end{equation}
where $m_{ij}$ represents the count of transitions from $\mathscr{C}_j$ to $\mathscr{C}_i$. Moreover, $m_j$ refers to the total count of transitions originating from $\mathscr{C}_j$, regardless of the destination:
\begin{equation}
m_j = \sum_{i=1}^K m_{ij}, \quad i, j = 1, \ldots, K.
\end{equation}
It is worth noticing that $m_{jj} = Q_{jj} = 0$ as only non-trivial transitions from $\mathscr{C}_j$ to $\mathscr{C}_i$ are considered.
The probabilities $Q_{ij}$ are stored into a matrix $\bm{Q} = \begin{bmatrix}Q_{ij}\end{bmatrix} \in \mathcal{R}^{K \times K}$. 

Consider a transition from $\mathscr{C}_j$ to $\mathscr{C}_i$, the residence time in the first cluster $\mathscr{C}_j$ is defined as:
\begin{equation}\label{eq:residence}
\tau_n = t_{n+1} - t_n,
\end{equation}
where $t_n$ is the time associated with the first snapshot entering the cluster $\mathscr{C}_j$, and $t_{n+1}$ is the time entering the next cluster.
The transition time is given as half the total residence time in both clusters:
\begin{equation}
\tau_{ij} = \frac{\tau_n + \tau_{n+1}}{2}.
\end{equation}
Consider all observed transitions from $\mathscr{C}_j$ to $\mathscr{C}_i$, the direct transition time $T_{ij}$ is defined as the mean of their transition times:
\begin{equation}
T_{ij} = \langle\tau_{ij}\rangle,
\end{equation}
where $\langle\cdot\rangle$ is the average operator. These averaged transition times are stored in a matrix $\bm{T} = \begin{bmatrix}T_{ij}\end{bmatrix} \in \mathcal{R}^{K \times K}$.

The CNM reconstruction is based on centroid visits at discrete times. 
The clusters, denoted as $k_0, k_1, k_2, \ldots$, are visited in sequence at the following times:
\begin{equation}
t_0 = 0, \quad t_1 = T_{k_1 k_0}, \quad t_2 = t_1 + T_{k_2 k_1} \ldots
\end{equation}
This visit sequence is in accordance with the transition probability and the transition times. 
In the following, continuous motion is assumed between these visits using a linear interpolation. However, it should be noted that applying splines could potentially result in a more smooth motion. Hence, the reconstructed field $\bm{q}^\approx(\bm{x}, t)$ for $t \in [t_n, t_{n+1}]$ is formulated as:
\begin{equation}\label{eq:reco}
\bm{q}^\approx(\bm{x}, t) = w_n(t) \bm{c}_{\!k(\bm{q}^n)}(x) + \left[1 - w_n(t)\right] \bm{c}_{\!k(\bm{q}^{n+1})}(x), \quad w_n = \frac{t_{n+1} - t}{t_{n+1} - t_n}.
\end{equation}
Following \citet{fernex2020sa}, it is possible to generalise the network model with a higher-order Markov chain. For a generalised $O$-order CNM, the direct transition probability is expressed as $Pr (k_{n+1} =i \vert \bm{c}_{\!k_n},...,\bm{c}_{\!k_{n-O-1}})$, which uses $O$ previous states to predict next state. The calculation of transition times is then adapted to suit this extended model. Note that this approach is equivalent to using time-delay coordinates.

\section{Orbital CNM}\label{sec:TCNM}
The CNM approaches consider a state vector composed of the realisations of flow field variables. Here, we propose a new approach that considers a state vector composed of orbits of the dynamical system.

The basic idea of orbital CNM (oCNM) is to cluster piecewise trajectories from a given dataset, dealing with data consisting of curves or functions. Using functional approaches, data clustering can be performed more efficiently and effectively, as curves can more accurately represent the shape and features of trajectories. Compared to CNM, where snapshot order does not influence the clustering, this allows a more meaningful and informative clustering of trajectories, leading to improved insights and decision-making.

\begin{figure}
	\centering
	\includegraphics[trim=1.5cm 0.8cm 1.2cm 0.8cm, clip,width=0.8\textwidth]{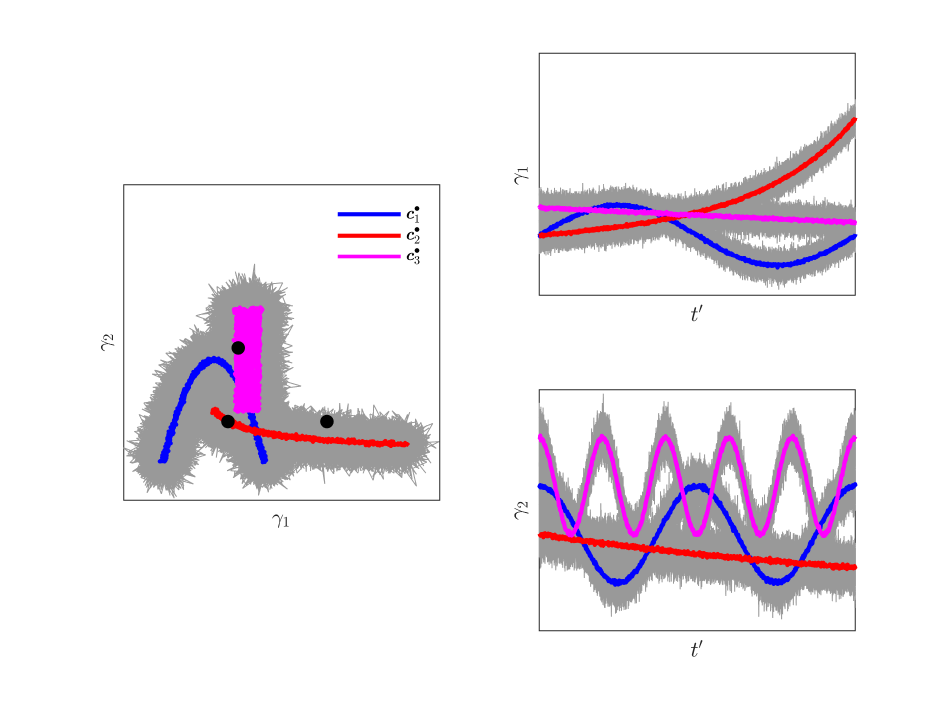}
	\caption{Sketch of functional clustering on an ensemble of $n$ two-dimensional trajectories with noise. The black markers denote phase space centroids, while the coloured lines indicate trajectory clusters centroids.}
	\label{fig:rawclust}
\end{figure}

To better understand the aim of orbital clustering, Figure~\ref{fig:rawclust} shows a functional clustering sketch. 
An ensemble of $n$ two-dimensional trajectories presenting noise is analysed and reported in grey in the figure. 
In the phase space clustering approach, a centroid would be a two-dimensional vector in the $(\gamma_1,\gamma_2)$ space, and in the figure are reported $3$ centroids as black dots. In the orbital framework, the clusters are no longer state vector realisations but functions/trajectories.

While working with functional data, one could assume that observations exist in an infinite-dimensional space, but we only have samples observed at a finite number of points. It is common to work with discrete observations $\bm{q}_{l,j}$ of each sample path $\bm{q}_l(t)$ in a finite set of knots, $j$. Therefore, the first step in functional data analysis frequently involves reconstructing the functional shape of the data from discrete observations \citep{Schmutz2020}. Depending on the shape reconstruction method, functional data analysis can be broadly classified into three categories: raw data approaches, filtering approaches, and adaptive methods \citep{Jacques2013,Jacques2014}. Raw data analysis involves analysing the functional data without any pre-processing or transformation; this approach is suitable for relatively smooth and continuous data over the entire domain. In the filtering approach, the curve shape reconstruction is addressed by assuming that the sample paths belong to a finite-dimensional space generated by a basis function. Finally, adaptive methods combine the previous ones depending on the data's local variability. In this work, the focus will be on the first two categories.

The first step of the analysis involves the segmentation of the original $d$-dimensional time series denoted with $\bm{q}(t)$, where $t$ represents the time defined in the interval $t\in[0, T]$. The dimensions $d$ usually correspond to the number of variables times the number of points of the spatial discretization. Let $\bm{q}_l(t)$ be the $l$-th segment of the time series, where $l=1,\ldots,L$. The segments can overlap with each other by employing a sliding window. It is possible to consider ${\bm{q}_1(t'),\ldots,\bm{q}_L(t')}$ to be a sample of $L$ trajectories, with $t'=t-t_{0l}$ and defined in $t' \in [0,T_l]$. In this domain $\bm{q}_l(t')$ is further discretised into $L_{\mathrm{traj}}$ points giving the discrete observations $\bm{q}_{l,j}$.
Considering the unbiased auto-correlation function defined as \cite{protas2015}:
\begin{equation}\label{Rdef}
R(\tau)=\frac{1}{T-\tau}\int_\tau^{T} \bm{q}(t-\tau)^{\intercal} \bm{q}(t) dt, \qquad \tau \in [0,T),
\end{equation}
$T_l$ is chosen to be proportional to $\tau^*$, with $\tau^*$ being the first time instance in which $R(\tau)=0$. An alternative approach is to select $T_l$ to be proportional to the period of the characteristic frequency.

\begin{figure}
\centering
\includegraphics[trim= 0cm 2cm 0cm 1cm,clip,width=1\textwidth]{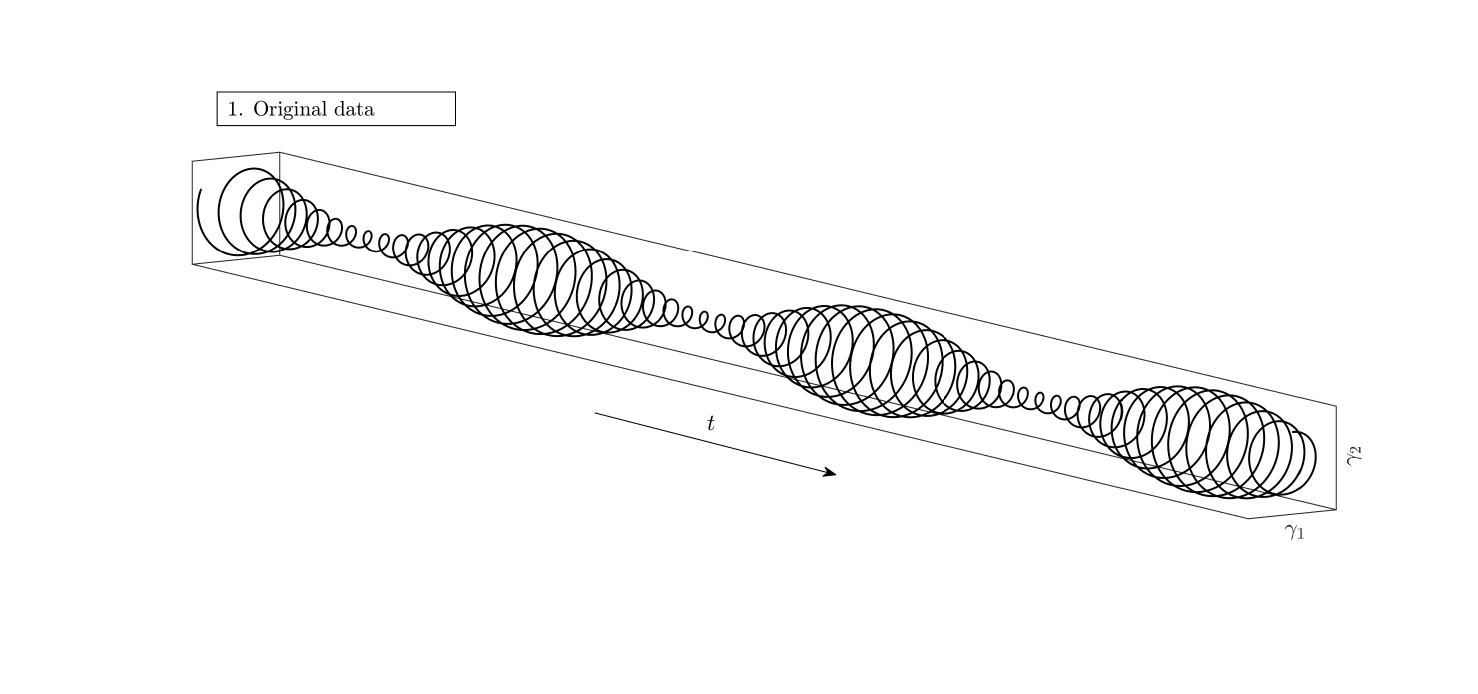}\\
\includegraphics[trim= 0cm 1.3cm 0cm 0.4cm,clip,width=1\textwidth]{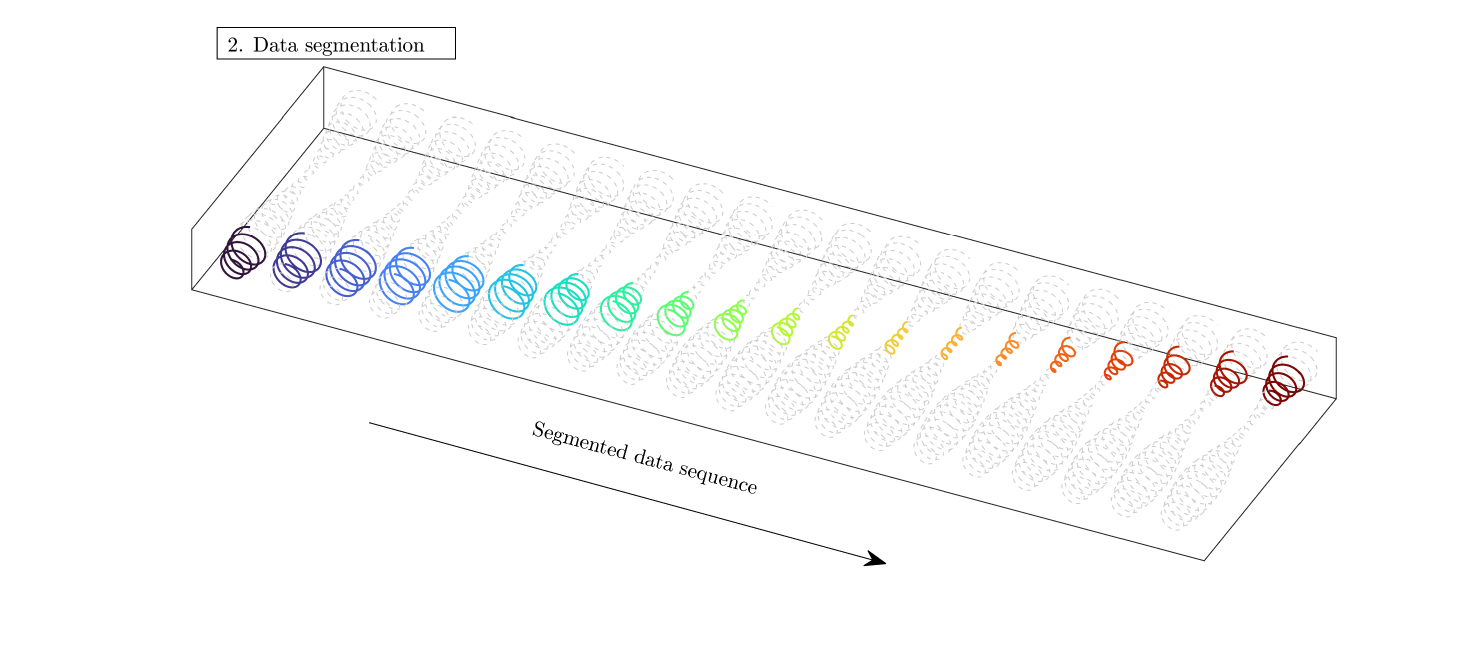}\\
\includegraphics[trim= 0cm 1.2cm 0cm 0.cm,clip,width=1\textwidth]{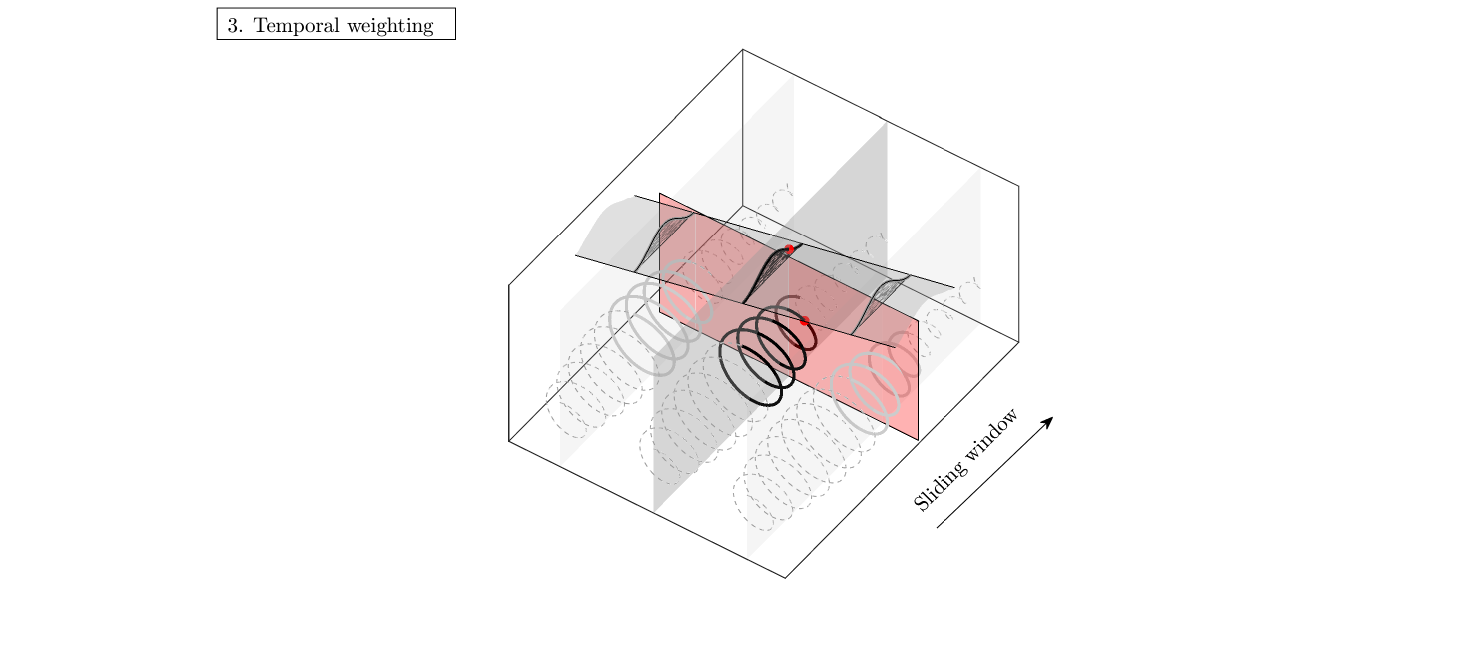}
\caption{Schematic overview of the data segmentation and temporal weighting preceding multivariate functional clustering. The original temporal data is segmented into pieces with a sliding weighted window satisfying the COLA (Constant Overlap-Add) constraint. 
}
\label{fig:weigths}
\end{figure}

Then, the data segments $\bm{q}_l(t')$, in the framework of functional principal component analysis (FPCA), can be approximated as a linear combination of $L$ basis functions, ${\bm{\phi}_1(t'),\ldots,\bm{\phi}_N(t')}$, such that
\begin{equation}\label{fpcaexp}
\bm{q}_l(t') \approx\bm{\mu}(t') + \sum_{i=1}^{L} c_{li} \bm{\phi}_i(t'), \qquad l=1,\ldots,L,
\end{equation}
where $\bm{\mu}(t') = E(\bm{q}_l(t'))$ is estimated by $\bm{\mu}(t')=\frac{1}{L}\sum_{l=1}^{L} \bm{q}_l(t')$ and represents the mean trajectory observed. The coefficients $c_{lk}$ are the principal component scores. 
It is worth noting that at this stage $\bm{\mu}(t')$ and $\bm{\phi}_i(t')$ are $d$-dimensional.
The assumption \eqref{fpcaexp} implies that the functional data lies in a finite-dimensional subspace spanned by the $\bm{\phi}_k$ basis functions.
Functions $\bm{\phi}_i$ can be chosen \emph{a priori} (filtering approaches) and based on raw data employing functional principal component analysis.  
\ref{appFPCA} reports the description of FPCA analysis. Figure~\ref{fig:weigths} presents a schematic representation of the process involving data segmentation and temporal weighting, which precedes the multivariate functional clustering.

\begin{figure}
\centering
\includegraphics[trim= 0cm 0.5cm 0cm 0.1cm,clip,width=1\textwidth]{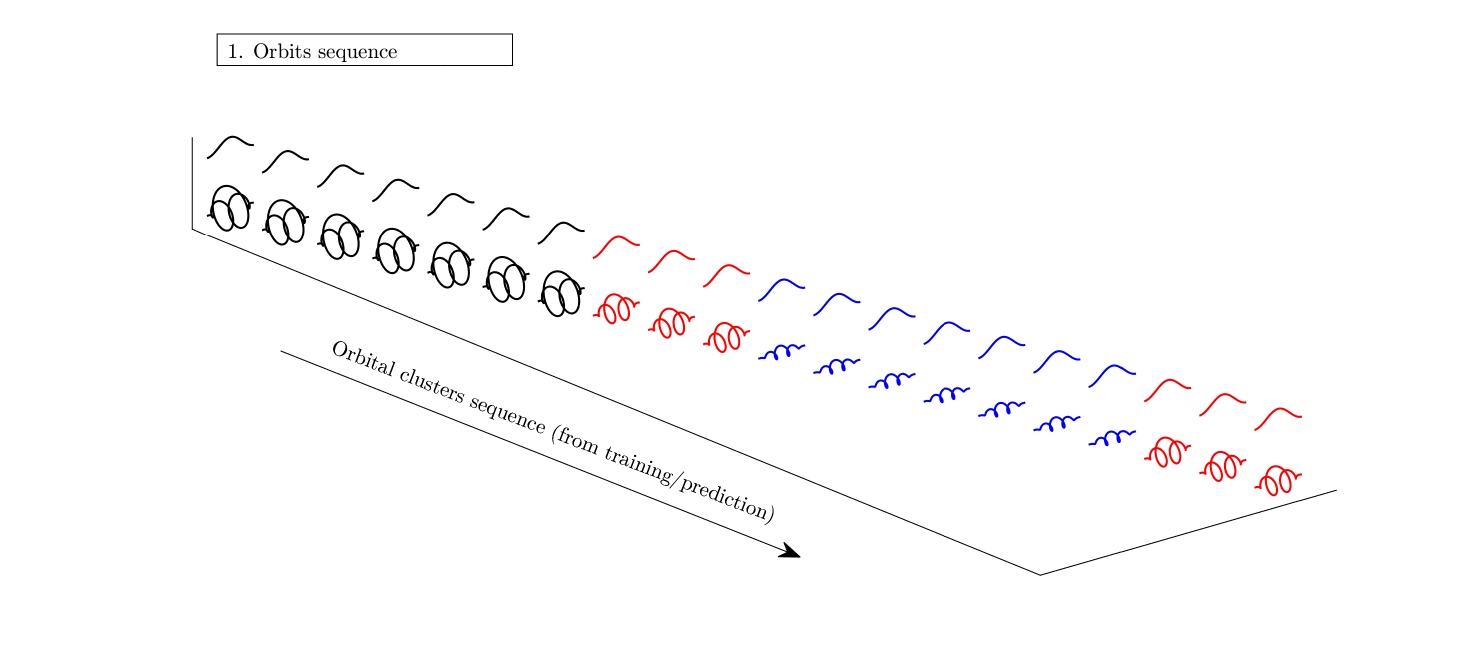}\\
\includegraphics[trim= 0cm 1cm 0cm 0.5cm,clip,width=1\textwidth]{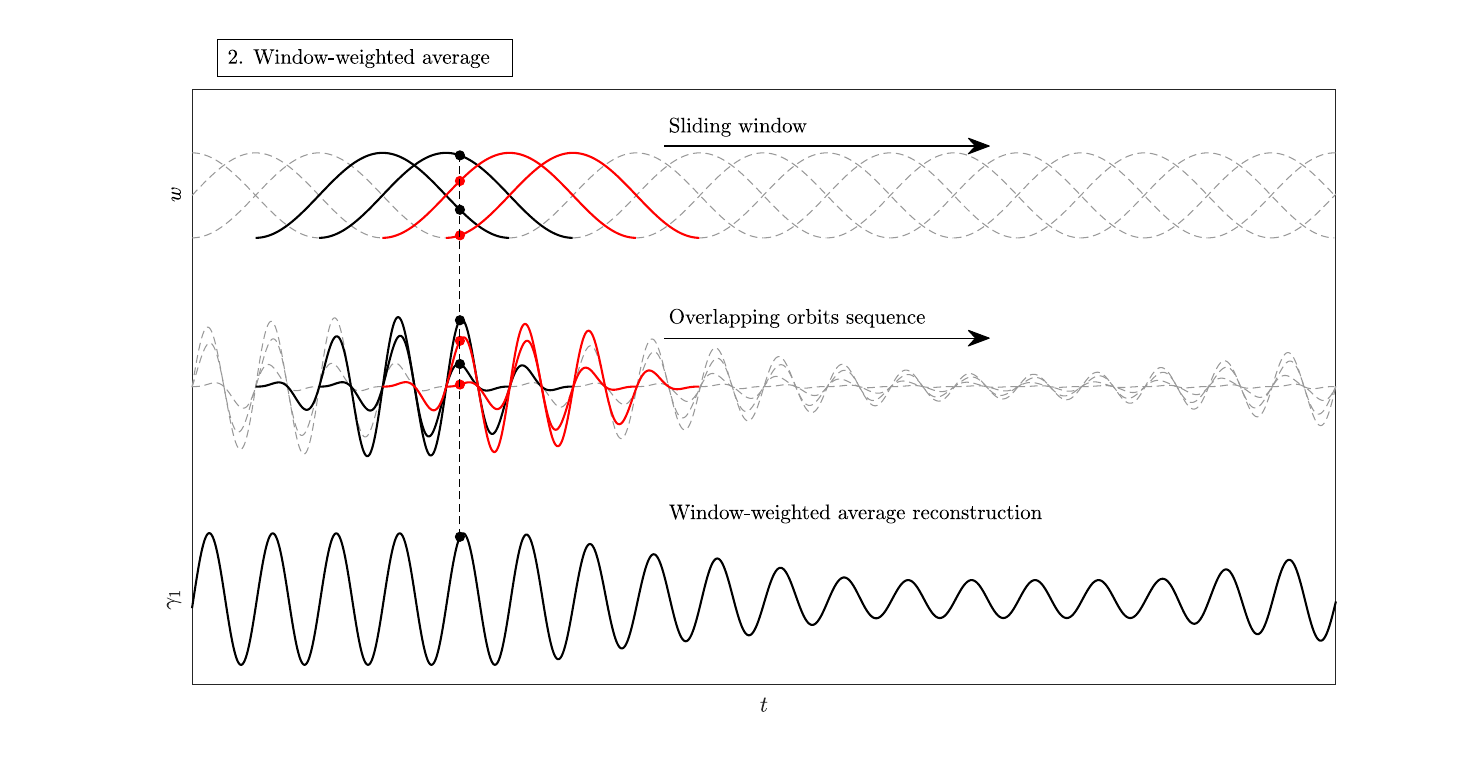}\\
\includegraphics[trim= 0cm 2cm 0cm 1.cm,clip,width=1\textwidth]{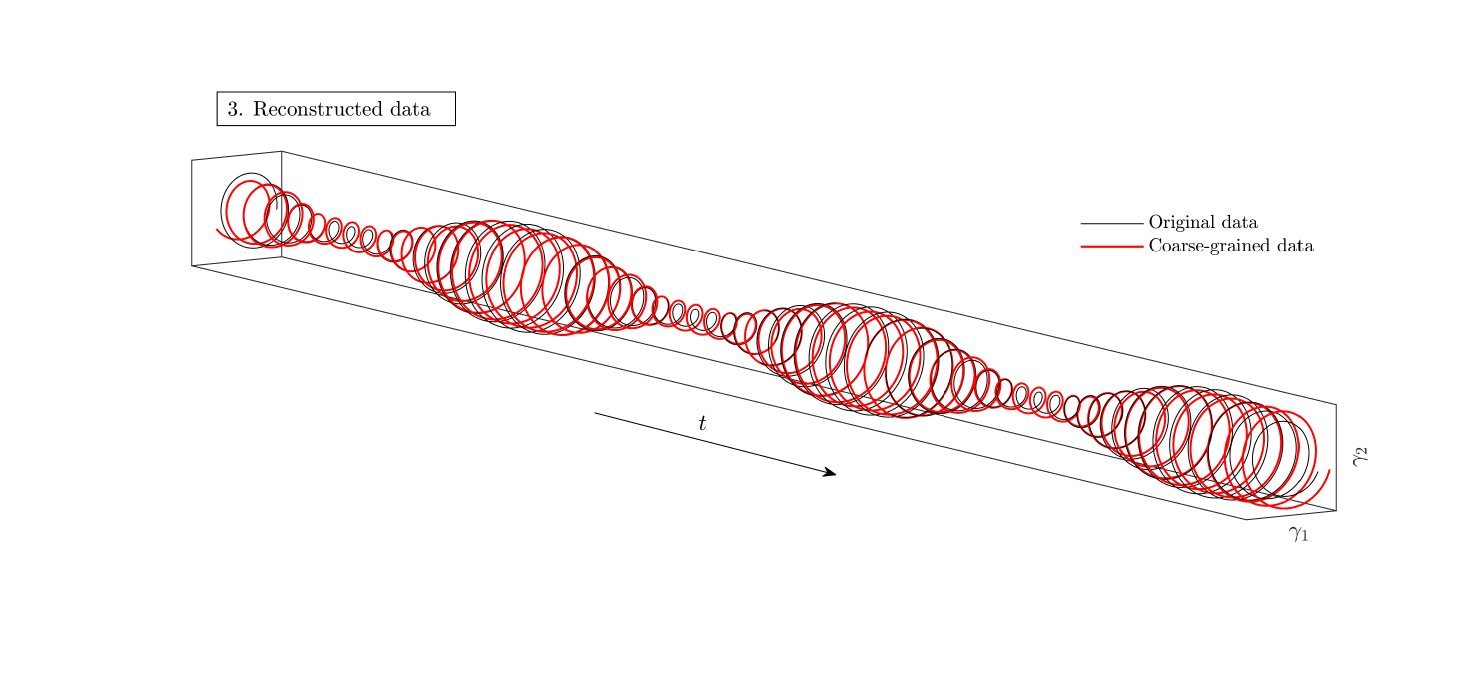}\\

\caption{Schematic overview of the reconstruction procedure from a sequence of orbital clusters. The orbit sequence is colour-coded according to the cluster affiliation. In the middle panel, the reconstruction procedure is sketched for a generic time instance by sliding window averaging.  By window-weighted reconstruction, the oCNM can reproduce the oscillatory trajectory with multiple frequencies.}
\label{fig:reconstruction}
\end{figure}

The truncation of the series \eqref{fpcaexp} to the order $P<L$
\begin{equation}
\bm{q}^\approx_l(t') =\bm{\mu}(t') + \sum_{i=1}^{P} c_{li} \bm{\phi}_i(t'),
\end{equation}
leads to a representation of the ensemble of trajectories containing the essential features of the data. Once the piecewise trajectories have been reconstructed, the segments can be merged to reconstruct the original data. Outside of their domain $t'=t-t_{0l} \in [0,T_l]$, the $\bm{q}^\approx_l(t')$ can be either zero-padded or periodically extended. The original data can be reconstructed with the introduction of the weights $w_{l}(t)$:
\begin{equation}\label{reconstr}
\bm{q}^\approx(t) =\sum_{l=1}^{L} w_{l}(t) \bm{ q}^\approx_l(t).
\end{equation}
The window-weighted average reconstruction is sketched in Figure~\ref{fig:reconstruction}.
The inspection of the equation~\eqref{reconstr} reveals that $w_{l}(t)$ must satisfy the condition
\begin{equation}\label{wcond}
\sum_{l=1}^{L} w_{l}(t)=1, \qquad \forall t.
\end{equation}
In particular the equation~\eqref{wcond} comes from the fact that equation~\eqref{reconstr} must be valid also for a perfect reconstruction involving not only the first $P$ eigenfunctions but the whole basis set $\bm{\phi}_k$, namely when $\bm{ q}^\approx_l \equiv\bm{ q}_l$.

\subsection{Functional clustering}\label{sec:clusteringtr}

The goal of functional clustering is to identify groups of functions that share similar characteristics, such as shape, amplitude, or frequency \citep{Jacques2013}. The first step to cluster functions consists of defining a metric to quantify the similarity or dissimilarity between pairs of functions. The choice of the distance metric can have a significant impact on the results of the clustering analysis; different data types or specific research goals can require different metrics. 

A common choice considers the Euclidean distance between two d-dimensional functions $\bm{f}(t)$ and $\bm{g}(t)$, which is defined as:
\begin{equation}\label{eudist}
d_E^2(\bm{f},\bm{g}) = \int_{a}^{b} (\bm{f}(t)-\bm{g}(t))^{\intercal} (\bm{f}(t)-\bm{g}(t)) dt.
\end{equation}

Once the distance matrix between all pairs of functions has been computed, it is possible to use standard clustering algorithms to group the functions, such as the hierarchical clustering \citep{Nielsen2016}, the spectral clustering \citep{Luxburg2007}, or the $k$-means clustering \citep{MacQueen1967}. 
Here, the $k$-means++ algorithm \citep{arthur2006}, minimising the intra-cluster variance, has been employed. In the framework of orbital clusters $\bm{c}^{\circ}_i (t)$, the intra-cluster functional variance is defined as:
\begin{equation}\label{innerJ}
J^\bullet(\bm{c}^{\circ}_1, \ldots, \bm{c}^{\circ}_K)=\frac{1}{L} \sum_{k=1}^K \sum_{\bm{q}_l \in \mathscr{C}_k}  \int_0^{T_l}\left\|\bm{q}_l(\tau)-\bm{c}^{\circ}_k(\tau) \right\|^2 d\tau,
\end{equation}
The minimisation of $J^\bullet$ gives the optimal orbital centroids ${\bm{c}^{\bullet}_{k=1,\ldots,K}}$ reads
\begin{equation}\label{eq:optCLU}
(\bm{c}^{\bullet}_1, \ldots, \bm{c}^{\bullet}_K) = \underset{\bm{c}^{\circ}_1, \ldots, \bm{c}^{\circ}_K}{\arg\min}\, J^\bullet(\bm{c}^{\circ}_1, \ldots, \bm{c}^{\circ}_K).
\end{equation}

\subsection{Filtering approaches}\label{sec:FACNM}

Filtering methods are a popular class of techniques used in functional data clustering to preprocess the raw data and reduce its complexity. In particular, they leverage on the fact that the functional data $\bm{q}(t)$, and the trajectories segments $\bm{q}_l(t')$, lie on a finite-dimensional space spanned by some temporal basis of functions,
\begin{equation}\label{basiexp}
\bm{q}_l(t') =\sum_{i=1}^{P} \bm{\alpha}_{li} f_i(t'),
\end{equation}
with $\bm{\alpha}_{lk}$ being $d$-dimensional. 

Depending on the choice of $f_k$, there are several types of filtering methods used in functional data clustering, including smoothing \citep{Ramsay2006}, B-splines \citep{Abraham2003}, wavelet-based methods \citep{Giacofci2013}, Fourier-based methods, and others. 
Each method has its strengths and weaknesses and is suitable for specific types of functional data.

\subsubsection{Short-time Fourier transform}\label{sec:STFTCNM}
The Short-Time Fourier Transform (STFT) based method is particularly suited for periodic and quasi-periodic datasets. In \ref{appSTFT}, STFT is introduced and described. In the STFT framework, referring to the equation~\eqref{basiexp}, each $l^{th}$ trajectory is modelled as a superposition of complex exponentials $e^{\mathrm{i}2\pi f t'}$. 
\begin{equation}\label{invSTFTdi}
\bm{q}_l(t') =\sum_{j=1}^{n_f} \hat{\bm{\alpha}}_{lj}e^{\mathrm{i}2\pi f_j t'},
\end{equation}
where $n_f$ is the number of frequencies involved in the expansion and $\hat{\bm{\alpha}}_{lk}\in \mathbb{C}^d$.

Complex valued coefficients $\hat{\bm{\alpha}}_{lk}$ (d-dimensional) are the discrete STFT component and are computed according to the equation~\eqref{diSTFT}. It is worth noticing that for the STFT case, the weights $w$ for the signal reconstruction  are already contained in the definition of $\hat{\bm{\alpha}}_{lk}$, see equation~\eqref{diSTFT}, and the reconstruction of the original data consists in the sum of the local trajectories:
\begin{equation}\label{invSTFTd}
\bm{q}^\approx(t) =\sum_{l=1}^L \bm{q}^\approx_l(t).
\end{equation}

For the clustering procedures, the Euclidean distance between two trajectories $l,j$ can be defined as:
\begin{equation}\label{distFT}
d_{lj}^2 = \sum_{k=1}^{n_f} (\bm{\xi}_{lk} - \bm{\xi}_{jk})^{\intercal}(\bm{\xi}_{lk} - \bm{\xi}_{jk}),
\end{equation}
where $\bm{\xi}_{lk}$ is 
\begin{equation}\label{eq:stateVSTFT}
\bm{\xi}_{lk} = [\Re(\hat \alpha_{lk});\Im(\hat \alpha_{lk})],
\end{equation}
with $\Re$ and $\Im$ being the real and imaginary part operators, respectively. As the standard clustering algorithms deal with real-valued data, the introduction of $\bm{\xi}$ allows employing such techniques for clustering complex-valued data. 

\subsubsection{B-splines}\label{sec:BSCNM}
Functional data clustering using B-splines is a common approach for analysing complex data sets where the observations are functions defined over a continuous domain \citep{Abraham2003}. B-splines are a type of basis function that can represent the functions in the data set, allowing for clustering based on the coefficients of the B-spline expansion. Each trajectory $\bm{q}_l(t')$ is represented as
\begin{equation}\label{bsplinesdef}
\bm{q}_l(t') =\sum_{i=0}^{p} \bm{\beta}_{li} B_i(t'),
\end{equation}
where $ \bm{\beta}_{li} \in \mathbb{R}^d$, $B_i(t')$ is the B-spline orthonormal basis and $p$ is the spline order. Further details on the derivation of the $\bm{\beta}_{li}$ and $B_i(t')$ can be found in \ref{appB}.

The most commonly used distance metric for this approach is the squared $L_2$ norm (Euclidean), which measures the difference between two functions as the sum of the squared differences between their corresponding B-spline coefficients:
\begin{equation}
d_{lj}^2 = \sum_{k=1}^{p} (\bm{\beta}_{lk} - \bm{\beta}_{jk})^{\intercal} (\bm{\beta}_{lk} - \bm{\beta}_{jk})
\end{equation}
This distance metric captures the difference between the shape of the two functions, regardless of their magnitude or location.

\subsubsection{Wavelets}\label{sec:WAVECNM}
Wavelets can be used as a powerful tool for data analysis in functional clustering, particularly for non-stationary signals and data with sharp changes or discontinuities. 
The wavelet transform is a mathematical technique used widely in modal analysis to decompose a signal into its frequency components while maintaining the time location of these components \citep{Farge1992,Schneider2010}. 
	
In this framework, we include a wavelet approach in cluster-based analysis, assuming each function $\bm{q}_i(t')$ can be represented as
\begin{equation}\label{waveexp}
\bm{q}_l(t') = \sum_{r=-N_{\mathrm{lev}}}^{N_{\mathrm{lev}}} \sum_{k=-N_{\mathrm{tra} } }^{N_{\mathrm{tra} }} \bm{\gamma}_{lrk} \psi\left(\frac{t'-2^r k}{2^r}\right),
\end{equation}
where $\bm{\gamma}_{lrk} \in \mathbb{R}^d$ are the coefficients of the wavelet expansion, $N_{\mathrm{lev}}$ is the number of scales levels, $N_{\mathrm{tra}}$ is the number of the translations, and $\psi(t')$ is the mother wavelet function. See \ref{appW} for a detailed derivation of the wavelet transform.

As with Fourier and B-spline methods, the squared $L_2$ norm (Euclidean distance) is often used as the distance metric, which measures the difference between two functions as the sum of the squared differences between their corresponding wavelet coefficients $\bm{\gamma}_{lrk}$:
\begin{equation}
d_{lj}^2 = \sum_{r=-N_{\mathrm{lev}}}^{N_{\mathrm{lev}}} \sum_{k=-N_{\mathrm{tra} } }^{N_{\mathrm{tra} }} (\bm{\gamma}_{lrk} - \bm{\gamma}_{jrk})^{\intercal} (\bm{\gamma}_{lrk} - \bm{\gamma}_{jrk}).
\end{equation}

This distance metric quantifies the difference in the frequency content and temporal location of the features of the two functions, making it particularly suitable for non-stationary and discontinuous signals. The clustering algorithm is then applied to the wavelet coefficients, which capture the most important features of the data. This approach is advantageous in many applications where the raw functional data are too complex to be directly analysed, and the key features of the data can be effectively captured by the wavelet transform. 

However, the choice of the mother wavelet function can significantly affect the clustering results, making it crucial to select a wavelet function that is appropriate for the analysed data. The Daubechies wavelets of order $10$ were used in the present study, as discussed in \ref{appW}.

%
%

\subsection{Orbital network model}
As already said, the clusters are no longer state vector realisations but orbits (functions/trajectories) in this framework. The clustering procedure results in a set of $K$ spatio-temporal clusters, denoted as $ \bm{c}^{\bullet}_k(t')$ with $k= 1, \ldots, K$. This approach offers a richer and more dynamic perspective on the system's evolution, capturing not only the instantaneous state but also the temporal trends and patterns over a specific interval.

\begin{figure}
\centering
\includegraphics[width=0.53\textwidth]{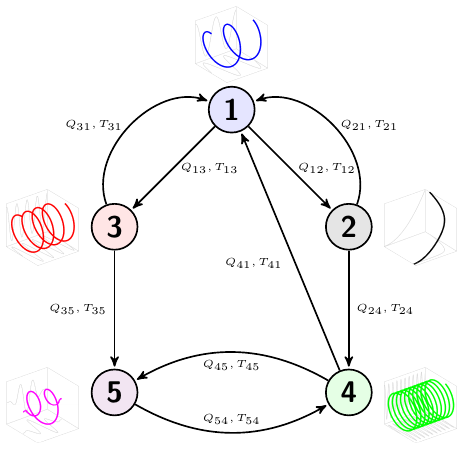}
\caption{Schematic representation of a directed network of clusters in the orbital-based CNM. The directed network represents the potential transitions between different trajectory clusters, each symbolising a distinct temporal behaviour of the system. The small curves drawn next to each node illustrate the general temporal behaviour of each cluster. The arrows between the clusters are labelled with the corresponding transition probabilities $Q_{ij}$ and mean transition times $T_{ij}$.}\label{fignetworksampletra}
\end{figure}

Similarly to the CNM, in the orbital-based Cluster Network Model, the transitions between these orbits or trajectories are modelled as a directed network, as sketched in Figure~\ref{fignetworksampletra}. 
The calculations of transition probabilities $Q_{ij}$ and mean transition times $T_{ij}$ between clusters follow a similar concept as in CNM. $Q_{ij}$ is calulated as the number of transitions from the centroid $ \bm{c}^{\bullet}_j$ to $ \bm{c}^{\bullet}_i$ over the total number of transitions departing from $ \bm{c}^{\bullet}_j$. The mean transition times $T_{ij}$ are the average time the system spends in one trajectory cluster before transitioning to another. 

According to $Q_{ij}$ and $T_{ij}$, a visitation sequence can be created, denoted as $ \bm{c}^{\bullet}_{k_i}$. The signal, similarly to \eqref{reconstr}, can be reconstructed through:
\begin{equation}\label{reconstrclust}
\bm{ q}^\approx(t) =\sum_{j=1}^{L} \alpha_{j}(t)  \bm{c}^{\bullet}_{k_j}(t),
\end{equation}
where $\alpha_{j}(t)$ takes into account the temporal weighting and interpolation coefficients for smoothness. Note that reconstruction of \eqref{reconstrclust} can create discrete jumps at overlapping trajectory times without interpolation.
In analogy to the CNM, this procedure can also be conveniently extended to an order $O$.  This involves considering transition probabilities that depend on the preceding $O$ trajectory clusters, expressed as $Pr (k_{n+1} =i \vert \bm{c}^{\bullet}_{k_n},...,\bm{c}^{\bullet}_{k_{n-O-1}})$. 

We emphasize that the centroids $\bm{c}^{\bullet}_{k}$ computed through linear clustering are not generally guaranteed to lie on the underlying nonlinear solution manifold. As a result, the interpolation in Eq.~\eqref{eq:reco}, \eqref{reconstr} and \eqref{reconstrclust} may introduce deviations from the manifold, particularly when the centroids are sparsely distributed or the manifold exhibits significant curvature. This limitation is partially mitigated by the functional approach of our reconstruction strategy, as illustrated in Fig.~\ref{fig:rawclust}. Nonetheless, ensuring accurate manifold approximation may require a sufficient number of clusters to capture local geometry effectively. Recent works, such as \cite{Kelshaw2025}, explore nonlinear clustering approaches to better preserve manifold structure. However, these methods are typically applied to state-space snapshots, while our approach leverages short-time trajectory segments that incorporate local dynamical information.

The actual flow computations are based on a lossless FPCA, as elaborated in \ref{appFPCA}. 

\subsection{Validation}\label{sec:validation}
The performance of the model is evaluated by the Root Mean Square Error (RMSE) of the auto-correlation function $R(\tau)$, the asymptotic cluster probability, and the representation error.

The auto-correlation function, as defined in \eqref{Rdef}, is used to avoid issues of comparing two trajectories directly due to phase mismatch \citep{Hou2022,fernex2020sa}. 
The Root Mean Square Error (RMSE) of the auto-correlation function is defined as:
\begin{equation}\label{eq:RMSE}
\mathrm{RMSE}=\sqrt{\frac{1}{N_R} \sum_{n=1}^{N_R}\left(R(\tau_n)-\hat{R}(\tau_n)\right)^2},
\end{equation}
where $\hat{R}$ is the model reconstruction and $N_R$ is the number of time lags $\tau_n$.

The asymptotic probability $p_i^{\infty}$ to be in a cluster $i$ can be estimated with
\begin{equation}
p_i^{\infty} = \frac{\sum \tau_i}{T_0},
\end{equation}
where $T_0$ is a sufficiently long time horizon simulated by the model, and $\sum \tau_i$ is the cumulative residence time from \eqref{eq:residence}. The vector containing all the $p_i^{\infty}$ indicates whether the predicted trajectories could populate the phase space similarly to the original data, thus providing insight into the transition error of the modelling method \citep{Hou2022}.

The representation error of the reconstructed dynamics, denoted by $E_r$, is defined by
\begin{equation}\label{eq:reprerr}
E_r=\frac{1}{M} \sum_{m=1}^M D_{\mathscr{T}}^m,
\end{equation}
where $D_{\mathscr{T}}^m$ is defined as the minimal distance from the snapshot $\bm{q}^m$ to all the states $\bm{q}^n$ of the reconstructed trajectory $\mathscr{T}$, which reads
\begin{equation}
D_{\mathscr{T}}^m=\min _{\bm{q}^n \in \mathscr{T}}\left\|\bm{q}^m-\bm{q}^n\right\|_{\Omega}.
\end{equation}
The representation error $E_r$ indicates how well the model reconstructs the dynamics.

A key parameter for the model is the number of clusters $K$. A model with a few clusters can effectively capture the dominant transition dynamics behaviour. However, larger representation errors are induced as fewer clusters represent the dynamics, and the snapshot details are eliminated. Conversely, a large $K$ allows the CNM to model the transition dynamics with more detail, but the transition relationships become more complex and often meaningless \citep{li2020jfm}. \cite{Nair2019jfm} proposed an optimal choice for the cluster number determined by the F-test \citep{Hand2012}, which considers the ratio of inter-cluster variance to the total variance. In the framework of orbital clusters $\bm{c}^{\bullet}_i (t)$ the inter-cluster variance $V^\bullet$ is defined as
\begin{equation}\label{interJ}
V^\bullet=\frac{1}{L} \sum_{k=1}^K L_k \int_0^{T_l} \left\|\bm{c}^{\bullet}_k (t)-\bm{\mu}(t)\right\|^2 dt,
\end{equation}
where $L_k$ is the number of segments pertaining to the cluster $k$. It is worth noticing that in the definitions \eqref{innerJ} and \eqref{interJ} when using filtering methods, the expansion coefficients $\bm{\alpha}_{ik}$ are employed. Except when otherwise specified, in this work, the minimum number of clusters has been chosen such that $V^\bullet/(V^\bullet+J^\bullet)>0.9$, which corresponds to resolving at least 90\% of the flow fluctuations after the cluster-coarse graining.

The performance of the proposed method relies on the assumption that the training dataset sufficiently resolves the invariant measure of the system \cite{Kac1960}. For ergodic systems, this implies that the trajectory used for training must be long enough to adequately sample the relevant regions of the attractor \cite{Bucci2023}. If the data fails to capture the full statistical structure of the dynamics—due to limited time span or insufficient exploration of phase space—then the clustering and local reconstructions may misrepresent the system’s true behavior. This is a common limitation of data-driven models, and it underlines the importance of collecting representative and ergodically rich datasets for reliable reduced-order modeling.

The source code for the oCNM method is available in the GitHub repository \cite{oCNM_repo}.
\section{Results}\label{sec:results}

The oCNM is first exemplified in Section \ref{sec:testcase} by Landau's equation \citep{Dusek1994,noack2003} and a synthetic multi-scale signal. In Section \ref{sec:pinball}, the method is also tested with force coefficients and flow fields of the fluidic pinball \citep{Deng2019,deng2021jfm}. By comparing the CNM and oCNM, we underline the applicability and usefulness of the method in dealing with complex fluid dynamics. These carefully designed test suites provide a clear understanding of the method and behaviour under known conditions.

\subsection{Orbital CNM of the Stuart-Landau oscillator}\label{sec:testcase}

The Landau equations, serving as a model system, exhibit the rich dynamics characteristic of many real-world phenomena \citep{Haken1978}. They are utilised in this work as a \textit{toy problem} to demonstrate the purpose and the capabilities of our orbital CNM (oCNM).
The governing equations are defined as:
\begin{equation}\label{eq:landauprob}
	\begin{aligned}
		\dot{a}_1 &= \sigma a_1 - \omega a_2, \\
		\dot{a}_2 &= \sigma a_2 + \omega a_1,
	\end{aligned}
\end{equation}
where $a_1$ and $a_2$ are the state variables, $\sigma$ is a damping term computed as 
\begin{equation}
	\sigma=r - a_1^2 - a_2^2,
\end{equation}
$\omega$ and $r$ indicate the angular frequency and the growth rate that influences the system behaviour. 
This system exhibits a range of rich dynamics, including sustained oscillations for certain parameter values, providing a comprehensive yet tractable test case for validating our oCNM. 
In this section, we set these parameters to $\omega = \pi$ and $r=0.1$.

\begin{figure}
	\centering
	\begin{minipage}{1\textwidth}
			\centering
		\subfloat[]{\includegraphics[trim= 0cm 0cm 0cm 0cm,clip,height=5.cm]{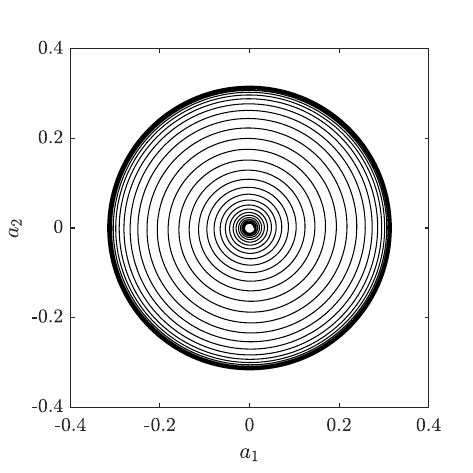}}
		\subfloat[]{\includegraphics[trim= 0cm 0cm 0cm 0cm,clip,height=5.cm]{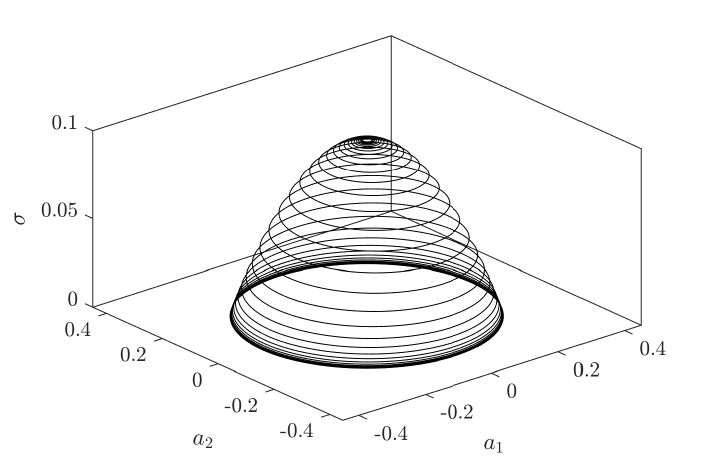}}\\
	\end{minipage}
	\caption{Panel (a): Phase space of the Landau equations with initial conditions $a_1(0)=0.01$ and $a_2(0)=0$, showing the trajectory spiralling out to a limit cycle. Panel (b): Three-dimensional plot of the trajectory in the $(\sigma$, $a_1$, $a_2)$ space. It illustrates that as the system settles into the limit cycle, forming a paraboloid manifold, the variable $\sigma$ approaches zero.}
	\label{fig:landaubase}
\end{figure}

Considering the initial conditions $a_1(0)=0.01$ and $a_2(0)=0$, it is possible to highlight a distinctive feature of these equations: the trajectory converges to a limit cycle with a radius equal to $\sqrt{r}$, indicating the tendency to oscillate. Figure~\ref{fig:landaubase} visually represents this behaviour. 
In Figure~\ref{fig:landaubase}(a), the phase space depicts the trajectory path toward the limit cycle. It is interesting to note how the system starts from nearly the origin and gradually spirals out to the limit cycle. Figure~\ref{fig:landaubase}(b) showcases a three-dimensional plot of the variables $\sigma$, $a_1$, and $a_2$, forming a paraboloid manifold. This indicates that the trajectory settles into the limit cycle as $\sigma$ approaches zero, reinforcing the notion of the system entering a state of sustained oscillation.

After obtaining the data from the Landau system, the CNM has been applied to it. For this analysis, we select a model order of $L=1$. $8000$ snapshots have been collected with a $\Delta t = 0.01$. We explore the performance of the model with different numbers of clusters $K$, specifically, $K=10$, $K=20$, and $K=100$. With $K=10$, the intercluster variance $V^\bullet/(V^\bullet+J^\bullet)$ is approximately $90\%$. 

\begin{figure}
	\centering
	\begin{minipage}{1\textwidth}
		\subfloat[$K=10$]{\includegraphics[trim= 0cm 0cm 0cm 0cm,clip,width=0.33\columnwidth]{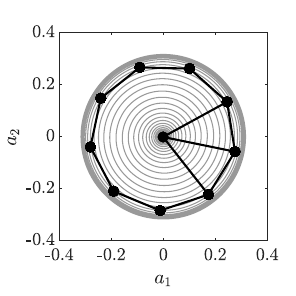}}
		\subfloat[$K=20$]{\includegraphics[trim= 0cm 0cm 0cm 0cm,clip,width=0.33\columnwidth]{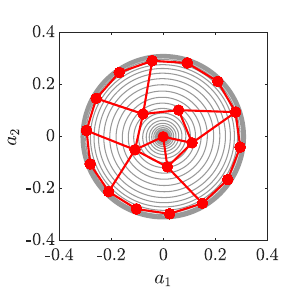}}
		\subfloat[$K=100$]{\includegraphics[trim= 0cm 0cm 0cm 0cm,clip,width=0.33\columnwidth]{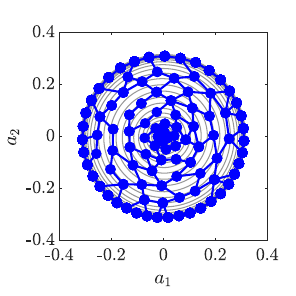}}
		\\
	\end{minipage}
	\caption{Phase space plots ($a_1$, $a_2$) with the true dynamics of the Landau system represented in grey. The model dynamics from the CNM with $K=10$, $K=20$, and $K=100$ are overlaid in black, red, and blue, respectively. Panel (a): $K=10$, Panel (b): $K=20$, Panel (c): $K=100$. It can be observed that the characteristic behaviour of amplitude selection is not well captured, even with the highest number of clusters.}
	\label{fig:landauCNM}
\end{figure}

Figure~\ref{fig:landauCNM} displays the phase space with the true dynamics represented in grey. 
The model dynamics with $K=10$, $K=20$, and $K=100$ are overlaid in black, red, and blue, respectively. The motion between cluster sequences (represented by continuous lines) is smoothed using equation~\eqref{eq:reco}. However, it is noteworthy that the CNM often leads to non-physical transitions, making it less reliable for accurately modelling the dynamics of Landau system. It is noticeable that even with the highest number of clusters ($K=100$), the CNM struggles to adequately capture the characteristic behaviour of amplitude selection in Landau equations.

Transitioning from CNM to orbital CNM (oCNM) improves capturing the intrinsic dynamics of the Landau system, including transitions and amplitude selection mechanisms. Each trajectory segment is composed of $L_{\mathrm{traj}}=200$ snapshots. 
For the spline basis, we choose a third-order spline, with comprehensive details available in \ref{appB}. For the STFT basis, a rectangular window is selected, refer to \ref{appSTFT}. The wavelet CNM employs a Daubechies wavelet, specifically 'db10', configured with $5$ levels; for an in-depth discussion on this, refer to \ref{appW}.

\begin{figure}
	\centering
	\begin{minipage}{1\textwidth}
		\subfloat[]{\includegraphics[trim= 0cm 0cm 0cm 0cm,clip,width=1\columnwidth]{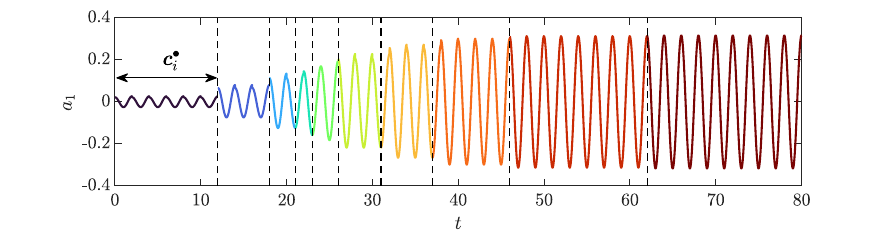}}\\
		\subfloat[]{\includegraphics[trim= 0cm 0cm 0cm 0cm,clip,width=1\columnwidth]{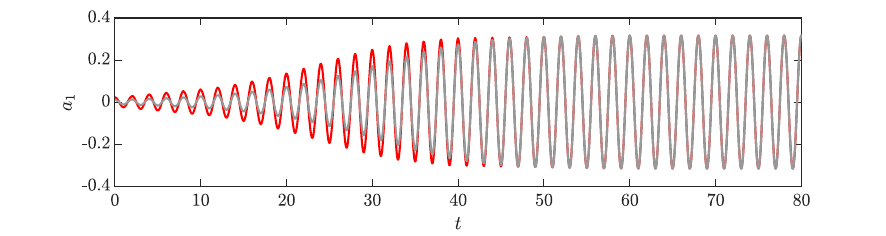}}
		\\
	\end{minipage}
	\caption{Demonstration of the oCNM with $K=10$ orbital clusters. Panel (a): sequence of jumps among the orbital cluster centroids (different colours) with the accompanying residence times denoted by vertical dashed lines. Panel (b): comparison of the original dynamics (in grey) with the smoothed representation (in red) using Equation~\eqref{reconstrclust}.}
	\label{fig:landauTCNM}
\end{figure}

Figure~\ref{fig:landauTCNM} demonstrates the proficiency of the oCNM using spline basis functions and adopting $10$ trajectory clusters. 
In Figure~\ref{fig:landauTCNM}(a), the oCNM coarse-graining is illustrated through a sequence of jumps between the orbital cluster centroids, represented by distinct colours. These jumps are characterised by the residence times in each cluster, which are visualised through vertical dashed lines. To transition smoothly between the discrete jumps of trajectory segments, Equation~\eqref{reconstrclust} is used. As a result of this smoothing technique, Figure~\ref{fig:landauTCNM}(b) shows the refined representation (in red) with the original data (in grey). Even with only 10 clusters, the oCNM captures the underlying dynamics of the system.

\begin{figure}
	\centering
	\begin{minipage}{1\textwidth}
		\subfloat[ CNM]{\includegraphics[trim= 0cm 0cm 0cm 0cm,clip,width=0.49\columnwidth]{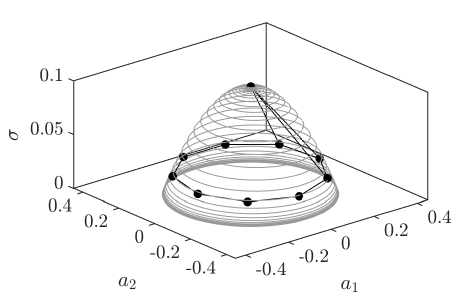}}
		\subfloat[Spline oCNM]{\includegraphics[trim= 0cm 0cm 0cm 0cm,clip,width=0.49\columnwidth]{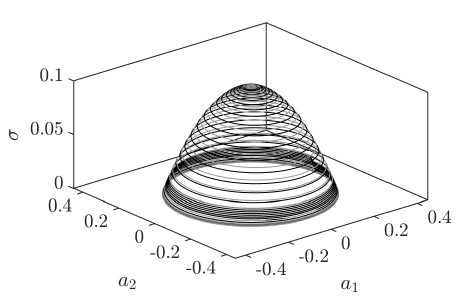}}\\
		\subfloat[STFT oCNM]{\includegraphics[trim= 0cm 0cm 0cm 0cm,clip,width=0.49\columnwidth]{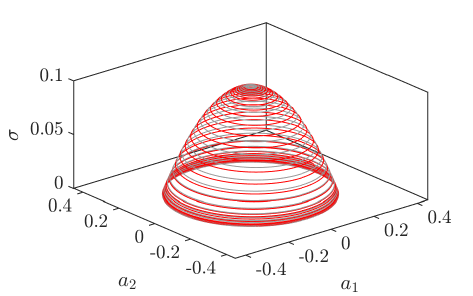}}
		\subfloat[Wavelets oCNM]{\includegraphics[trim= 0cm 0cm 0cm 0cm,clip,width=0.49\columnwidth]{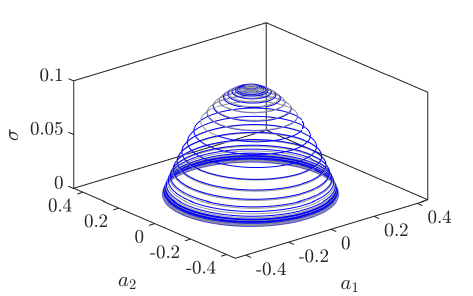}}\\
		\\
	\end{minipage}
	\caption{Comparison of different CNM variants in the space $(a_1, a_2, \sigma)$ for Landau's system. The true data is represented in grey across all panels. Panel (a) showcases the  CNM (solid-point black), Panel (b) illustrates the spline oCNM (black), Panel (c) demonstrates the STFT CNM (red), and Panel (d) displays the Wavelets oCNM (blue). This visual representation underscores the enhanced capability of the oCNM in its various forms over the CNM.}
	\label{fig:comparisonCNM}
\end{figure}

In Figure~\ref{fig:comparisonCNM}, a three-dimensional plot provides a comparative perspective on the performance of different CNM variants in capturing the temporal dynamics of the Landau system. 
Panel (a) represents the  CNM, while panels (b), (c), and (d) depict the spline oCNM, STFT oCNM, and Wavelets oCNM, respectively. A cluster number of $K=10$ has been used for all these representations. The true dynamics of the system are displayed in grey across all panels, providing a baseline for comparison. It becomes evident that the orbital CNM, across its various adaptations, demonstrates superior capability in accurately capturing this system's temporal behaviour and amplitude selection mechanism, as compared to the CNM.

\begin{figure}
	\centering
	\includegraphics[trim= 0cm 0cm 0cm 0cm,clip,width=1\columnwidth]{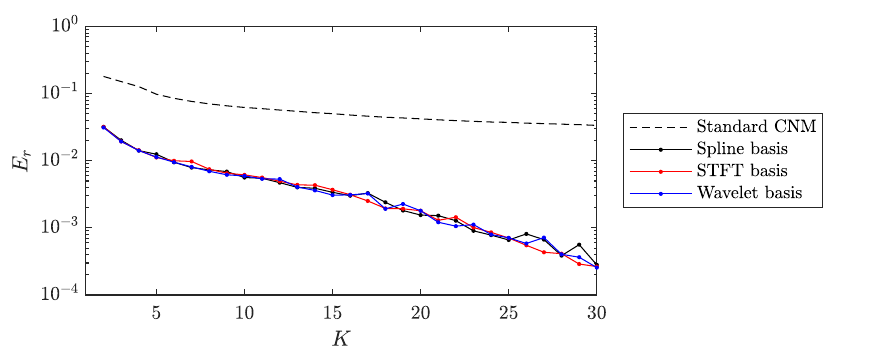}
	\caption{Comparison of representation error $E_r$ for different CNM variants and different values of $K$.}
	\label{fig:errorComparison}
\end{figure}

The distinctions between the CNM and the oCNMs are accentuated when examining Figure~\ref{fig:errorComparison}. 
The representation error $E_r$, as defined in equation~\eqref{eq:reprerr}, is plotted in this figure. It becomes evident that the oCNMs consistently yield a representation error at least an order of magnitude smaller than the  CNM. Such a dramatic reduction in error underscores the efficacy and superiority of the trajectory-based approaches in capturing the inherent dynamics of the system.

To further underscore the strengths of the oCNM, it is instructive to analyse the following time-modulated signal:
\begin{equation}\label{eq:signalb1}
	b_1 = A(t) \cos(2\pi f_1 t + \theta(t))
\end{equation}
\begin{equation}\label{eq:signalb2}
	b_2 = A(t) \sin(2\pi f_1 t + \theta(t))
\end{equation}
where
\begin{equation}
	A(t) = \cos(2\pi f_2 t),
\end{equation}
\begin{equation}
	\theta(t) = c_0 2\pi \cos(2\pi f_3 t)
\end{equation}
and $c_0$ is a constant. The chosen parameters are $f_1=100$, $f_2=2$, $f_3=0.25$. A number of samples $n_t=10^6$ has been saved within a simulation time $t_{\text{fin}}=100$, and with a $\Delta t = 10^{-4}$. 

\begin{figure}
	\centering
	\includegraphics[trim= 0cm 0cm 0cm 0cm,clip,width=0.22\columnwidth]{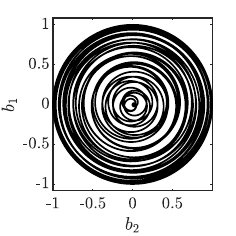}\includegraphics[trim= 0cm 0cm 0cm 0cm,clip,width=0.77\columnwidth]{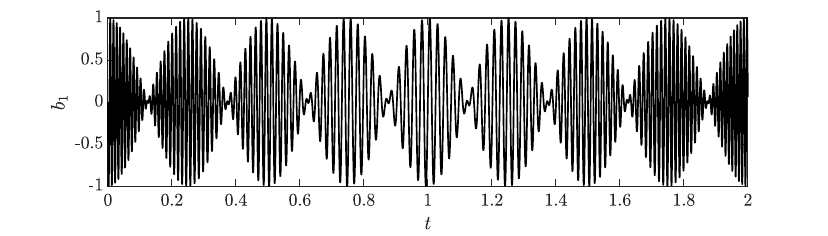}
	\caption{Analysis of the time-modulated signal. The phase space representation (left) illustrates the evolution of the system in the $(b_1, b_2)$ plane. The right panel depicts the variation of the signal $b_1$ over time, revealing its temporal complexities. This signal offers insights into the behaviour of Landau's equations \eqref{eq:landauprob} with time-dependent parameters.} 
	\label{fig:modulatedSignal}
\end{figure}

This particular signal is characterised by both frequency and amplitude modulation. It can be interpreted as a post-transent solution of the Landau's equations (\ref{eq:landauprob}) if one assumes that the parameters $\omega$, and $l$ vary over time. Figure~\ref{fig:modulatedSignal} showcases the phase space representation of this signal and the temporal evolution of $b_1(t)$.

\begin{figure}
	\centering
	\includegraphics[trim= 0cm 0cm 0cm 0cm,clip,width=0.9\columnwidth]{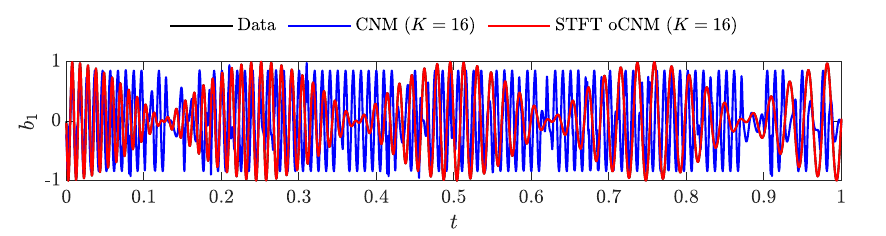}
	\caption{Comparison of the time evolution of $b_1$: the original data (in black), the CNM (in blue), and the STFT oCNM (in red).
		Both CNM methods employ $K=16$ clusters.}
	\label{fig:CNMcomparisonreco}
\end{figure}

Following the representation in Figure~\ref{fig:modulatedSignal}, we investigate the application of both the CNM and the oCNM. For this analysis, the STFT oCNM is chosen. Both methods employ $K=16$ clusters. For the STFT oCNM, segment length is set at $L_{\mathrm{traj}}=5000$, with segment overlaps accounting for $75\%$ of the segment length. A Hann window is also employed, ensuring that the COLA (Constant Overlap-Add) constraint is satisfied. Figure~\ref{fig:CNMcomparisonreco} showcases the original $b_1$ data in black, the output from the CNM in blue, and the STFT oCNM results in red.
Notably, while the CNM struggles to accurately capture the dynamics along with the amplitude and frequency modulation, the STFT oCNM, even with only $K=16$, is indistinguishable from the original data.

\begin{figure}
	\centering
	\includegraphics[trim= 0cm 0cm 0cm 0cm,clip,width=0.8\columnwidth]{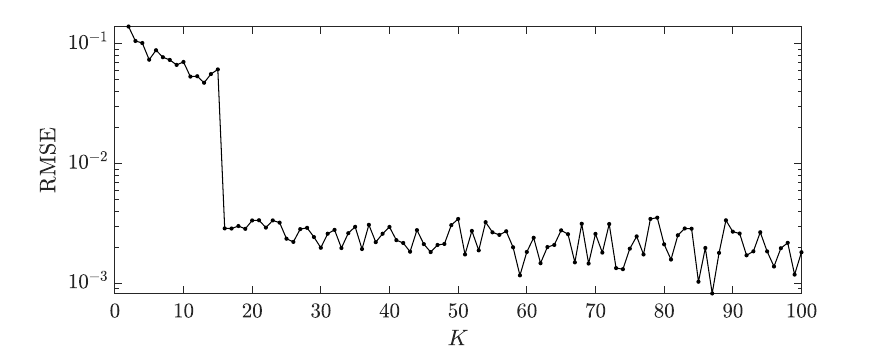}\
	\caption{RMSE values vs $K$. The sharp decrease at $K=16$ underscores its optimal choice for this analysis.}
	\label{fig:RMSEplot}
\end{figure}

The selection of $K=16$ is due to the Root Mean Square Error (RMSE) as defined in equation~\eqref{eq:RMSE}. Figure~\ref{fig:RMSEplot} presents the RMSE values across a range of $K$. 
It is evident that at $K=16$, there is a significant drop in RMSE, which is then followed by a gradual decrease as $K$ is further incremented.

\begin{figure}
	\centering
	\begin{minipage}{1\textwidth}
		\includegraphics[trim= 0cm 0.85cm 0cm 0.2cm,clip,width=1\columnwidth]{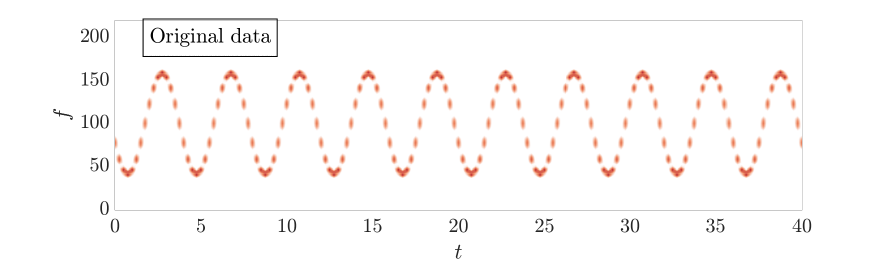}\\
		\includegraphics[trim= 0cm 0.85cm 0cm 0.2cm,clip,width=1\columnwidth]{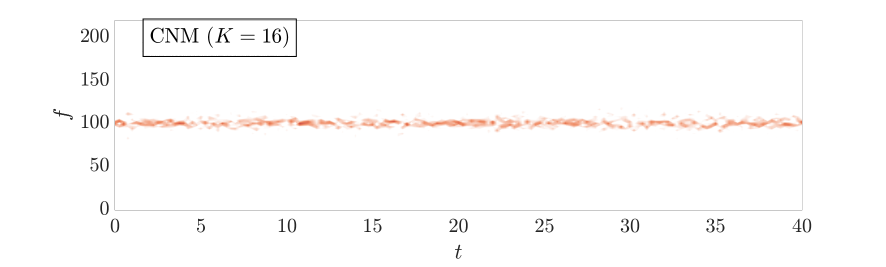}\\
		\includegraphics[trim= 0cm 0cm 0cm 0.2cm,clip,width=1\columnwidth]{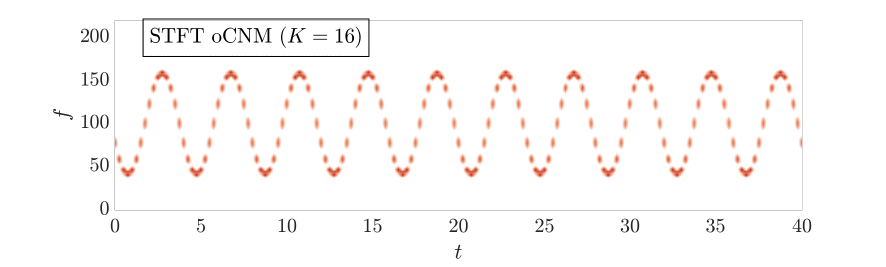}\\
	\end{minipage}
	\caption{Comparison of spectrograms for the original $b_1$ signal, top panel, the CNM, middle panel, and the STFT oCNM, bottom panel.}
	\label{fig:spectrogramComparison}
\end{figure}

As previously mentioned, the signal defined by equations~\eqref{eq:signalb1} and \eqref{eq:signalb2} exhibits frequency modulation over time. Insight into the temporal behaviour of the signal can be obtained from its spectrogram, that is, the amplitude of $\bm{\xi}_{ik}$, where the index $i$ denotes the frequencies and $k$ spans time. Figure~\ref{fig:spectrogramComparison} depicts the spectrogram of the original $b_1$, compared to those from the CNM and the STFT oCNM. This representation shows that the STFT oCNM perfectly captures the frequency modulation of the original signal, while the CNM cannot achieve this purpose. 

\begin{figure}
	\centering
	\includegraphics[trim= 0cm 0cm 0cm 0.7cm,clip,width=0.8\columnwidth]{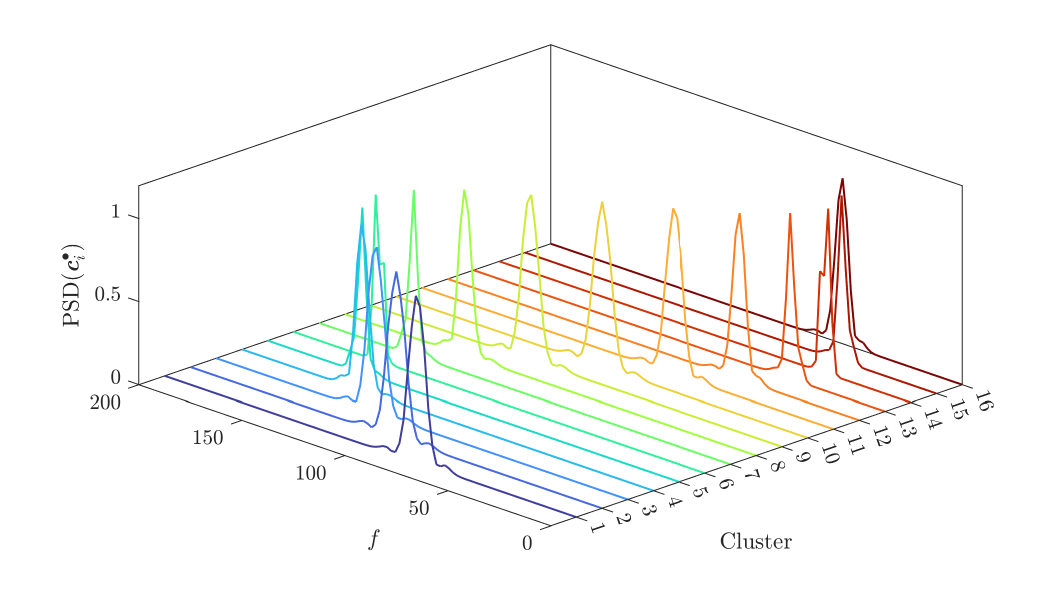}
	\caption{Normalised PSD of each orbital cluster centroid $\mathrm{PSD}(\bm{c}^{\bullet}_i)$. }
	\label{fig:centroidPSD}
\end{figure}

It is instructive to examine the trajectory cluster centroids, $\bm{c}^{\bullet}_i$. 
The spectral content intrinsic to each $\bm{c}^{\bullet}_i$ can be discerned from its power spectral density (PSD). Figure~\ref{fig:centroidPSD} presents the PSD of each centroid, normalised with respect to its respective peak value. This representation clearly illustrates that each centroid is associated with a distinctive frequency. In this context, the centroids are presented in the sequence corresponding to their manifestation in the data.

\begin{figure}
	\setlength\tabcolsep{1pt} 
	\begin{tabular}{l l} 
		\hline
		\vspace{-10pt}	& \\
		$\bm{c}^{\bullet}_2$ & $\bm{c}^{\bullet}_4$\\
		\includegraphics[trim= 0.0cm 0cm 0.0cm 0cm,clip,width=0.1425\textwidth]{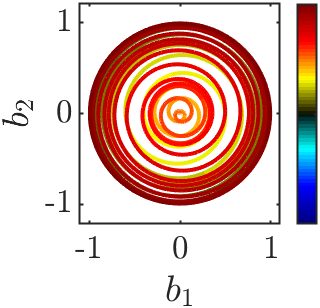} \includegraphics[trim=0.0cm 0cm 0.0cm 0cm,clip,width=0.3325\textwidth]{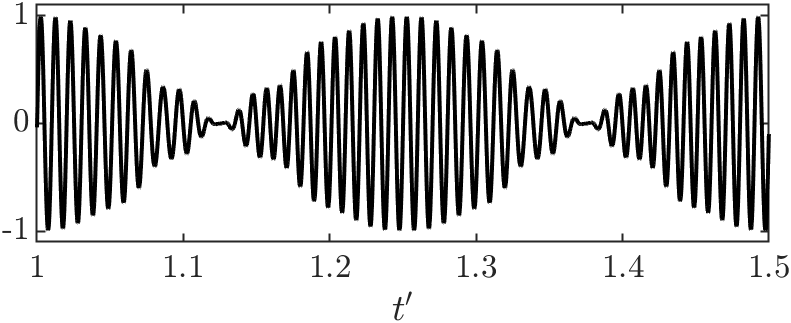} & \includegraphics[trim=0.0cm 0cm 0.0cm 0cm,clip,width=0.1425\textwidth]{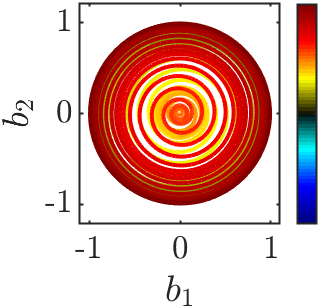} \includegraphics[trim= 0.0cm 0cm 0.0cm 0cm,clip,width=0.3325\textwidth]{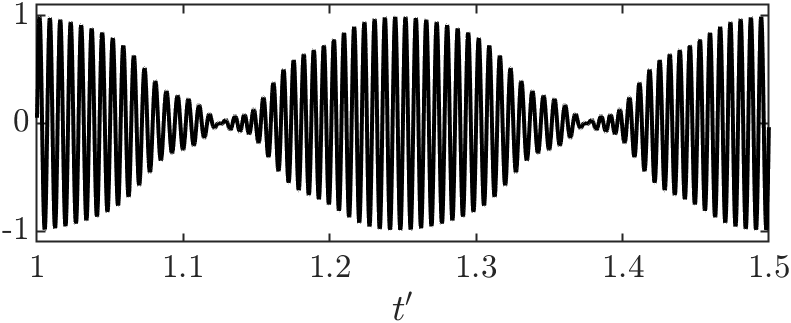} \\
		$\bm{c}^{\bullet}_6$ & $\bm{c}^{\bullet}_8$\\
		\includegraphics[trim= 0.0cm 0cm 0.0cm 0cm,clip,width=0.1425\textwidth]{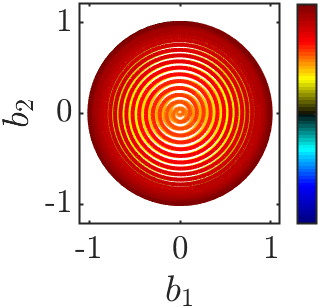} \includegraphics[trim=0.0cm 0cm 0.0cm 0cmcm 0cm 0.5cm 0cm,clip,width=0.3325\textwidth]{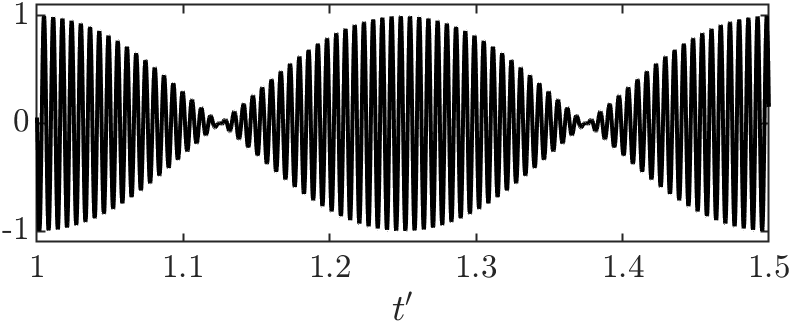} & \includegraphics[trim= 0.0cm 0cm 0.0cm 0cm,clip,width=0.1425\textwidth]{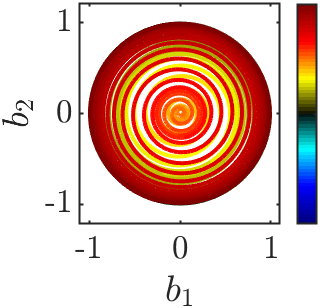} \includegraphics[trim=0.0cm 0cm 0.0cm 0cm,clip,width=0.3325\textwidth]{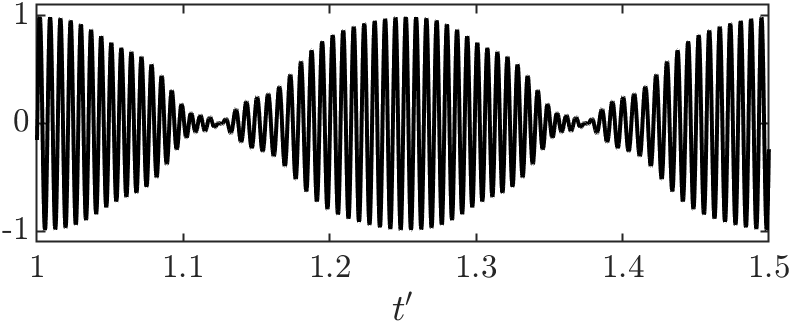} \\
		$\bm{c}^{\bullet}_{10}$ & $\bm{c}^{\bullet}_{12}$\\
		\includegraphics[trim= 0.0cm 0cm 0.0cm 0cm,clip,width=0.1425\textwidth]{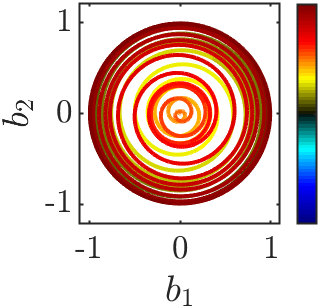} \includegraphics[trim= 0.0cm 0cm 0.0cm 0cm,clip,width=0.3325\textwidth]{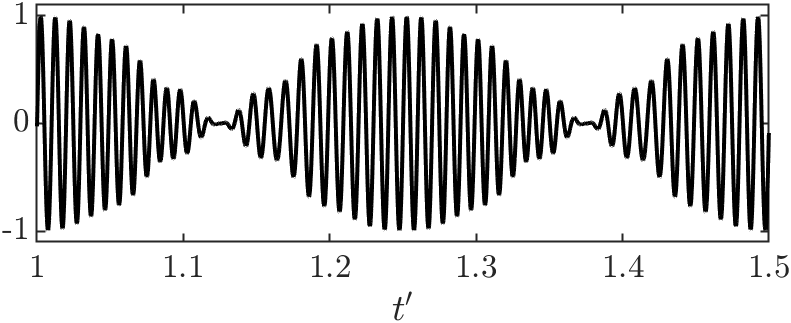} & \includegraphics[trim= 0.0cm 0cm 0.0cm 0cm,clip,width=0.1425\textwidth]{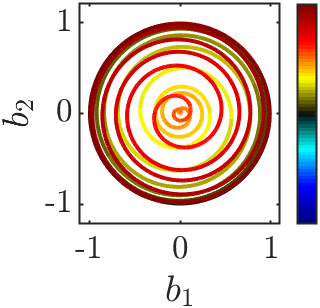} \includegraphics[trim= 0.0cm 0cm 0.0cm 0cm,clip,width=0.3325\textwidth]{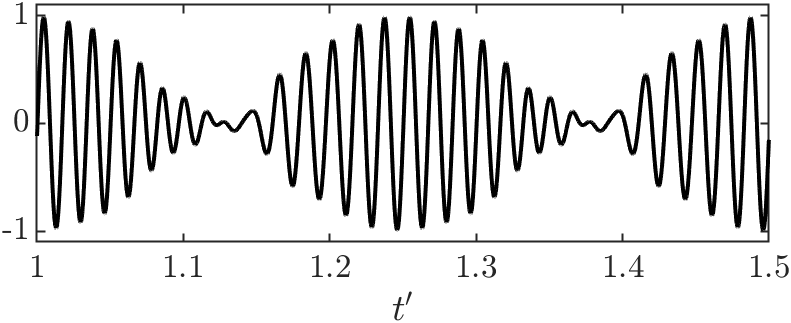}\\
		$\bm{c}^{\bullet}_{14}$ & $\bm{c}^{\bullet}_{16}$\\
		\includegraphics[trim= 0.0cm 0cm 0.0cm 0cm,clip,width=0.1425\textwidth]{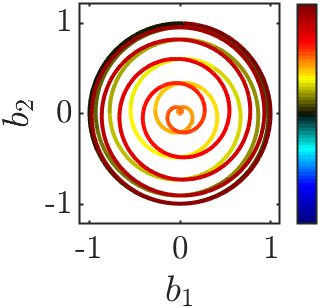} \includegraphics[trim=0.0cm 0cm 0.0cm 0cm,clip,width=0.3325\textwidth]{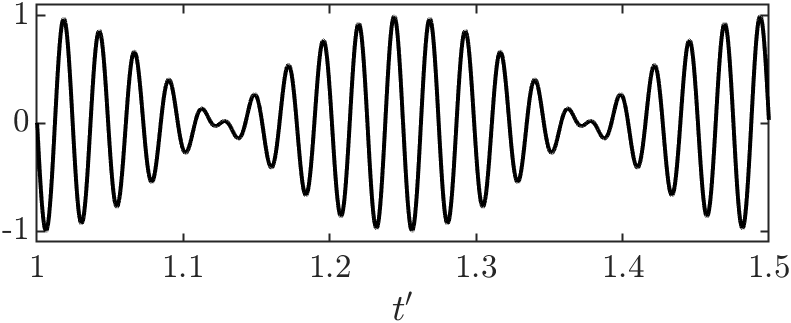} & \includegraphics[trim= 0.0cm 0cm 0.0cm 0cm,clip,width=0.1425\textwidth]{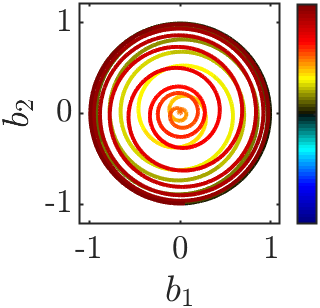} \includegraphics[trim= 0.0cm 0cm 0.0cm 0cm,clip,width=0.3325\textwidth]{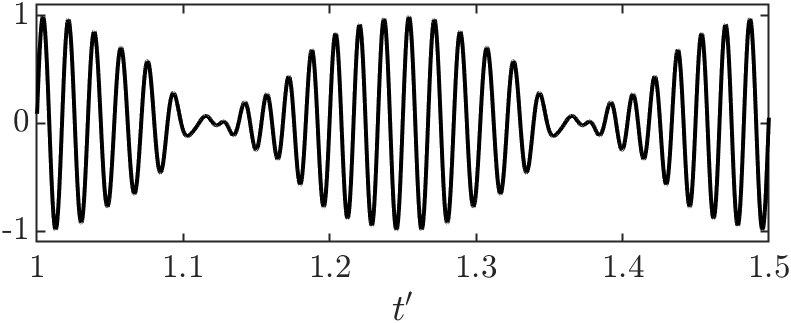}\\
		
		\hline
	\end{tabular}
	
	\caption{Temporal dynamics and the corresponding phase space for even-numbered centroids $\bm{c}^{\bullet}_i$. The colour bar adjacent to the phase spaces indicates the progression of local time $t'$.}
	\label{fig:temporalAndPhaseSpace}
\end{figure}

The temporal dynamics within a given segmentation interval and the phase space corresponding to the centroids $\bm{c}^{\bullet}_i$ are depicted in Figure~\ref{fig:temporalAndPhaseSpace}. 
Specifically, even-numbered centroids are showcased in panels from (a) to (h). Each characteristic trajectory intrinsically encapsulates the amplitude modulation. This is because a full cycle of the amplitude oscillation is inherently contained within each segment. Moreover, the distinct frequency characteristics of these centroids can be readily observed. Transitioning between these clusters offers a coherent description that accurately reflects the frequency modulation in the original dataset.

\begin{figure}
	\centering
	\begin{minipage}{1\textwidth}
		\subfloat[ CNM]{\includegraphics[trim= 0cm 0.cm 1cm 0.cm,clip,width=0.5\columnwidth]{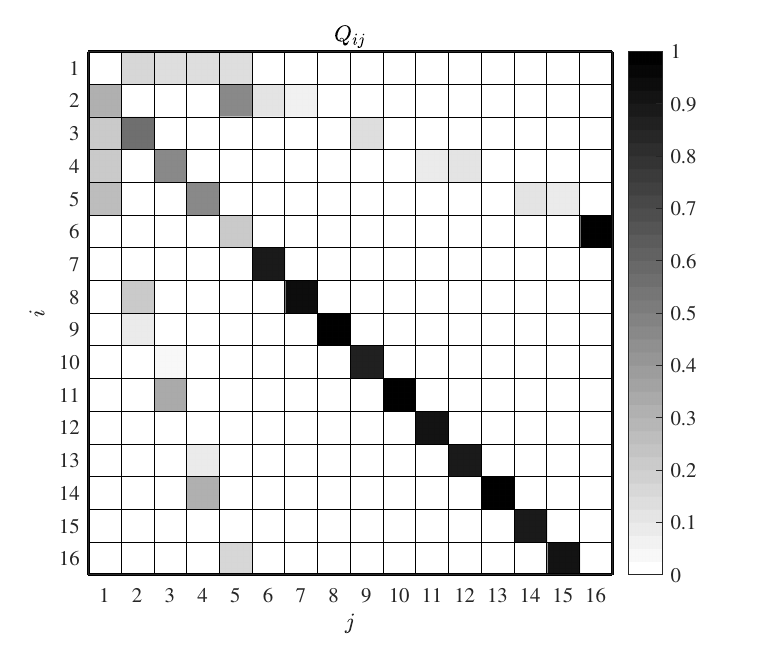}   }
		\subfloat[STFT oCNM]{\includegraphics[trim= 0cm 0.cm 1cm 0.cm,clip,width=0.5\columnwidth]{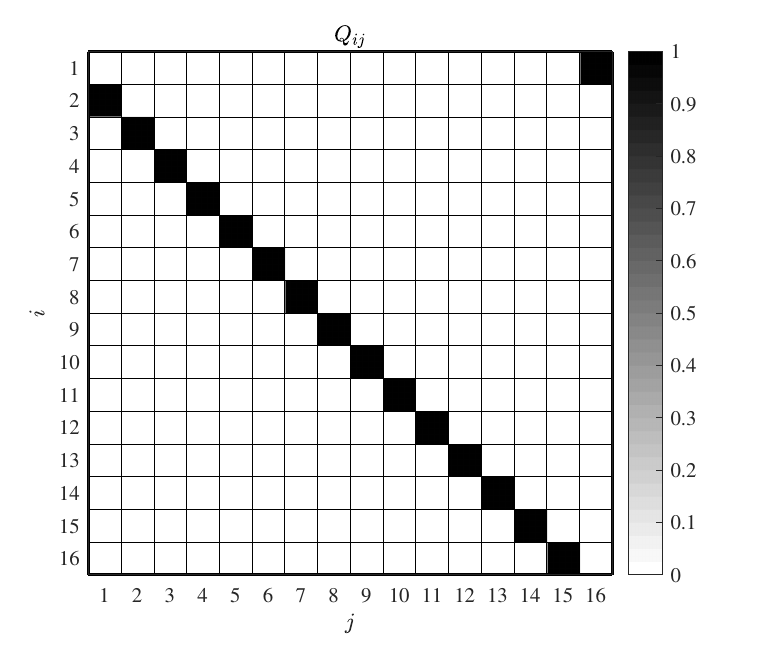} }\\
	\end{minipage}
	\caption{Direct transition probability matrix $\bm{Q}$ for CNM in panel (a) and STFT oCNM in panel (b).}
	
	\label{fig:Qmatrix}
\end{figure}

The frequency modulation's limit cycle-like behaviour is evident in the structure of the direct transition probability matrix $ \bm{Q} $. Figure~\ref{fig:Qmatrix} illustrates $ \bm{Q} $ for the CNM in panel (a) and the STFT oCNM in panel (b). The $ \bm{Q} $ matrix in the STFT oCNM scenario notably showcases the characteristic shape of a limit cycle, as discussed in \citet{Hou2022}.

\begin{figure}
	\centering
	\includegraphics[trim= 0cm 0cm 0cm 0cm,clip,width=1\columnwidth]{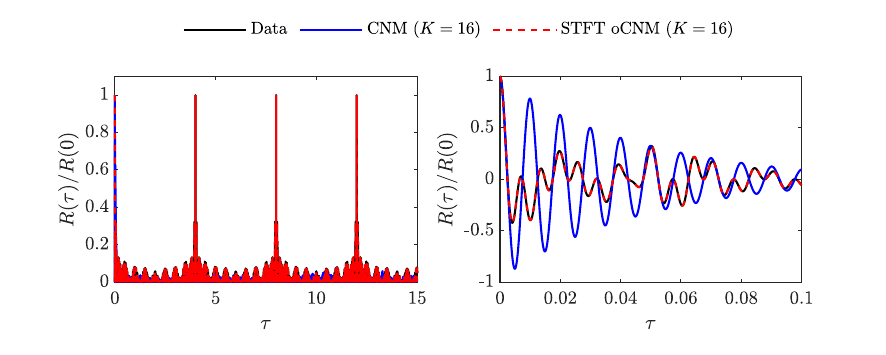}\
	\caption{Comparison of the normalised unbiased auto-correlation function $R$: original data in black, CNM in blue, and STFT oCNM in red dashed lines. The right panel provides a zoomed-in view for small time lags.}
	\label{fig:toyR}
\end{figure}

It is interesting to compare the unbiased auto-correlation function $R$ for the two CNM methodologies to provide deeper insights into the statistical properties of the original and modelled data. The function $R$ is formally defined in Equation~\eqref{Rdef}. Figure~\ref{fig:toyR} presents the normalised $R$ on the left, with a zoomed view targeting short-time lags on the right. 
The visualisations represent the original data in black, the CNM in blue, and the STFT oCNM in red dashed lines. A close examination of this figure further emphasises the superior ability of the oCNM to effectively capture the complex dynamics inherent in the data.

Thus far, oCNM has demonstrated its superiority in effectively capturing complex temporal behaviours exhibited by toy applications, like Landau's problem and its associated amplitude and frequency modulation.


\subsection{Numerical methods for the Fluidic Pinball}\label{sec:pinball}

The fluidic pinball flow has emerged as an intriguing and efficient setup for investigating the application of machine learning in fluid flow control \citep{brunton2020arfm,CornejoMaceda2021jfm, Farzamnik2023}. This system undergoes various flow patterns, starting with a stable flow and transitioning into a symmetric oscillatory flow characterised by a Hopf bifurcation \citep{deng2021jfm,Deng2023}. It further evolves into asymmetric vortex shedding following a pitchfork bifurcation, eventually reaching a state of chaos, all by manipulating the Reynolds number \citep{Deng2019}. 

\begin{figure}
\centering
\subfloat[Numerical grid]{\centering\includegraphics[trim= 0.8cm 2.5cm 0.8cm 2.5cm,clip,width=0.45\columnwidth]{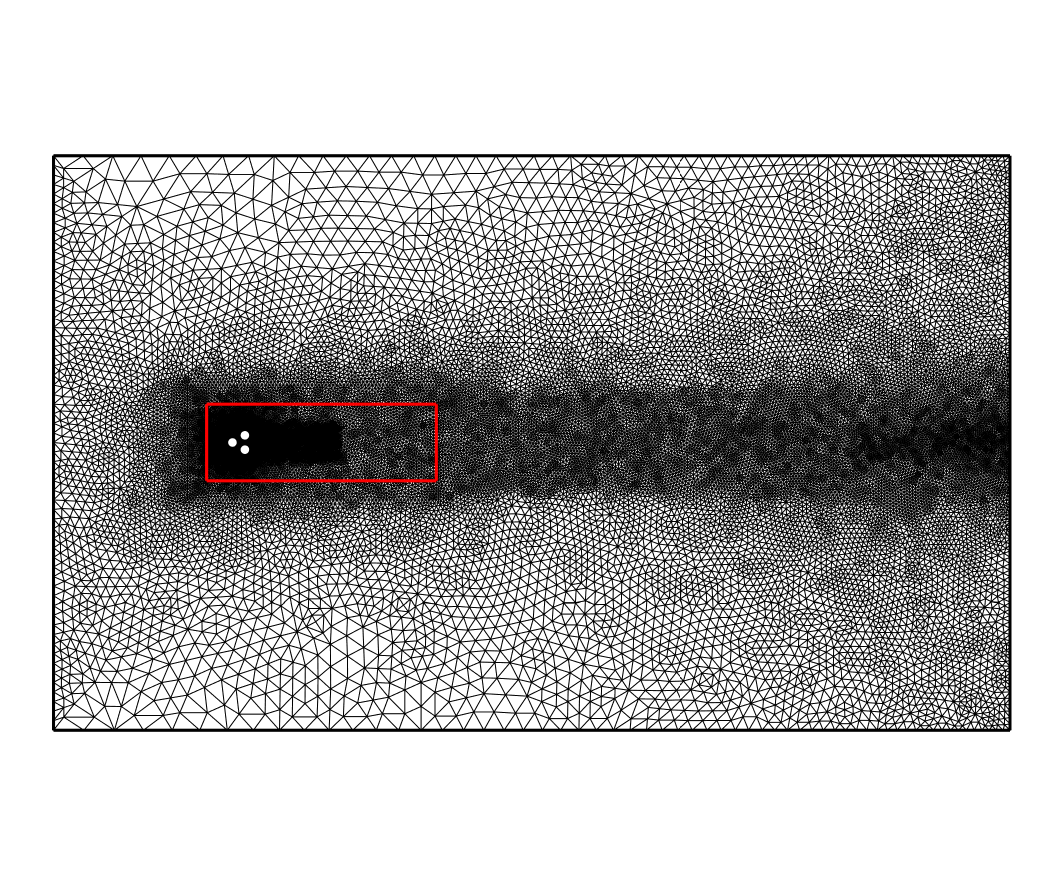}   }
\subfloat[Vorticity snapshot $\Rey=160$]{\raisebox{0.8cm}{\centering\includegraphics[trim= 0cm 0.cm 0cm 0.0cm,clip,width=0.45\columnwidth]{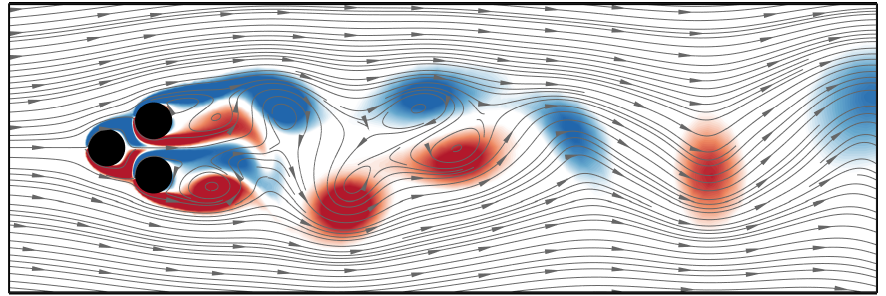} }}
\caption{Numerical grid and a sample vorticity snapshot for fluidic pinball at $\Rey=160$. In all the panels, the abscissa and ordinate are the dimensionless axial coordinate $x$ and transversal coordinate $y$, respectively, omitted for clarity. In panel (a) $-20<x<80$ and $-30<y<30$ while in panel (b) $-4<x<20$ and $-4<y<4$.
The vorticity distribution has been normalised to its maximum.
}
\label{fig:pinballlayout}
\end{figure}

The geometric layout, as illustrated in Figure~\ref{fig:pinballlayout}, comprises three cylinders with a diameter of $D$, positioned at the vertices of an equilateral triangle orientated upstream, with each side measuring $3D/2$ in the $(x, y)$ plane. 
The inlet of the domain is subjected to a uniform velocity $U_\infty$ for the incoming flow, without any external forces acting on the cylinders throughout this study. We enforce a no-slip condition on the cylinders, and we assume that the velocity in the far wake region remains at $U_\infty$. The Reynolds number is defined as $\Rey = U_\infty D/\nu$, where $\nu$ is the kinematic viscosity of the fluid. At the output region of the domain, a no-stress condition is applied. 

The numerical data presented here has been generated using a solver based on a second-order finite-element discretization method of the Taylor–Hood type, as outlined by \cite{Taylor1973}. This solver operates on an unstructured grid consisting of $61\,032$ triangles and $30\,826$ vertices in a computational domain of size $\left[-20, 80\right] \times \left[-30, 30\right]$, employing implicit integration of the third order in time. 
The instantaneous flow field is determined through a Newton–Raphson iteration process, persisting until the residual reaches a predefined, small tolerance. 
This approach is also used to compute the steady-state solution, which stems from the steady Navier–Stokes equations. The direct Navier–Stokes solver employed in this work has undergone validation in prior studies by \citet{noack2003,Deng2019}, with comprehensive technical documentation available in \citet{noack2017report}.
For further insights into the simulation methodology, please refer to \citet{Deng2019}.
A refined observation zone in size of $\left[-4, 20\right] \times \left[-4, 4\right]$ is chosen to visualise the wake flow, as shown in Figure~\ref{fig:pinballlayout}. 

In subsequent subsections, the orbital CNM (oCNM) is applied to the force coefficients and velocity fields of the fluidic pinball at different $\Rey$.

\subsection{oCNM applied on force coefficients}\label{sec:cdcl}
\begin{figure}
	\centering
	\begin{minipage}{1\textwidth}
		\centering
		\begin{minipage}[t]{0.24\textwidth}
			\subfloat[$\Rey=60$]{
				\includegraphics[trim= 0cm 0.cm 0cm 0.cm,clip,width=1\columnwidth]{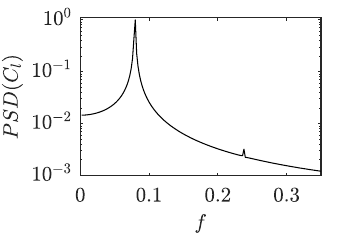}}
		\end{minipage}
		\begin{minipage}[t]{0.24\textwidth}
			\subfloat[$\Rey=90$]{\includegraphics[trim= 0cm 0.cm 0cm 0.cm,clip,width=1\columnwidth]{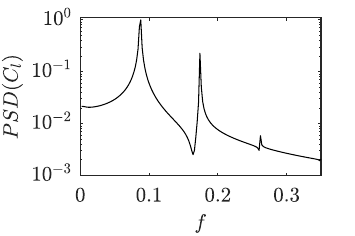} }
		\end{minipage}
		\begin{minipage}[t]{0.24\textwidth}
			\subfloat[$\Rey=120$]{	\includegraphics[trim= 0cm 0.cm 0cm 0.cm,clip,width=1\columnwidth]{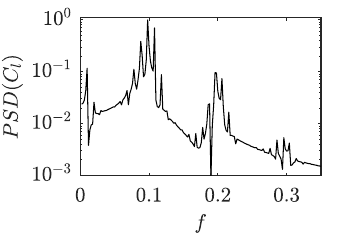}   }
		\end{minipage}
		\begin{minipage}[t]{0.24\textwidth}
			\subfloat[$\Rey=160$]{\includegraphics[trim= 0cm 0.cm 0cm 0.cm,clip,width=1\columnwidth]{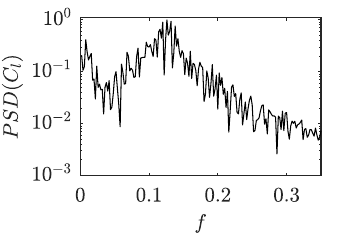} }
		\end{minipage}
		
	\end{minipage}
	\caption{Power spectral density of lift coefficient at different values of $\Rey$.}
	\label{fig:COEFFSfft}
\end{figure}

\begin{figure}
\centering
\begin{minipage}{1\textwidth}
\centering
\includegraphics[trim= 0cm 7.5cm 0cm 0.0cm,clip,width=0.7\columnwidth]{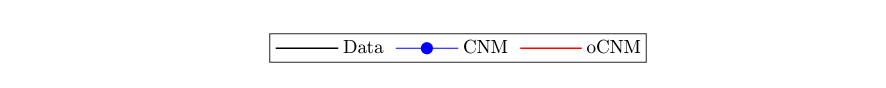} \\

\begin{minipage}[t]{0.24\textwidth}
	\subfloat[$\Rey=60$]{
		\includegraphics[trim= 0cm 0.cm 0cm 0.65cm,clip,width=1\columnwidth]{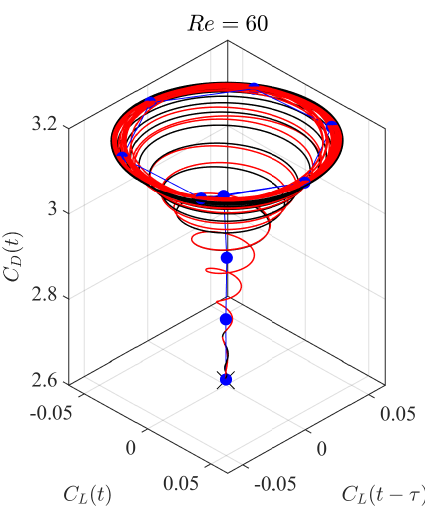}   }
\end{minipage}
\begin{minipage}[t]{0.24\textwidth}
\subfloat[$\Rey=90$]{\includegraphics[trim= 0cm 0.cm 0cm 0.65cm,clip,width=1\columnwidth]{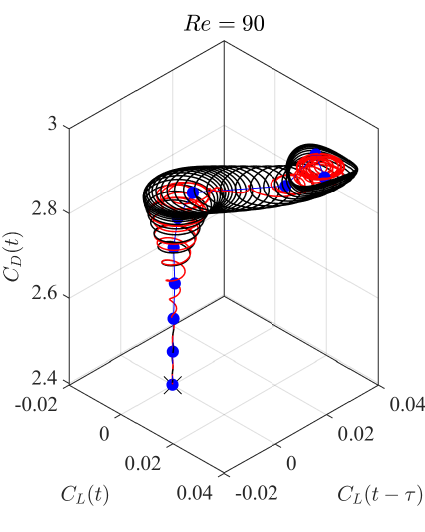} }
\end{minipage}
\begin{minipage}[t]{0.24\textwidth}
\subfloat[$\Rey=120$]{\includegraphics[trim= 0cm 0.cm 0cm 0.65cm,clip,width=1\columnwidth]{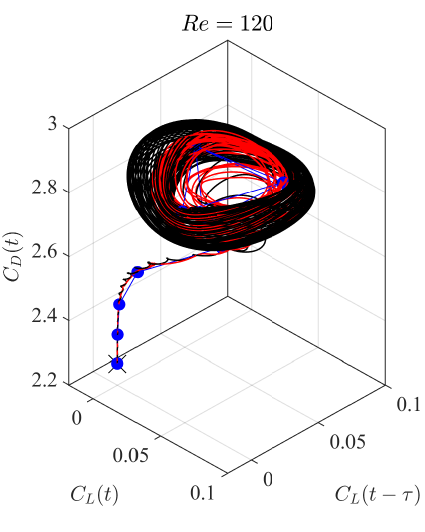}   }
\end{minipage}
\begin{minipage}[t]{0.24\textwidth}
\subfloat[$\Rey=160$]{\includegraphics[trim= 0cm 0.cm 0cm 0.65cm,clip,width=1\columnwidth]{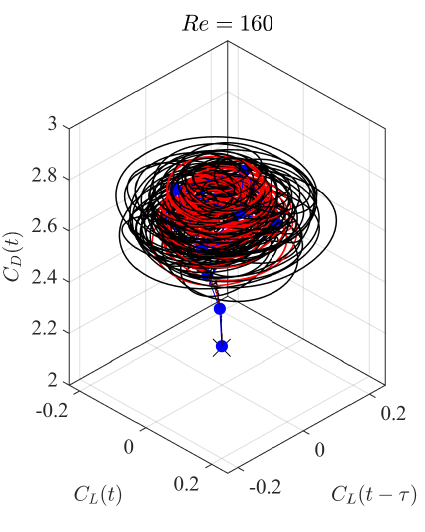} }
\end{minipage}

\end{minipage}
\caption{Clustered phase portrait of force coefficients at different values of $\Rey$. Both in CNM and STFT oCNM the number of clusters is $K=10$.}
\label{fig:COEFFSPspaceclust}
\end{figure}

The fluidic pinball exhibits diverse behaviours depending on the Reynolds number \citep{noack2017report}. Figure \ref{fig:COEFFSfft} illustrates the power spectral density (PSD) of the lift coefficient in the post-transient solution across increasing Reynolds numbers.
 At low Reynolds numbers, a single dominant frequency is observed, whereas at higher Reynolds numbers, the spectrum becomes broadband.
In Figure~\ref{fig:COEFFSPspaceclust}, we present the 3D phase portraits of the force dynamics $[C_L(t),C_L(t-\tau),C_D(t)]^{\intercal}$ at the different values of $\Rey$ in black solid lines.
The time delay for these representations is fixed at $\tau=\frac{1}{4f_c}$, with $f_c$ representing the characteristic frequency for each case. Figure~\ref{fig:COEFFSPspaceclust}(a) demonstrates the system's oscillatory behaviour as it converges toward a limit cycle (Hopf bifurcation). In Figure~\ref{fig:COEFFSPspaceclust}(b), we observe the transition of the limit cycle into instability, marking a supercritical pitchfork bifurcation. Figures~\ref{fig:COEFFSPspaceclust}(c,d) display instances of quasi-periodic and fully chaotic dynamics, respectively \citep{deng2021jfm}.

The application of the novel CNM technique to the force coefficients of the fluidic pinball illustrates a significant improvement in the representation of the system's dynamics over the CNM approach. Figure~\ref{fig:COEFFSPspaceclust} presents the clustered phase portraits for the force coefficients using the CNM and the STFT-enhanced orbital CNM, with a fixed number of clusters $K=10$. 
The state vector for the CNM is composed of $[C_L(t),C_L(t-\tau),C_D(t)]^{\intercal}$, with the time-delay $\tau$ equal to a quarter of the characteristic period, while for the oCNM only of $[C_L(t),C_D(t)]^{\intercal}$. For the latter method, segment length $L_{\mathrm{traj}}$ is set at $4/(f_c \Delta t)$, $f_c$ being the leading frequency. Hann window has been employed with overlaps accounting for $75\%$ of the segment length. 

\begin{figure}
\centering 
\setlength\tabcolsep{2pt} 
\begin{tabular}{c c c c } 
 \hline 
 & & & \\
$\bm{c}^{\bullet}_1$
& \multicolumn{1}{m{2.7cm}}{\includegraphics[trim=0.2cm 0.0cm 0.1cm 0.15cm, clip,height=0.12\textwidth]{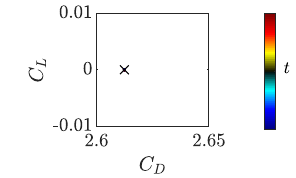}}

&\multicolumn{1}{m{2.4cm}}{\includegraphics[trim=0.2cm 0.0cm 1.0cm 0.0cm, clip,height=0.12\textwidth]{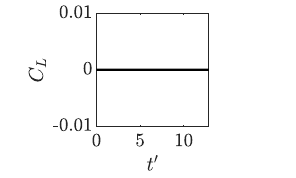}}
&\multicolumn{1}{m{2.4cm}}{\includegraphics[trim=0.2cm 0.0cm 1.0cm 0.0cm, clip,height=0.12\textwidth]{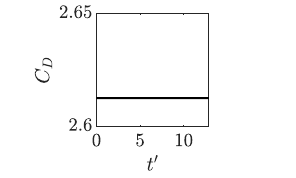}} \\
$\bm{c}^{\bullet}_2$
& \multicolumn{1}{m{2.7cm}}{\includegraphics[trim=0.2cm 0.0cm 0.1cm 0.15cm, clip,height=0.12\textwidth]{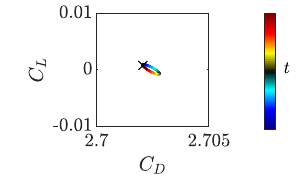}}

&\multicolumn{1}{m{2.4cm}}{\includegraphics[trim=0.2cm 0.0cm 1.0cm 0.0cm, clip,height=0.12\textwidth]{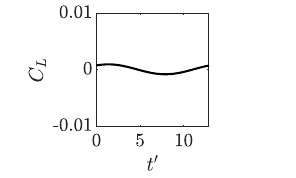}}
&\multicolumn{1}{m{2.4cm}}{\includegraphics[trim=0.2cm 0.0cm 1.0cm 0.0cm, clip,height=0.12\textwidth]{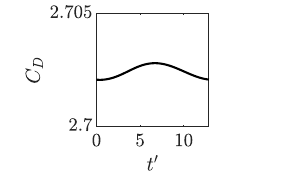}}\\
$\bm{c}^{\bullet}_3$
& \multicolumn{1}{m{2.7cm}}{\includegraphics[trim=0.2cm 0.0cm 0.1cm 0.15cm, clip,height=0.12\textwidth]{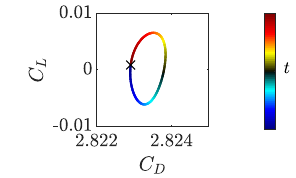}}

&\multicolumn{1}{m{2.4cm}}{\includegraphics[trim=0.2cm 0.0cm 1.0cm 0.0cm, clip,height=0.12\textwidth]{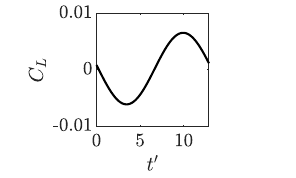}}
&\multicolumn{1}{m{2.4cm}}{\includegraphics[trim=0.2cm 0.0cm 1.0cm 0.0cm, clip,height=0.12\textwidth]{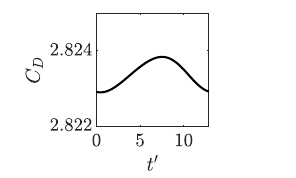}}\\
$\bm{c}^{\bullet}_4$
& \multicolumn{1}{m{2.7cm}}{\includegraphics[trim=0.2cm 0.0cm 0.1cm 0.15cm, clip,height=0.12\textwidth]{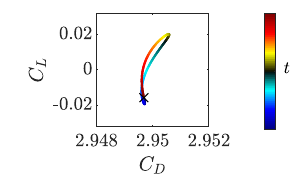}}

&\multicolumn{1}{m{2.4cm}}{\includegraphics[trim=0.2cm 0.0cm 1.0cm 0.0cm, clip,height=0.12\textwidth]{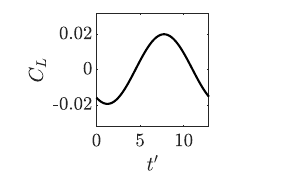}}
&\multicolumn{1}{m{2.4cm}}{\includegraphics[trim=0.2cm 0.0cm 1.0cm 0.0cm, clip,height=0.12\textwidth]{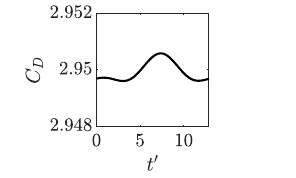}}\\
$\bm{c}^{\bullet}_5$
& \multicolumn{1}{m{2.7cm}}{\includegraphics[trim=0.2cm 0.0cm 0.1cm 0.15cm, clip,height=0.12\textwidth]{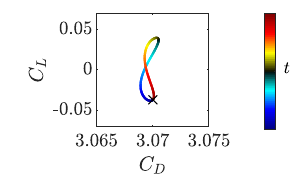}}

&\multicolumn{1}{m{2.4cm}}{\includegraphics[trim=0.2cm 0.0cm 1.0cm 0.0cm, clip,height=0.12\textwidth]{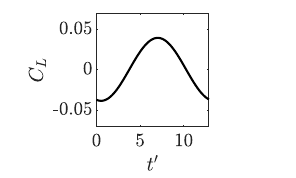}}
&\multicolumn{1}{m{2.4cm}}{\includegraphics[trim=0.2cm 0.0cm 1.0cm 0.0cm, clip,height=0.12\textwidth]{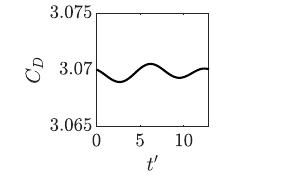}}\\
$\bm{c}^{\bullet}_6$
& \multicolumn{1}{m{2.7cm}}{\includegraphics[trim=0.2cm 0.0cm 0.1cm 0.15cm, clip,height=0.12\textwidth]{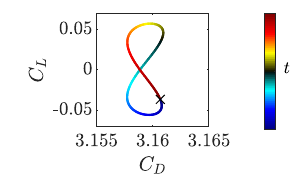}}

&\multicolumn{1}{m{2.4cm}}{\includegraphics[trim=0.2cm 0.0cm 1.0cm 0.0cm, clip,height=0.12\textwidth]{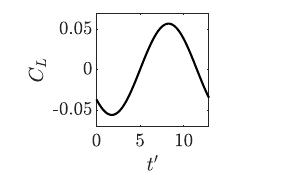}}
&\multicolumn{1}{m{2.4cm}}{\includegraphics[trim=0.2cm 0.0cm 1.0cm 0.0cm, clip,height=0.12\textwidth]{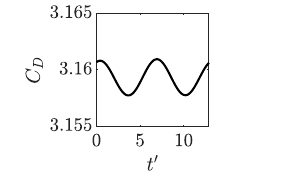}}\\
$\bm{c}^{\bullet}_7$
& \multicolumn{1}{m{2.7cm}}{\includegraphics[trim=0.2cm 0.0cm 0.1cm 0.15cm, clip,height=0.12\textwidth]{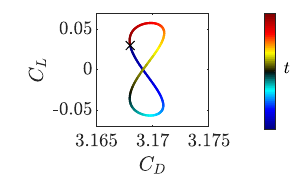}}

&\multicolumn{1}{m{2.4cm}}{\includegraphics[trim=0.2cm 0.0cm 1.0cm 0.0cm, clip,height=0.12\textwidth]{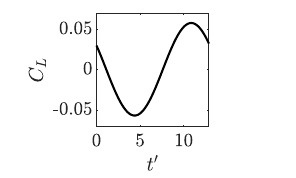}}
&\multicolumn{1}{m{2.4cm}}{\includegraphics[trim=0.2cm 0.0cm 1.0cm 0.0cm, clip,height=0.12\textwidth]{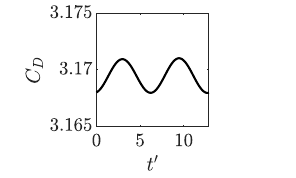}}\\
$\bm{c}^{\bullet}_8$
& \multicolumn{1}{m{2.7cm}}{\includegraphics[trim=0.2cm 0.0cm 0.1cm 0.15cm, clip,height=0.12\textwidth]{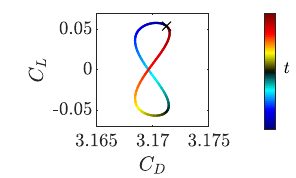}}

&\multicolumn{1}{m{2.4cm}}{\includegraphics[trim=0.2cm 0.0cm 1.0cm 0.0cm, clip,height=0.12\textwidth]{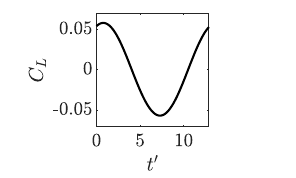}}
&\multicolumn{1}{m{2.4cm}}{\includegraphics[trim=0.2cm 0.0cm 1.0cm 0.0cm, clip,height=0.12\textwidth]{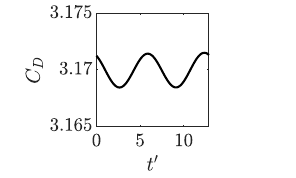}}\\
$\bm{c}^{\bullet}_9$
& \multicolumn{1}{m{2.7cm}}{\includegraphics[trim=0.2cm 0.0cm 0.1cm 0.15cm, clip,height=0.12\textwidth]{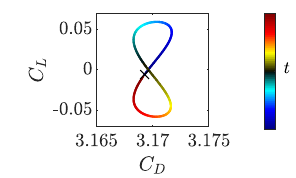}}

&\multicolumn{1}{m{2.4cm}}{\includegraphics[trim=0.2cm 0.0cm 1.0cm 0.0cm, clip,height=0.12\textwidth]{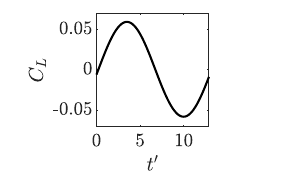}}
&\multicolumn{1}{m{2.4cm}}{\includegraphics[trim=0.2cm 0.0cm 1.0cm 0.0cm, clip,height=0.12\textwidth]{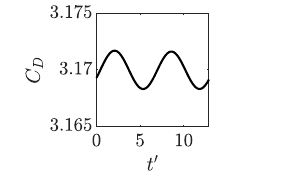}}\\
$\bm{c}^{\bullet}_{10}$
& \multicolumn{1}{m{2.7cm}}{\includegraphics[trim=0.2cm 0.0cm 0.1cm 0.15cm, clip,height=0.12\textwidth]{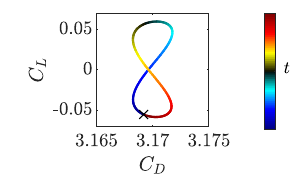}}

&\multicolumn{1}{m{2.4cm}}{\includegraphics[trim=0.2cm 0.0cm 1.0cm 0.0cm, clip,height=0.12\textwidth]{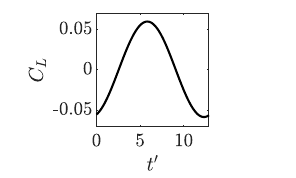}}
&\multicolumn{1}{m{2.4cm}}{\includegraphics[trim=0.2cm 0.0cm 1.0cm 0.0cm, clip,height=0.12\textwidth]{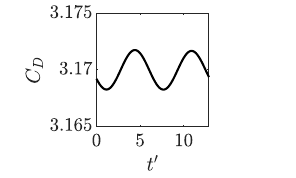}}\\
\hline 
\end{tabular}
\caption{Orbital cluster centroids of force coefficients at $\Rey=60$. The figure illustrates the temporal evolution of the drag ($C_D$) and lift ($C_L$) coefficients within each cluster. 
The colour bar indicates the progression of local time $t'$.}
\label{fig:re60sluster}
\end{figure}

The comparison in Figure~\ref{fig:COEFFSPspaceclust} highlights the enhanced capability of the oCNM in capturing the underlying dynamical features of the fluidic pinball system. 
Particularly, by employing the short-time Fourier transform as a functional basis within the CNM framework, we can achieve a more accurate depiction of the amplitude selection mechanism, especially evident in the post-Hopf bifurcation scenario at $\Rey=60$ in panel (a) and $\Rey=90$ in panel (b). For the first configuration, a detailed analysis of the cluster centroids shape offers insights into the system dynamics. Figure~\ref{fig:re60sluster} showcases orbital cluster centroids of force coefficients. 
The first centroid corresponds to the steady cluster, characterised by fixed values of $C_L$ and $C_D$. 
An examination of the phase space for cluster $i=3$, associated with the transient phase, reveals that the $C_D$ coefficient oscillates at the same frequency as the $C_L$ coefficient.
A noteworthy observation from this analysis is the emergence of a distinct temporal pattern in the force coefficients centroids in post-transient: the drag coefficient exhibits a frequency that is notably double that of the lift coefficient.

\begin{figure}
\centering
\includegraphics[trim= 0cm 0.cm 0cm 0.cm,clip,width=1\columnwidth]{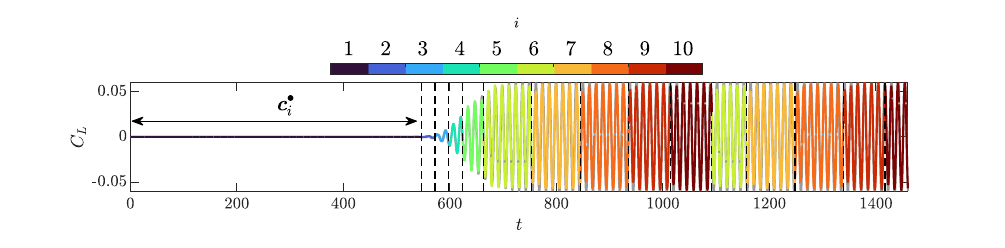}
\caption{Sequence of orbital clusters centroids for the configuration at $\Rey=60$. $K=10$.}\label{fig:re60sequenza}
\end{figure}

This analysis underscores the amplitude selection mechanism within this regime, as illustrated in Figure~\ref{fig:re60sequenza}, which presents the cluster sequence for this specific configuration. 
It is worth noting that in this figure, whereby cluster affiliation is colour-coded according to the cluster index, the sequence has been smoothed according to the Equation~\eqref{reconstrclust} has been shown. 

\begin{figure}
\centering
\subfloat[$\Rey=120$]{\includegraphics[trim= 0cm 0.cm 3.3cm 0.0cm,clip,height=3.5cm]{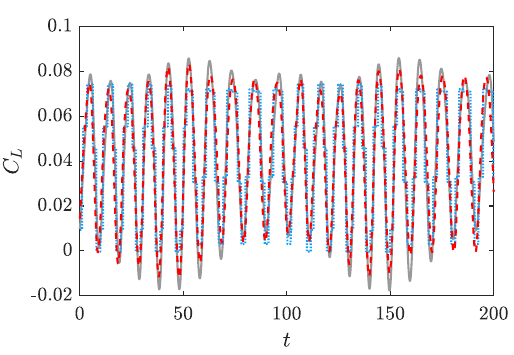}
\includegraphics[trim= 0cm 0.cm 0cm 0.0cm,clip,height=3.5cm]{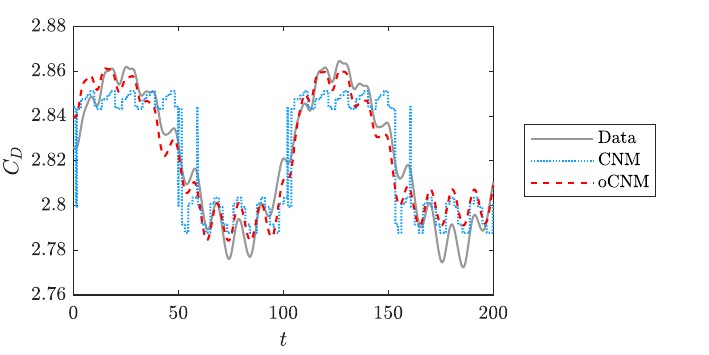} 
}\\
\subfloat[$\Rey=160$]{
\includegraphics[trim= 0cm 0.cm 3.3cm 0.0cm,clip,height=3.5cm]{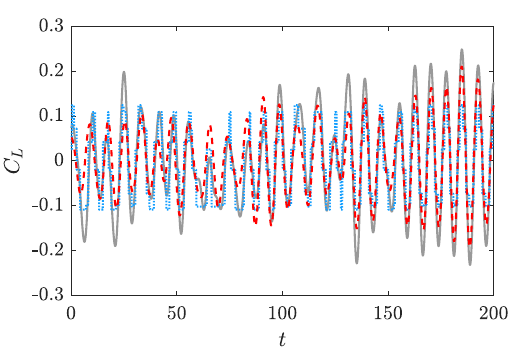}
\includegraphics[trim= 0cm 0.cm 0cm 0.0cm,clip,height=3.5cm]{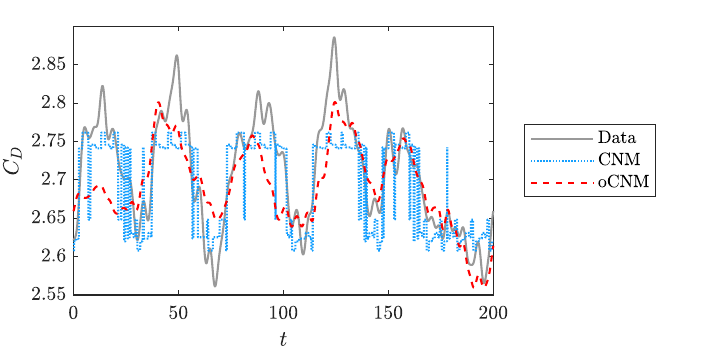}
}\\
\caption{Comparison of force coefficient predictions between CNM and oCNM models at higher Reynolds numbers. Panel~(a) shows predictions for $\Rey=120$ and panel~(b) for $\Rey=160$.}
\label{fig:Re0120160STAZreco}
\end{figure}

As the system progresses to higher Reynolds numbers ($\Rey=120$ and $\Rey=160$), entering regimes of quasi-periodic and chaotic behaviour, the oCNM maintains its fidelity in capturing the complex dynamics. This contrasts with the CNM, which, while still capable of identifying distinct dynamical regions, lacks the same level of detail and accuracy in representing the transitions between these states. This distinction becomes evident when examining Figure~\ref{fig:Re0120160STAZreco}, which compares the force coefficient predictions by the CNM and oCNM (with Fourier basis) models for $\Rey=120$ in panel (a) and $\Rey=160$ in panel (b). 
The prediction of the oCNM model closely aligns with the observed behaviours, reflecting the nuanced changes in dynamics as the system transitions from quasi-periodic to fully chaotic flow regimes. Conversely, the CNM model, though identifying the overarching trends, fails to replicate the subtleties and the exact transition points with similar accuracy.

\subsection{oCNM applied to flow fields}\label{sec:flowfields}
Here, the input data are the velocity component fields, specifically in the quasi-periodic and chaotic regimes of the fluidic pinball system. The objective is to elucidate the underlying dynamics and spatial structures characterising these intricate flow states, using the CNM and oCNM methodologies for a detailed exploration. The dataset, as outlined in \S~\ref{sec:pinball}, comprises both velocity components. After the transient, $5\,000$ snapshots were collected, each sampled at $\Delta t = 0.1$. For the oCNM analysis, the number of clusters was set to $K = 10$ to ensure that the variance ratio $V^\bullet / (V^\bullet + J^\bullet) >0.9$. Additionally, the elbow method was applied to the RMSE of the autocorrelation function $R$ between the training and reconstructed datasets, as described in Section~\ref{sec:RMSE}.

\begin{figure}
\centering
\begin{minipage}{1\textwidth}
\subfloat[]{\includegraphics[trim= 0cm 0.cm 1cm 0.cm,clip,width=0.45\columnwidth]{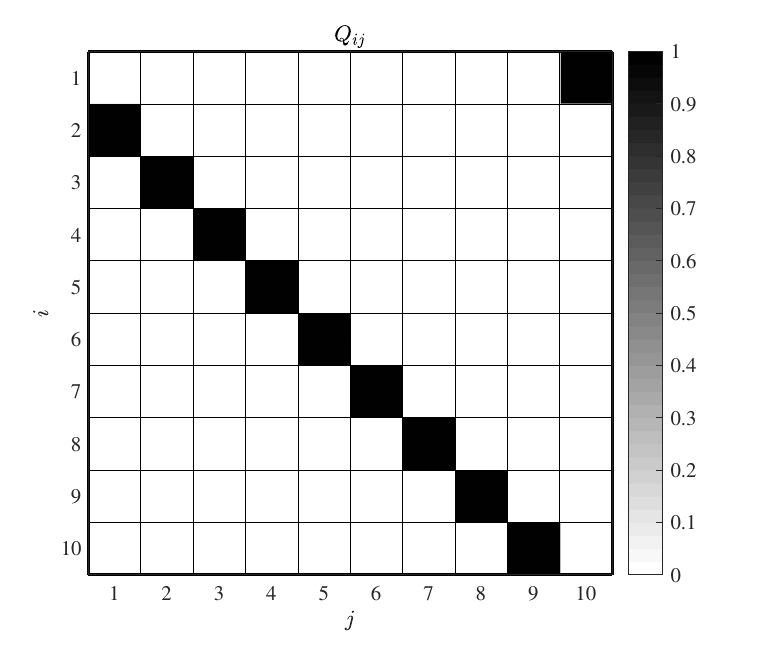}   }
\subfloat[]{\includegraphics[trim= 0cm 0.cm 1cm 0.cm,clip,width=0.45\columnwidth]{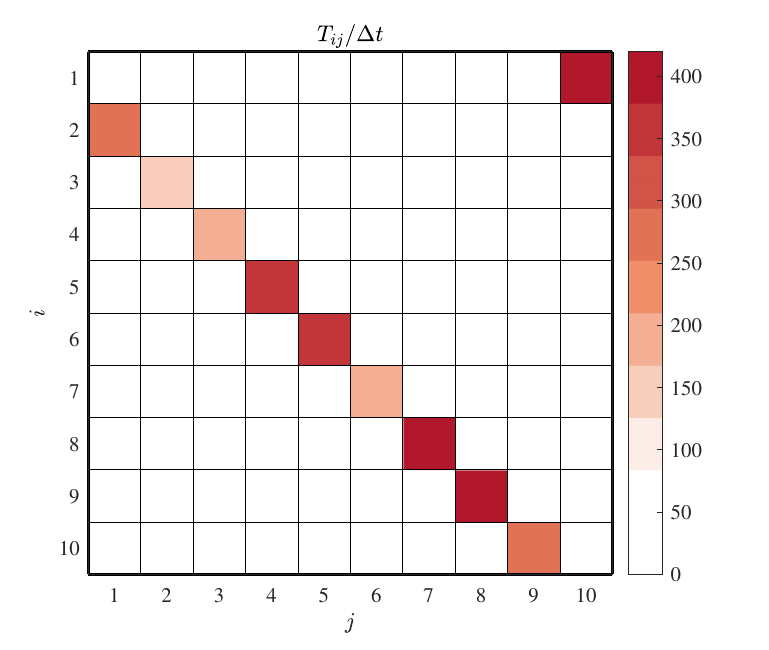} }\\
\end{minipage}
\caption{The direct transition probabilities ($Q_{ij}$) and times ($T_{ij}$) matrices of the orbital CNM for the case $\Rey=120$.}
\label{fig:Qmatrixre120}
\end{figure}

\begin{figure}
\centering
\begin{minipage}{1\textwidth}

\subfloat[]{\includegraphics[trim= 0cm 0.cm 0cm 0.cm,clip,width=0.5\columnwidth]{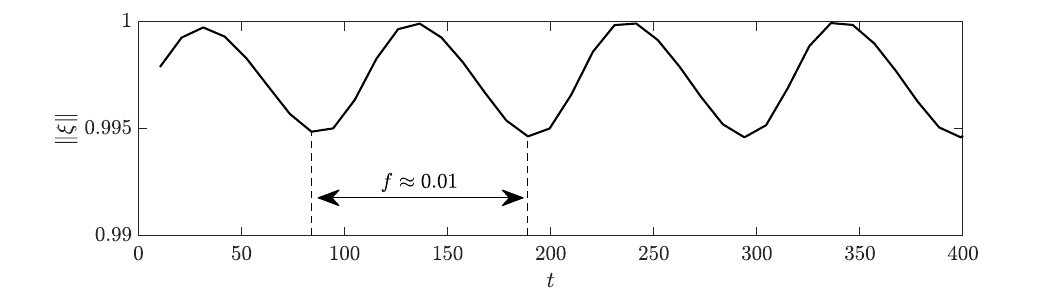}}\\
\vspace{-3.2cm}

\subfloat[]{\includegraphics[trim= 0cm 0.cm 0cm 0.cm,clip,width=0.5\columnwidth]{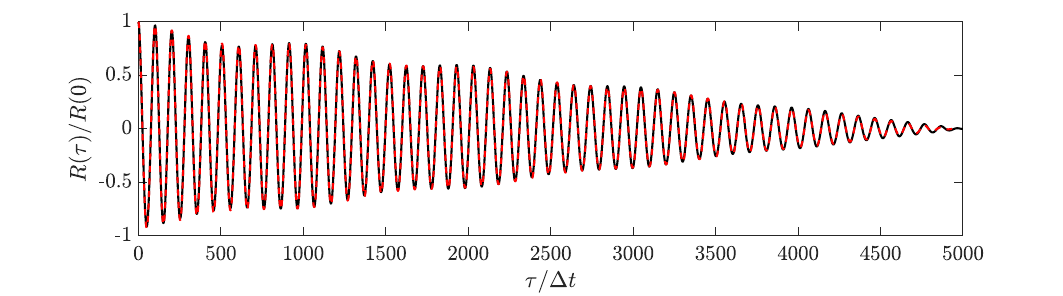}}
\subfloat[]{\includegraphics[trim= 0cm 0.cm 0cm 0.cm,clip,width=0.5\columnwidth]{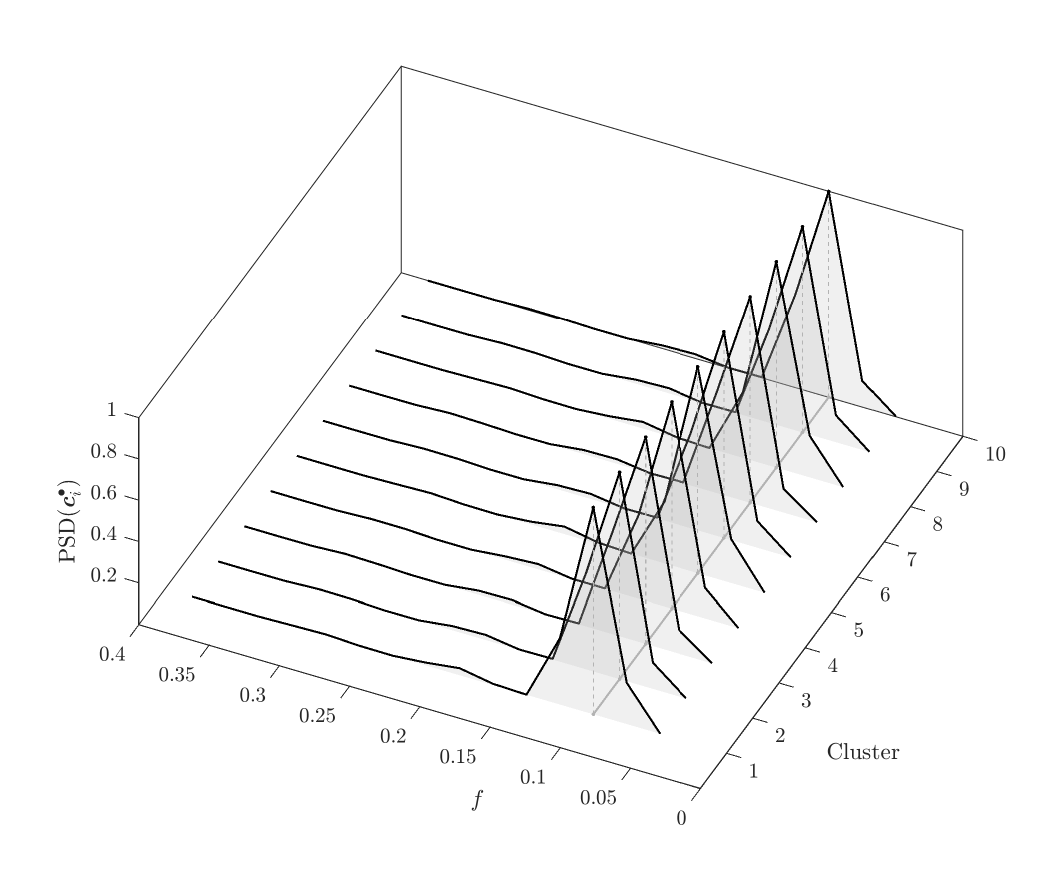}}
\end{minipage}
\caption{Orbital CNM frequency behaviour and validation. Panel (a): slow timescale behaviour of $\|\bm{\xi}\|$. Panel (b): auto-correlation function for the original data (black) and the oCNM output (red dashed). Panel (c): normalised PSD of each trajectory cluster centroid, $\mathrm{PSD}(\bm{c}^{\bullet}_i)$. Case with $\Rey=120$.}
\label{fig:STFTspectrumre120}
\end{figure}

The oCNM effectively distinguishes between fast and slow timescales by modelling the slow dynamics and encapsulating the rapid fluctuations within clusters of short-term trajectories. 
For the $\Rey=120$ scenario, the length of each trajectory segment is determined to be
\begin{equation}\label{eq:ni}
    L_{\mathrm{traj}}=\frac{4}{f_c \Delta t}, 
\end{equation}
incorporating a $75\%$ overlap between segments and utilising a Hann window for temporal weights.
The direct transition probability ($Q_{ij}$) and transition time ($T_{ij}$) matrices for this case are illustrated in Figure~\ref{fig:Qmatrixre120}. 
Notably, the $Q_{ij}$ matrix exhibits a periodic behaviour. Additionally, an estimation of the leading frequency is obtained from the transition times matrix $T_{ij}$ as:
\begin{equation}
f_{\rm slow}\approx \frac{1}{T_{2,1}+T_{3,2}+\ldots+T_{1,10}} \approx 0.01,
\end{equation}
highlighting the slow timescale periodic dynamics of the system. This observation is further corroborated by examining the norm of the vector $\bm{\xi}$, as defined in Equation~\eqref{eq:stateVSTFT}, shown in Figure~\ref{fig:STFTspectrumre120}(a). 
This result confirms that the low-frequency modulations are present when the vortex shedding has developed to a certain degree \citep{deng2022jfm}. Figure~\ref{fig:STFTspectrumre120}(b) presents the auto-correlation function for the original data and the oCNM with Fourier basis output. The auto-correlation function of the data reveals the two dominant frequencies within the dynamics, which the oCNM successfully identifies and captures. 
Figure~\ref{fig:STFTspectrumre120}(c) displays the normalised power spectral density (PSD) of the cluster centroids, indicating a consistent peak frequency and bandwidth across all clusters. 

\begin{figure}
\centering 
\setlength\tabcolsep{5pt} 
\begin{tabular}{l l} 
 \hline 
$\bm{c}^{\bullet}_1$ & \\
\includegraphics[trim= 1.9cm 0.9cm 3cm 0.5cm,clip,width=3.cm]{Re0120cluster_ni420_tau_120_ik_1_i0_1.pdf} &
\includegraphics[trim= 1.9cm 0.9cm 3cm 0.5cm,clip,width=3.cm]{forwardRe0120cluster_ni420_tau_120_ik_1_i0_1}\\
$\bm{c}^{\bullet}_2$ & \\
\includegraphics[trim= 1.9cm 0.9cm 3cm 0.5cm,clip,width=3.cm]{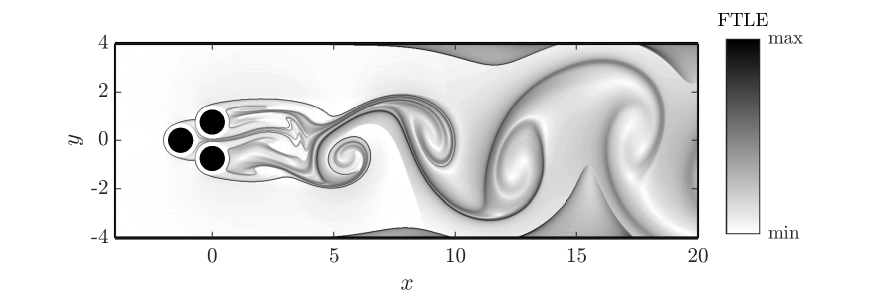} &
\includegraphics[trim= 1.9cm 0.9cm 3cm 0.5cm,clip,width=3.cm]{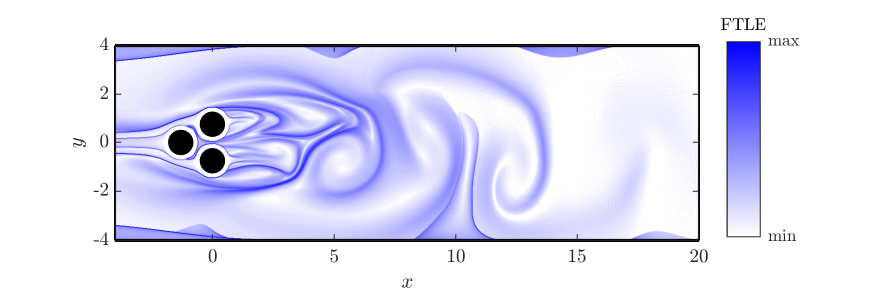}\\
$\bm{c}^{\bullet}_3$ & \\
\includegraphics[trim= 1.9cm 0.9cm 3cm 0.5cm,clip,width=3.cm]{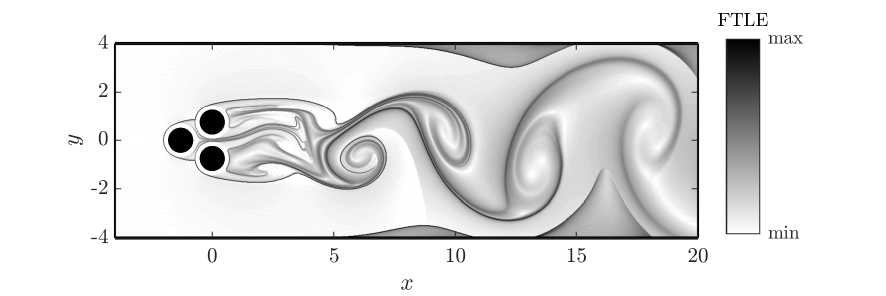} &
\includegraphics[trim= 1.9cm 0.9cm 3cm 0.5cm,clip,width=3.cm]{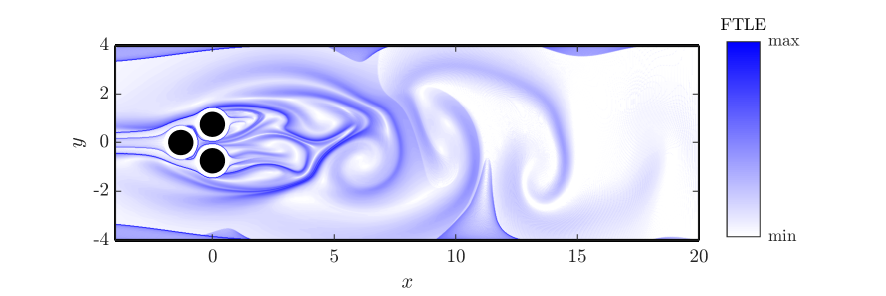}\\
$\bm{c}^{\bullet}_4$ & \\
\includegraphics[trim= 1.9cm 0.9cm 3cm 0.5cm,clip,width=3.cm]{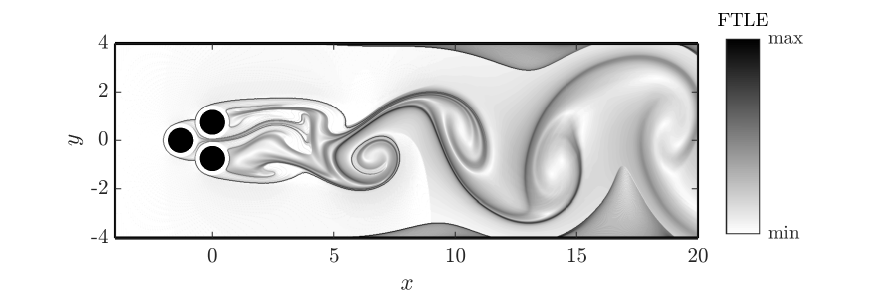} &
\includegraphics[trim= 1.9cm 0.9cm 3cm 0.5cm,clip,width=3.cm]{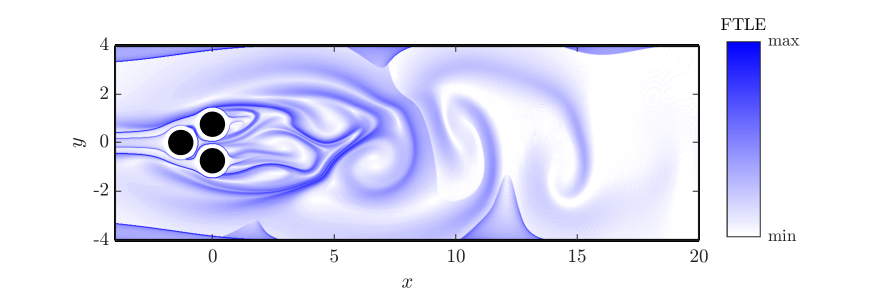}\\
$\bm{c}^{\bullet}_5$ & \\
\includegraphics[trim= 1.9cm 0.9cm 3cm 0.5cm,clip,width=3.cm]{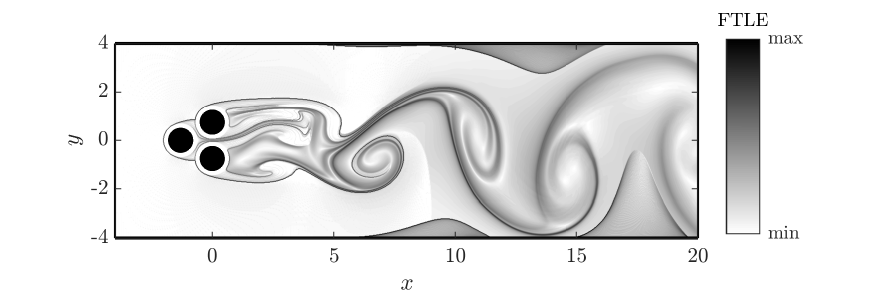} &
\includegraphics[trim= 1.9cm 0.9cm 3cm 0.5cm,clip,width=3.cm]{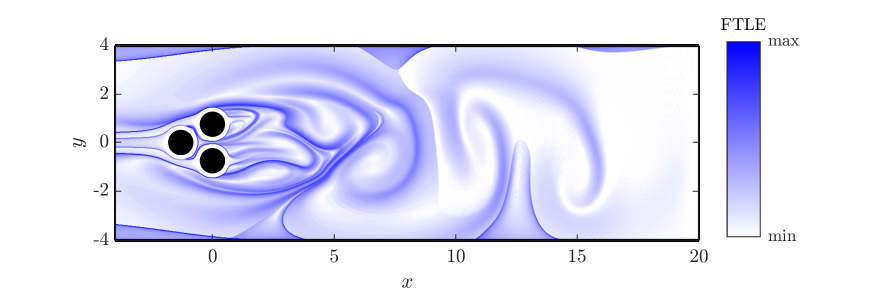}\\
$\bm{c}^{\bullet}_6$ & \\
\includegraphics[trim= 1.9cm 0.9cm 3cm 0.5cm,clip,width=3.cm]{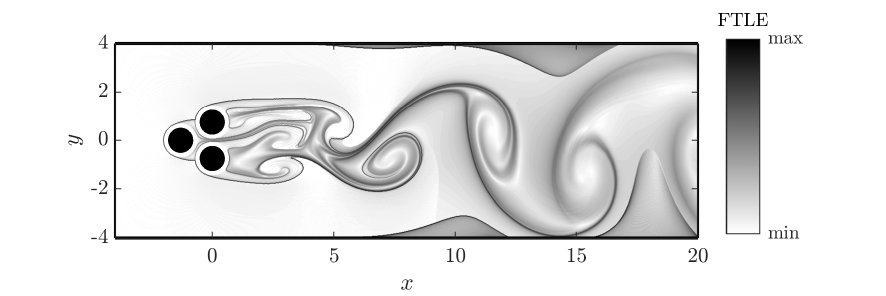} &
\includegraphics[trim= 1.9cm 0.9cm 3cm 0.5cm,clip,width=3.cm]{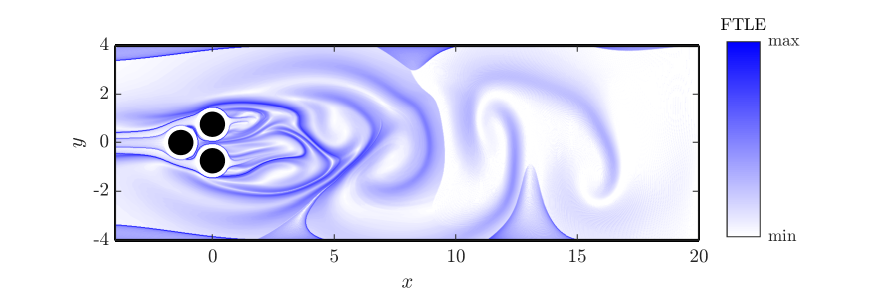}\\
$\bm{c}^{\bullet}_7$ & \\
\includegraphics[trim= 1.9cm 0.9cm 3cm 0.5cm,clip,width=3.cm]{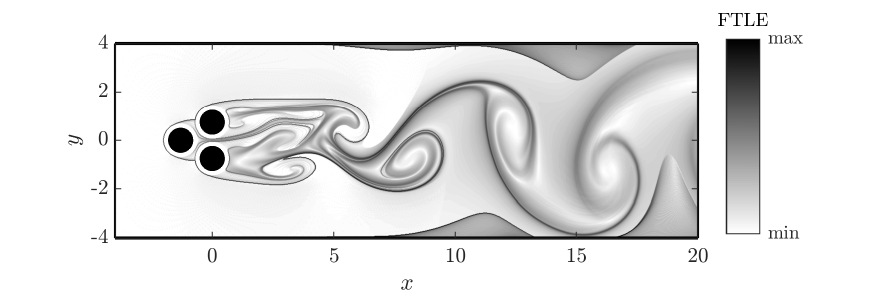} &
\includegraphics[trim= 1.9cm 0.9cm 3cm 0.5cm,clip,width=3.cm]{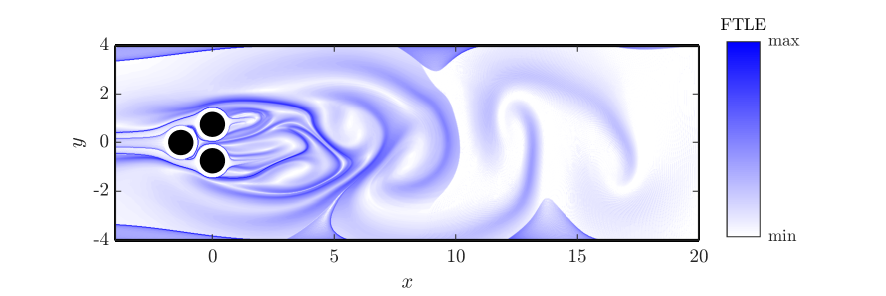}\\
$\bm{c}^{\bullet}_8$ & \\
\includegraphics[trim= 1.9cm 0.9cm 3cm 0.5cm,clip,width=3.cm]{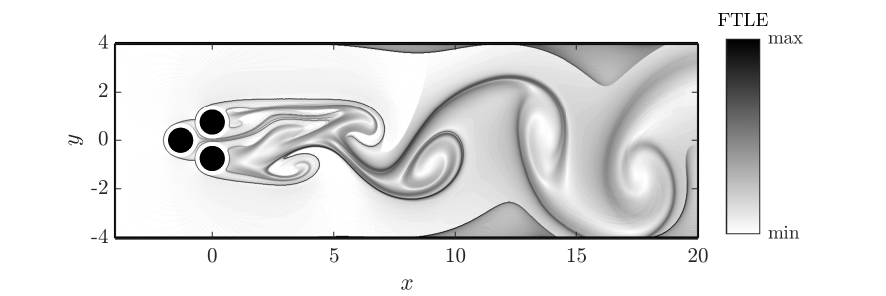} &
\includegraphics[trim= 1.9cm 0.9cm 3cm 0.5cm,clip,width=3.cm]{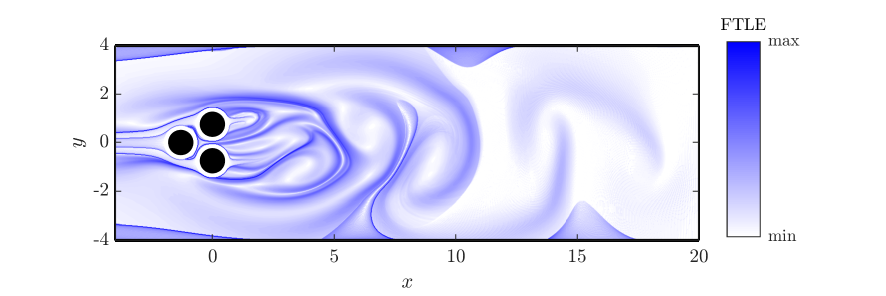}\\
$\bm{c}^{\bullet}_9$ & \\
\includegraphics[trim= 1.9cm 0.9cm 3cm 0.5cm,clip,width=3.cm]{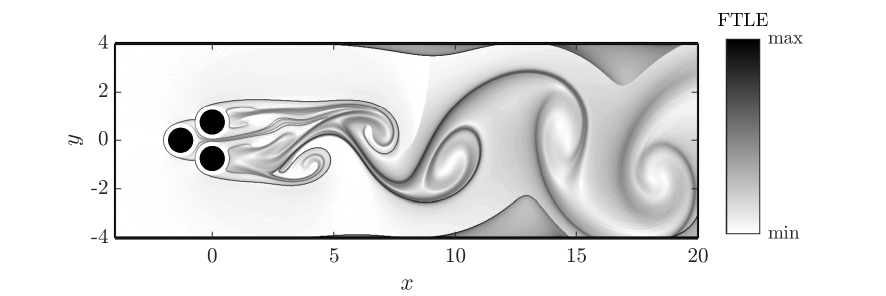} &
\includegraphics[trim= 1.9cm 0.9cm 3cm 0.5cm,clip,width=3.cm]{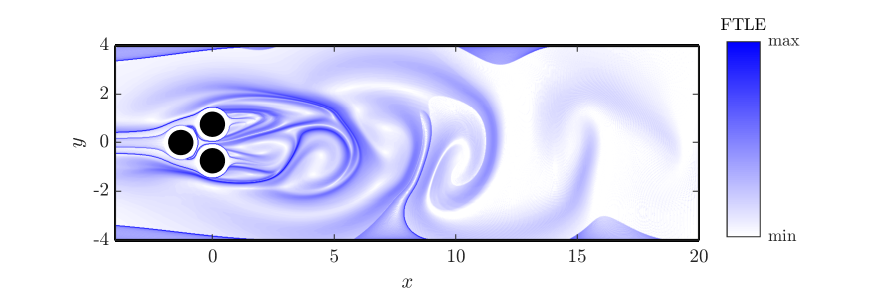}\\
$\bm{c}^{\bullet}_{10}$ & \\
\includegraphics[trim= 1.9cm 0.9cm 3cm 0.5cm,clip,width=3.cm]{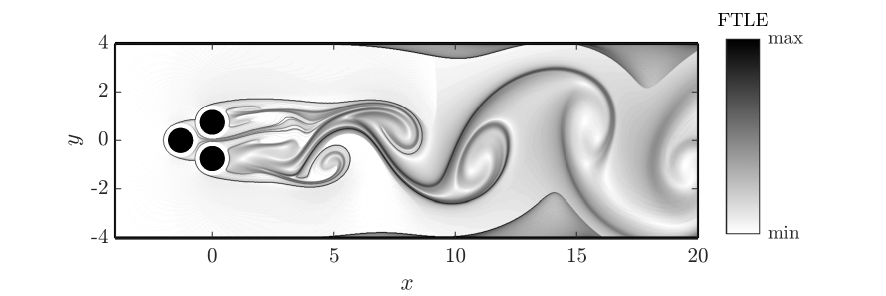} &
\includegraphics[trim= 1.9cm 0.9cm 3cm 0.5cm,clip,width=3.cm]{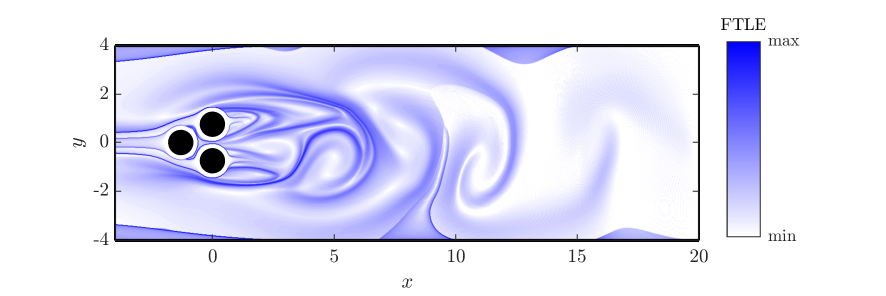}\\
\hline 
\end{tabular}
\caption{Spatial distribution of the backward (black) and forward (blue) Finite-Time Lyapunov Exponent (FTLE) for velocity centroids of oCNM at $\Rey=120$. In all the panels, the abscissa and ordinate are the dimensionless axial coordinate $-4<x<20$ and transversal coordinate $-4<y<4$, respectively, omitted for clarity. Each distribution has been normalised to its maximum.}
\label{fig:FTLEre120}
\end{figure}

The spatio-temporal cluster centroids $\bm{c}^{\bullet}_i(t)$, representing the velocity components $u$ and $v$, are examined to uncover inherent patterns and dynamics. 
Using the backward/forward Finite-Time Lyapunov Exponent (FTLE), a Lagrangian metric detailed in \ref{sec:FTLE}, the coherent structures within the fluid flows are identified, with a specific focus on those influenced by the velocity centroids. 
In particular, given a sequence of $L$ spatio-temporal cluster centroids $\bm{c}^{\bullet}_i(t)$  belonging to the same cluster, the reconstruction in \eqref{reconstr} is performed, and the FTLE is subsequently computed.
For this analysis, the integration time $\nu$, either backward or forward, is determined by the inverse of the characteristic frequency, $1/f_c$. 
Figure~\ref{fig:FTLEre120} illustrates the FTLE computations for the $\Rey=120$ centroids, colour-coded in black for the backward FTLE and in blue for the forward FTLE, effectively highlighting structures associated with vortex shedding. 
Notably, when focussing on the forward FTLE, the clusters primarily differ in the shape of the base-bleeding jet around its deflected position.

\begin{figure}
\centering
\resizebox{\textwidth}{!}{%
\begin{tikzpicture}[node distance=4.5cm, auto]

    \tikzstyle{block} = [rectangle, draw=none, fill=none,text width=2.8cm, rounded corners, minimum height=1cm,inner sep=0pt]

    \node [block] (block1) {$\bm{c}^{\bullet}_1$ \\ \includegraphics[trim= 1.9cm 0.9cm 3cm 0.5cm,clip,height=1.cm]{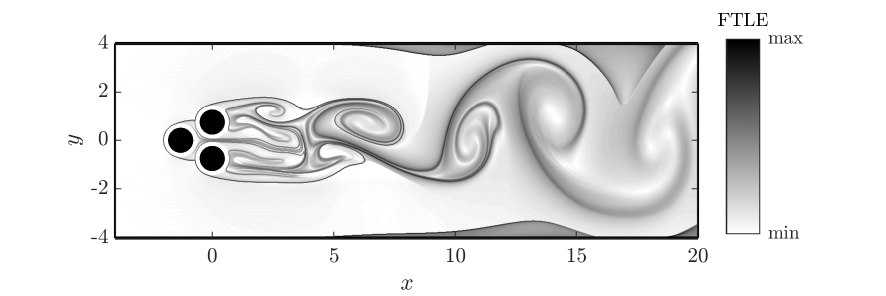}};
    \node [block, below  of=block1,xshift=-1.6cm,yshift=0cm, node distance=1.7cm] (block2) {$\bm{c}^{\bullet}_2$ \\ \includegraphics[trim= 1.9cm 0.9cm 3cm 0.5cm,clip,height=1.cm]{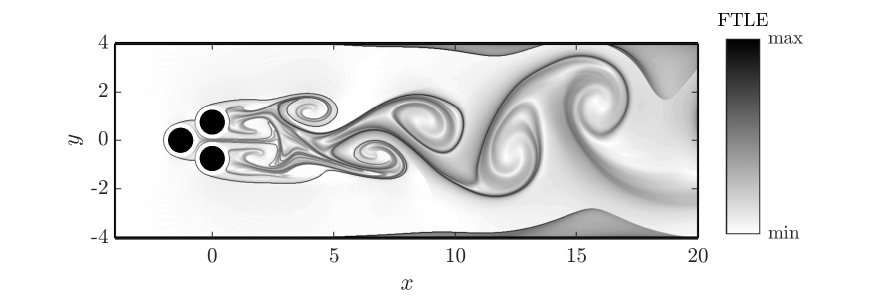}};
    \node [block, below of=block2, node distance=2cm] (block3) {$\bm{c}^{\bullet}_3$ \\ \includegraphics[trim= 1.9cm 0.9cm 3cm 0.5cm,clip,height=1.cm]{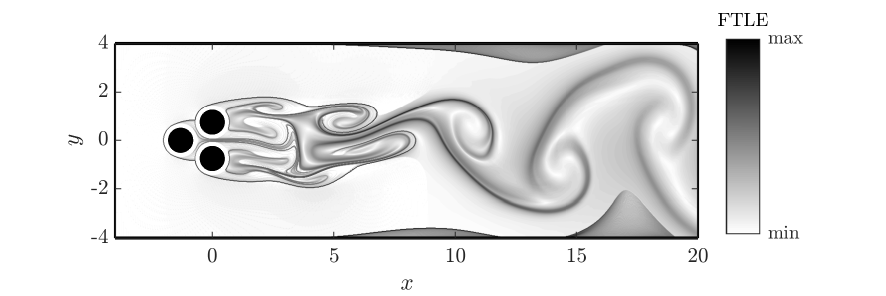}};
    
    \node [block, below of=block3, node distance=1.7cm,xshift=1.6cm,yshift=0cm] (block4) {$\bm{c}^{\bullet}_4$ \\ \includegraphics[trim= 1.9cm 0.9cm 3cm 0.5cm,clip,height=1.cm]{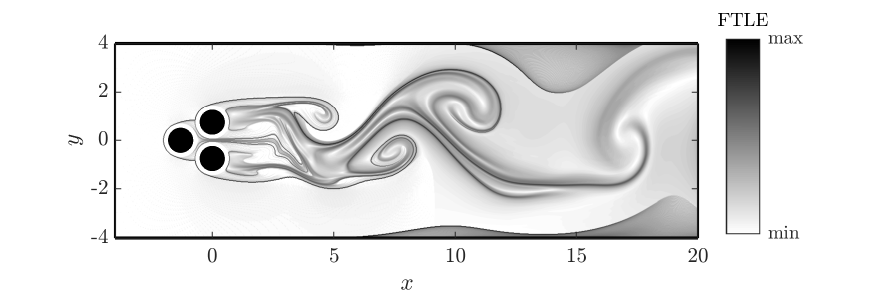}};
    
    \node [block, above of=block4,xshift=1.6cm,yshift=0cm, node distance=1.7cm] (block5) {$\bm{c}^{\bullet}_5$ \\ \includegraphics[trim= 1.9cm 0.9cm 3cm 0.5cm,clip,height=1.cm]{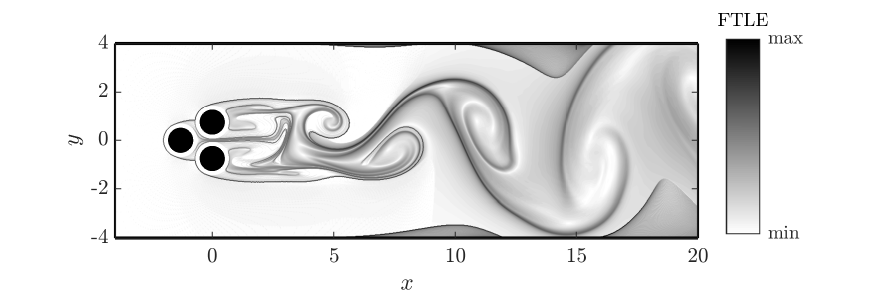}};
	\node [block, above of=block5, node distance=2cm] (block6) {$\bm{c}^{\bullet}_6$ \\ \includegraphics[trim= 1.9cm 0.9cm 3cm 0.5cm,clip,height=1.cm]{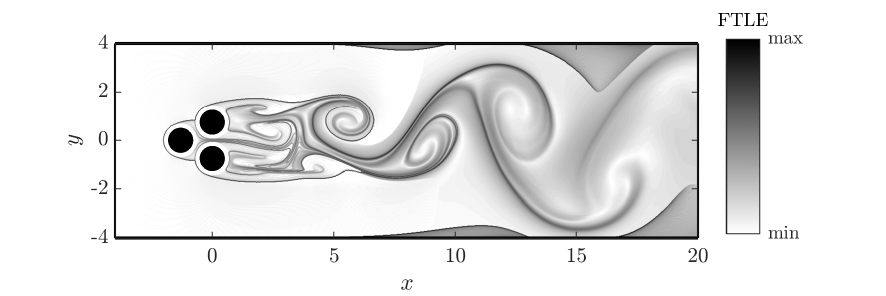}};

  \path[-{Latex[scale=1.15]}]
    ($(block1.west)+(0,-0.2)$) edge[out=190, in=80]  ($(block2.base)+(-0.6,-0.2)$);  
  \path[-{Latex[scale=1.15]}]
    ($(block2.south)+(-0.72,0)$) edge[out=260, in=100]  ($(block3.base)+(-0.72,-0.2)$);
  \path[-{Latex[scale=1.15]}]
    ($(block3.south)+(-0.6,0)$) edge[out=280, in=170]  ($(block4.west)+(0,-0.2)$);
  \path[-{Latex[scale=1.15]}]
    ($(block4.east)+(0,-0.2)$) edge[out=10, in=260]  ($(block5.south)+(0.78,-0)$);
  \path[-{Latex[scale=1.15]}]
    ($(block5.base)+(0.9,-0.2)$) edge[out=80, in=280]  ($(block6.south)+(0.9,-0)$);
  \path[-{Latex[scale=1.15]}]
    ($(block6.base)+(0.78,-0.2)$) edge[out=100, in=-10] ($(block1.east)+(0,-0.2)$);

 \node [block, draw=black, right of=block6,xshift=-0.1cm,yshift=-1cm, node distance=4.cm,text width = 4cm,inner sep=4pt] (blockQT) { \includegraphics[trim= 0cm 0.cm 0.85cm 0.cm,clip,width=1\columnwidth]{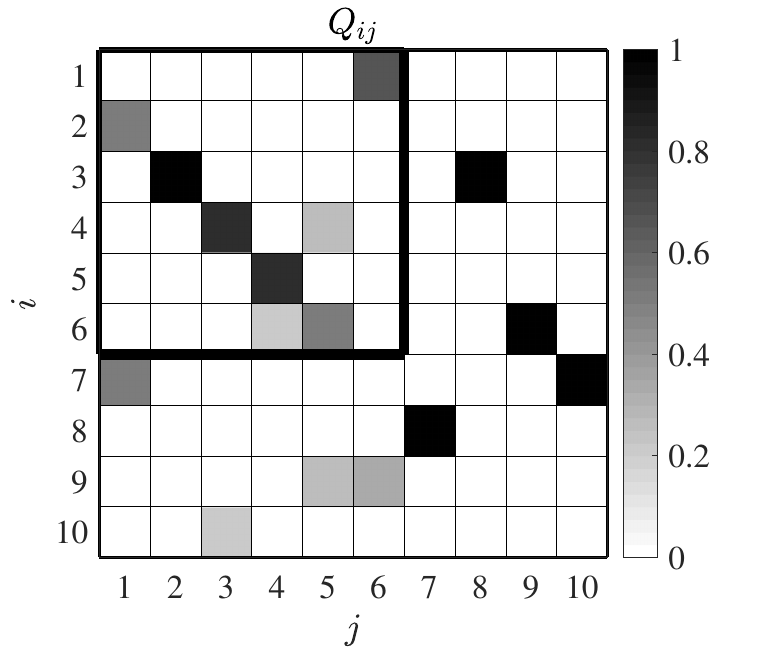}\\ \includegraphics[trim= 0cm 0.cm 0.85cm 0.cm,clip,width=1\columnwidth]{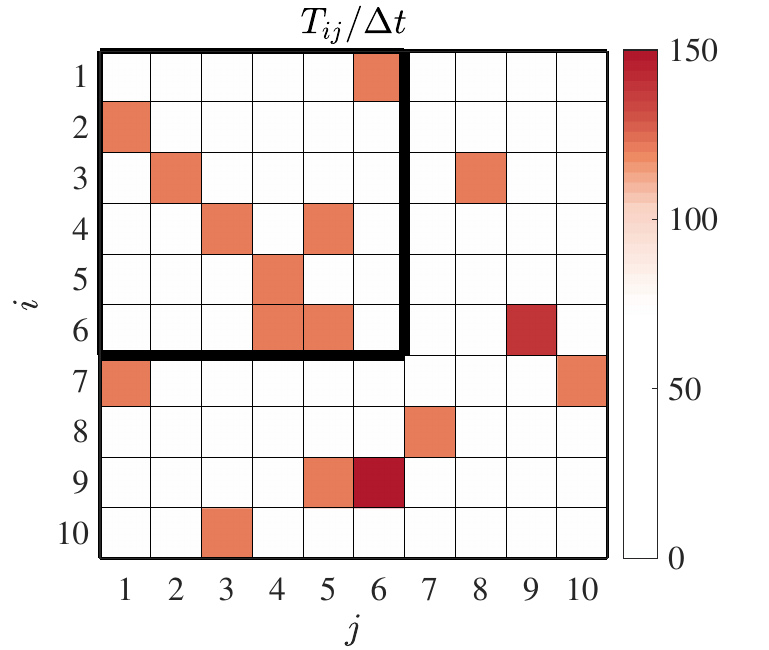}};

  \node [block, right of=block1, node distance=9.5cm] (block7) {$\bm{c}^{\bullet}_7$ \\ \includegraphics[trim= 1.9cm 0.9cm 3cm 0.5cm,clip,height=1.cm]{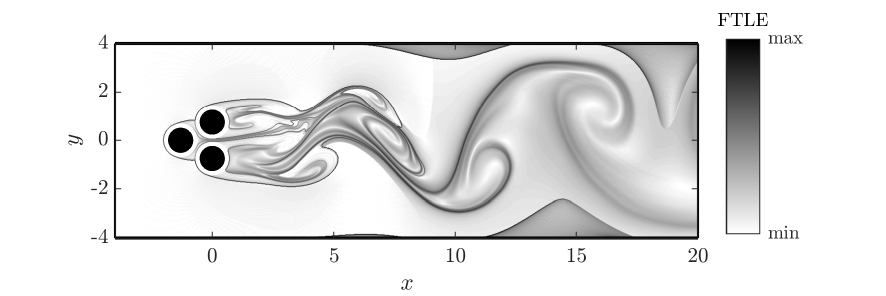}};
  \node [block, below of=block7, node distance=1.8cm] (block8) {$\bm{c}^{\bullet}_8$ \\ \includegraphics[trim= 1.9cm 0.9cm 3cm 0.5cm,clip,height=1.cm]{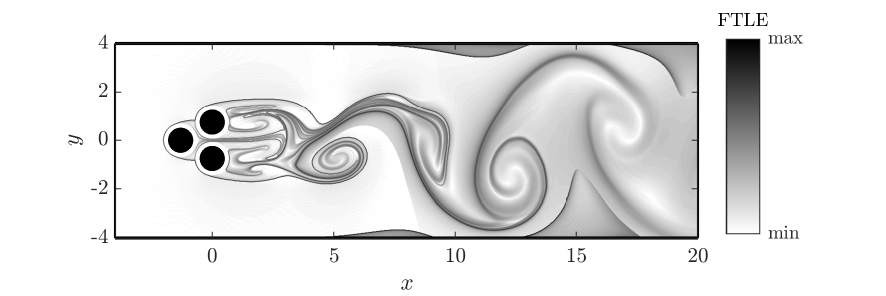}};
   \node [block, below of=block8, node distance=1.8cm] (block9) {$\bm{c}^{\bullet}_9$ \\ \includegraphics[trim= 1.9cm 0.9cm 3cm 0.5cm,clip,height=1.cm]{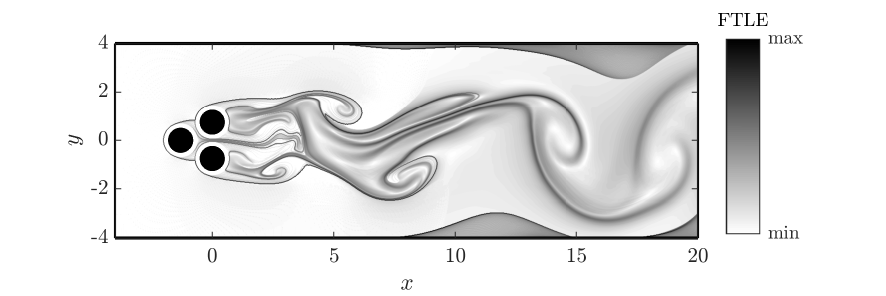}};
    \node [block, below of=block9, node distance=1.8cm] (block10) {$\bm{c}^{\bullet}_{10}$ \\ \includegraphics[trim= 1.9cm 0.9cm 3cm 0.5cm,clip,height=1.cm]{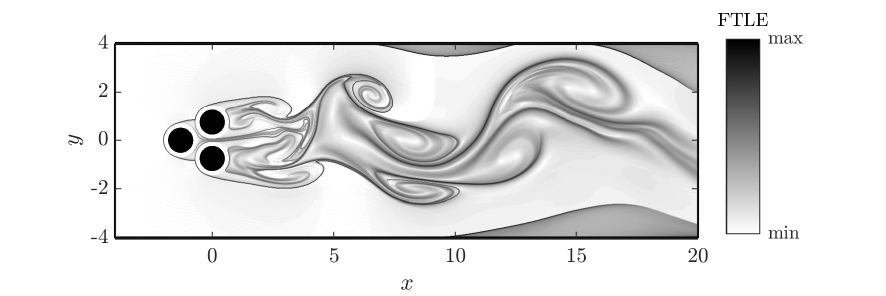}}; 

\end{tikzpicture}
}
\caption{Orbital CNM of the chaotic dynamics of the fluidic pinball at $\Rey=160$. The FTLE distributions have been normalised as in Figure~\ref{fig:FTLEre120}.}
\label{fig:Qmatrixre160}
\end{figure}

As the Reynolds number increases, the fluidic pinball transitions into a chaotic regime. 
The parameters for the oCNM at $\Rey=160$ mirror those used for the earlier configuration, with the segment length calculated as per Equation~\eqref{eq:ni}. Figure~\ref{fig:Qmatrixre160} presents the $Q_{ij}$ and $T_{ij}$ for the oCNM at this Reynolds number together with backward/forward FTLE for the centroids of trajectory clusters.
This network model uncovers that even within the fully chaotic state, a primary cycle of clusters exists that corresponds to periodic vortex shedding. This is interspersed with random transitions to clusters featuring stochastic disorder in the wake flow. This behaviour is corroborated by analysing the backward/forward FTLE for the centroids of trajectory clusters, illustrated on the left and right side of Figure~\ref{fig:Qmatrixre160}. 
Clusters 1 through 6 exhibit periodic vortex shedding, similar to the behaviour observed at lower Reynolds numbers. In contrast, clusters 7 through 10 are characterised by chaotic motion in the wake, featuring multiple vortice emissions.

\begin{figure}
\centering
\begin{minipage}{1\textwidth}

\subfloat[]{\includegraphics[trim= 0cm 0.cm 0cm 0.cm,clip,width=0.5\columnwidth]{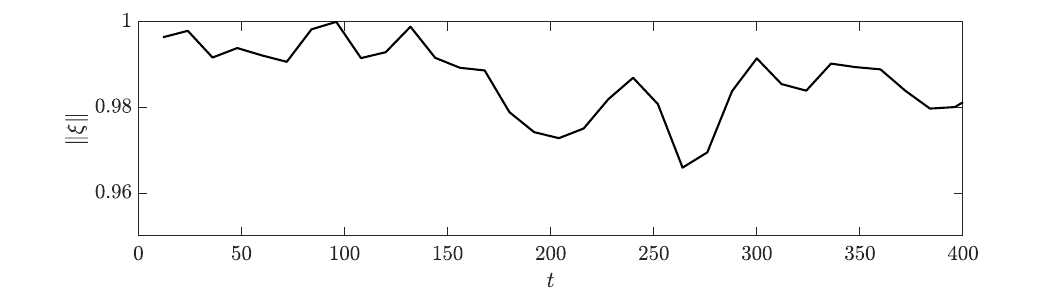}}\\
\vspace{-3.2cm}

\subfloat[]{\includegraphics[trim= 0cm 0.cm 0cm 0.cm,clip,width=0.5\columnwidth]{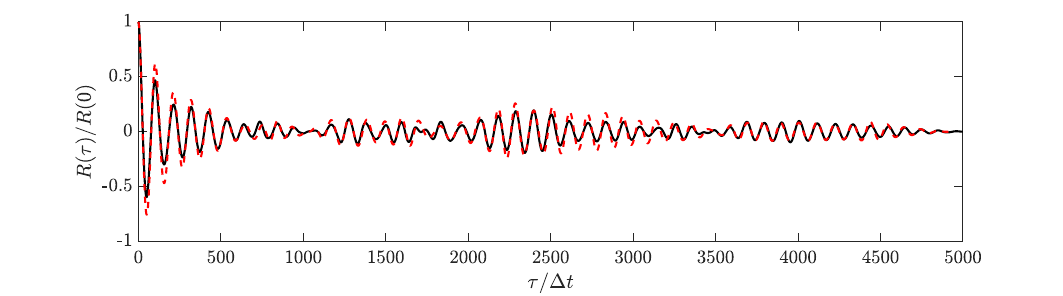}}
\subfloat[]{\includegraphics[trim= 0cm 0.cm 0cm 0.cm,clip,width=0.5\columnwidth]{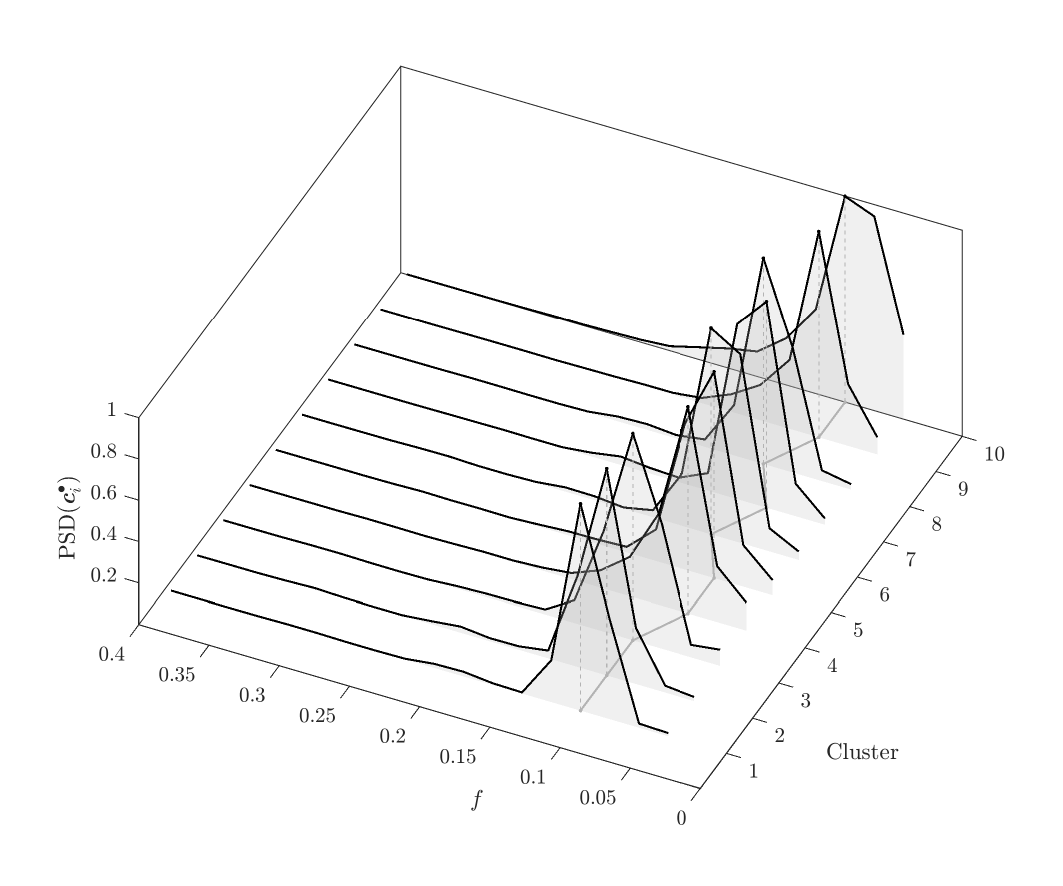}}
\end{minipage}
\caption{Orbital CNM frequency behaviour and validation. Panel (a): slow timescales behaviour of $\|\bm{\xi}\|$. Panel (b): auto-correlation function for the original data (black) and the oCNM output (red dashed). Panel (c): normalised PSD of each trajectory cluster centroid, $\mathrm{PSD}(\bm{c}^{\bullet}_i)$. Case with $\Rey=160$.}
\label{fig:STFTspectrumre160}
\end{figure}

A prominent cycle emerges in the chaotic regime, marked by random transitions from and to chaotic clusters. An examination of panel (a) in Figure~\ref{fig:STFTspectrumre160} indicates the absence of coherent slow timescale motion, while panel (c) observes variability in peak frequencies and bandwidths among the clusters.
Despite these complexities, the oCNM adeptly captures the auto-correlation function of the data, as shown in panel (b) of the same figure.

In this section, only the results related to the Fourier basis of the oCNM are shown, as other methods offer comparable performances, as shown in \ref{sec:testcase} with toy models.

\section{Conclusions}\label{sec:conclusion}

In this work, we proposed a novel data-driven reduced-order modelling approach, namely orbital Cluster-based Network Modelling (oCNM).
In contrast to CNM's clustering of instantaneous states, the oCNM applies functional clustering and considers state orbits/short-term trajectories of dynamical systems.
The resulting oCNM can effectively distinguish between fast and slow timescales by encapsulating the rapid dynamics in clusters while modelling the slow dynamics in a network. 
This functional clustering used in the oCNM effectively resolves the issues related to multi-scale behaviours like frequency modulation and amplitude selection of complex dynamics, especially for the transient dynamics and chaotic states.

The oCNM was first demonstrated using Landau’s equation and its post-transient solution with time-varying parameters, featuring a synthetic multi-scale signal. This analysis has explored the different choices for the functional spaces, namely the Fourier, the B-splines, and the wavelets decompositions of the short-term trajectories functional representation. 
Then, the oCNM was validated on a high-dimensional flow system, the fluidic pinball, for the quasi-periodic and chaotic flow regimes at different Reynolds numbers.
When applied to the force dynamics of the fluidic pinball, the oCNM reveals significant improvements in predicting the force coefficient, as compared to the CNM.
Moreover, when employing a Fourier basis for the representation of trajectories, this approach proves to be particularly effective in describing the nonlinear saturation mechanism in the transient and post-transient dynamics at lower Reynolds numbers and in predicting the quasi-periodic and chaotic dynamics on the attractor at higher Reynolds numbers.
When applied to the velocity fields of the fluidic pinball, the oCNM delineates the complex dynamics inherent in quasi-periodic and chaotic regimes. This approach demonstrates its strength in identifying and quantifying the transition through different spatio-temporal flow states, as evidenced by the direct transition probabilities and times, along with the finite-time Lyapunov exponent analysis, showcasing its capacity to reveal the nuanced behaviours of vortex shedding and chaotic dynamics across various Reynolds numbers.

This work paves a new path to capture and understand the behaviours of complex dynamical systems, enhancing the capability to describe transient dynamics and broadband frequency behaviour, especially in fluid dynamics. The non-physical diffusion of the probability distribution in phase-space modelling is overcome by analysing piecewise trajectories instead of instantaneous states. 
By integrating the functional representation of short-time trajectories/orbits with CNM, the oCNM enables a new ability to incorporate the sophisticated dynamics characteristic of complex systems, analogous to how Spectral Proper Orthogonal Decomposition (SPOD) relates to POD.

Looking ahead, exploring adaptive methods for stochastic functional representation emerges as a promising frontier, which has the potential to dynamically adjust to the evolving characteristics of fluid flows, offering a more precise depiction of complex systems.
Additionally, developing alternative strategies for constructing network models presents an opportunity to refine our understanding of transitional flow states further. 
Investigating diverse methodologies to define function distances based on state predictions will also contribute to the sophistication of our models, enabling a deeper and more comprehensive understanding of fluid dynamics.

\FloatBarrier

\section*{Acknowledgments}
The authors appreciate the valuable discussions with Steven L. Brunton, Guy Yoslan Cornejo Maceda, Hou Chang, Stefano Discetti, Andrea Ianiro, Fran\c{c}ois Lusseyran, Moritz Sieber and Kilian Oberleithner

\section*{Funding}
This work was supported by the National Natural Science Foundation of China under Grants 12172109 and 12202121,
the China Postdoctoral Science Foundation under Grants 2023M730866 and 2023T160166,
the Guangdong Basic and Applied Basic Research Foundation under Grant 2022A1515011492, 
and the Shenzhen Science and Technology Program under Grants JCYJ20220531095605012, KJZD20230923115210021, and 29853MKCJ202300205.

\section*{Declaration of interests}
The authors report no conflict of interest.

\section*{Author ORCIDs}
A. Colanera, https://orcid.org/0000-0002-4227-1162; 
N. Deng, https://orcid.org/0000-0001-6847-2352; 
M. Chiatto, https://orcid.org/0000-0002-5080-7756; 
L. de Luca; https://orcid.org/0000-0002-1638-0429; 
B.~R. Noack, https://orcid.org/0000-0001-5935-1962.

\appendix
\section{Functional principal component analysis}\label{appFPCA}
Functional Principal Component Analysis (FPCA) is a method used in functional data analysis to reduce the dimensionality of the data by representing it in terms of a small number of principal components. FPCA provides a representation of the functional data as a linear combination of a small number of basis functions, where the coefficients correspond to the principal components. 

As shown in \S~\ref{sec:TCNM}, for the analysis, the raw original time series $\bm{q}(t)$ is split in $L$ trajectories ${\bm{q}_1(t'),\ldots,\bm{q}_L(t')}$,with $t'=t-t_{0l}$ and defined in $t' \in [0,T_l]$. The functional data $\bm{q}_l(t')$ can be approximated as a linear combination of $P$ basis functions, $\bm{\phi}_1(t'),\ldots,\bm{\phi}_p(t')$, such that
\begin{equation}\label{fpcaexpApp}
\bm{q}_l(t') \approx\bm{\mu}(t') + \sum_{i=1}^{P} c_{li} \bm{\phi}_i(t'), \qquad i=1,\ldots,L.
\end{equation}
where $\bm{\mu}(t') = E(\bm{q}_l(t'))$ is the mean trajectory observed. To calculate the spatio-temporal modes $\bm{\phi}_i$ it is useful to define the covariance matrix:
\begin{equation}
\bm{C}(s,t) = E\big[(\bm{q}(s)-\bm{\mu}(s))(\bm{q}(t)-\bm{\mu}(t))^{\intercal}\big],
\end{equation}
that is estimated using:
\begin{equation}
\bm{C}(s,t) \approx \frac{1}{L-1} \sum_{l=1}^L ((\bm{q}_l(s)-\bm{\mu}(s))(\bm{q}_l(t)-\bm{\mu}(t))^{\intercal}).
\end{equation}
where $\bm{C}(s,t)$ is a $d \times d$ matrix-valued function. 
The basis functions $\bm{\phi}_i$ are the eigenfunctions of the integral operator:
\begin{equation}\label{covop}
\int_0^{T_l} \bm{C}(s,t) \bm{\phi}_i(t) dt = \lambda_i \bm{\phi}_i(s),
\end{equation}
where $\lambda_i$ are the eigenvalues of the covariance operator and the first $p$ eigenfunctions of the covariance operator, ${\bm{\phi}_i(t)}$, are the basis functions used in FPCA \eqref{fpcaexp}. 
The leading eigenfunctions capture most of the variation in the data, and the remaining eigenfunctions with small eigenvalues are considered noise. Note that the $\bm{\phi}_i$ functions are orthonormal, as
\begin{equation}
\int_0^{T_l} \bm{\phi}_i(\tau)^{\intercal} \bm{\phi}_l(\tau) d\tau = \delta_{il},
\end{equation}
where $\delta_{il}$ is the Kronecker delta. It is worth noticing that the discrete counterpart of Equation~\eqref{covop} consists of an eigenvalue problem, the dimensions of which are $(d\times L_{\mathrm{traj}})\times(d\times L_{\mathrm{traj}})$.  This procedure can be done numerically using standard techniques, such as singular value decomposition. Notice that in Equation~\eqref{covop}, the covariance matrix $\bm{C}$ can be also substituted by nonlinear kernels $\bm{K}(s,t)$, as in \citet{Schoelkopf1998}.

Once the eigenfunctions are estimated, the functional data can be represented in terms of the principal component scores $c_{li}$, which are projections of $\bm{q}_{l}(t')-\bm{\mu}(t')$ onto the $\bm{\phi}_i$ eigenfunctions:
\begin{equation}\label{cij}
c_{li} = \int_0^{T_l} (\bm{q}_{l}(\tau)-\bm{\mu}(\tau) )^{\intercal} \bm{\phi}_i(\tau) d\tau. \end{equation}
The method involves the representation of the functional data as a linear combination of a small number of basis functions, which are chosen to capture the essential features of the data. The principal component scores can be used to visualise the data or to perform further analysis, such as clustering. 

In filtering methods, the matrices containing the STFT $\bm{\xi}_{lj}$, spline $\bm{\beta}_{lj}$ and wavelets $\bm{\gamma}_{lrk}$ coefficients can be quite large. Despite the possibility of sparsity, these matrices can still be computationally expensive to cluster due to their size.

Consider a generic functional expansion basis for the signal $\bm{q}_l(t)$ as shown in Equation~\eqref{basiexp}. In this scenario, it's also possible to expand the function $\bm{\phi}_i(t')$ as follows:
\begin{equation}\label{basiexpphi}
\bm{\phi}_i(t') =\sum_{j=1}^{P} \bm{b}_{ij} f_j(t').
\end{equation}

By substituting Equations~\eqref{basiexp} and \eqref{basiexpphi} into Equation~\eqref{covop}, it is possible to derive the coefficients $\bm{b}_{ij}$. Depending on the chosen basis, the coefficients $c_{li}$ can be obtained from Equation~\eqref{cij}.

In the work presented here, the number of expansion basis functions $P$ has been set equal to the number of trajectories $L$, thus avoiding any loss in the approximation process.

\section{Basis expansions}\label{appbasis}
In this section, several types of basis expansions that serve as crucial tools in signal processing are discussed. \S~\ref{appSTFT} reports the Short-Time Fourier Transform (STFT), an adaptable method used to balance time and frequency information. \S~\ref{appB} introduces B-spline basis expansion, a versatile tool notable for its capability to create smooth and continuous functions. Lastly, \S~\ref{appW} explains the Wavelet Transform, a technique suited to multi-resolution analysis.

\subsection{Short-time Fourier transform}\label{appSTFT}
The short-time Fourier transform (STFT) is a signal processing technique used to analyse the frequency content of a signal over time; it provides the spectral content of a signal in short overlapping time intervals or windows. The STFT is computed by performing the Fourier transform of a windowed segment of the signal and then repeating the process for each segment obtained by sliding the window along the signal. This results in a time-frequency representation of the signal, reporting its frequency content as a function of time.

Considering a generic vectorial function $\bm{q}(t)$ it can be expressed as:
\begin{equation}\label{contSTFT}
\hat F (\omega,t)=\int_{-\infty}^{\infty} w(t-\tau) \bm{q}(\tau) e^{-\mathrm{i}\omega\tau} d\tau,
\end{equation}
where $\omega=2\pi f$ represent the frequency and $w(t-\tau)$ is the chosen temporal window. The size and overlap of the window used for the STFT can be adjusted to control the trade-off between frequency and time resolution.

The discrete counterpart of \eqref{contSTFT} can be written as:
\begin{equation}\label{diSTFT}
 \bm{\hat\alpha}_{lj} =\sum_{n=-\infty}^\infty \bm{q}(t_n) w(n-R_sl) e^{-\mathrm{i}2\pi f_j n \Delta t}
\end{equation}
where $R_s$ is the sliding parameter equal to the difference between the window size $n_w$ and overlapping $n_{ov}$. The inverse STFT is calculated as follows:
\begin{equation}\label{invSTFTco}
\bm{q}(t)=\frac{1}{2\pi}\int_{-\infty}^{\infty}\int_{-\infty}^{\infty} \hat F (\omega,\tau) e^{\mathrm{i}\omega t} d\tau d\omega,
\end{equation}
that in discrete formulation reads:
\begin{equation}\label{invSTFTdi2}
\bm{q}(t_n) =\sum_{l=-\infty}^\infty\sum_{j=-\infty}^\infty \hat \alpha_{lj}e^{\mathrm{i}2\pi f_j n \Delta t}.
\end{equation}
By substituting \eqref{diSTFT} into \eqref{invSTFTdi}, it is possible to recover the Constant Overlap-Add (COLA) Constraint:
\begin{equation}\label{invSTFTdi3}
\sum_{l=-\infty}^\infty w(n-R_sl)=1 \quad \forall n,
\end{equation}
that assures a correct reconstruction.

\subsection{B-splines decomposition}\label{appB}
B-splines are a popular choice for modelling functional data, as they provide a flexible and computationally efficient way to represent functions/trajectories over a continuous domain $t \in [a,b]$. Each trajectory $\bm{q}_l$ is sampled over a set of knots ${t_1, t_2, ..., t_{L_{\mathrm{traj}}}}$, where $a=t_1 \leq t_2 \leq ... \leq t_{L_{\mathrm{traj}}} \leq b$, with $a,b$ generally different between trajectories. By defining a set of basis functions ${B_1(t), B_2(t), ..., B_p(t)}$, where $p={L_{\mathrm{traj}}}-o-1$ ($o$ being the degree of the polynomials) is the order of the B-spline basis, it is possible to represent each function $\bm{q}_l$ using such basis as follows:
\begin{equation}
\bm{q}_l(t) = \sum_{j=0}^{p} \bm{\beta}_{l,j} B_j(t),
\end{equation}
where $\bm{\beta}_{l,j}$ are the coefficients of the B-spline expansion. The B-spline basis functions can be recursively defined as follows:

\begin{equation}
B_{i,0}(t) =
\begin{cases}
1 \quad \text{ if } t_i \leq t < t_{i+1} \\
0 \quad \text{ otherwise},
\end{cases}
\end{equation}
\begin{equation}
B_{i,j}(t) = \frac{t-t_i}{t_{i+j}-t_i}B_{i,j}(t) + \frac{t_{i+j+1}-t}{t_{i+j+1}-t_{i+1}}B_{i+1,j-1}(t),
\end{equation}
for $i=1,2,...,L_{\mathrm{traj}}-p$ and $j=2,3,...,p$. To fit the B-spline basis to each function $\bm{q}_l$, we can use a least squares regression to estimate the coefficients $\bm{\beta}_{l,j}$. 
It is worth noticing that $\sum_{i} B_{i,j}(t)=1$, $\forall t$ and the B-spline basis functions form a complete and orthonormal basis.

\subsection{Wavelet Decomposition}\label{appW}

Wavelets are mathematical functions that divide data into various components, each of which is studied with a resolution that matches its scale. This feature makes wavelets particularly suitable for analysing physical scenarios where the signal comprises discontinuities or sharp spikes.

The Continuous Wavelet Transform (CWT) of a signal, $\bm{q}(t)$, is mathematically expressed as:

\begin{equation}
CWT_{\bm{q}}(a,b) = \frac{1}{\sqrt{|a|}} \int_{-\infty}^{\infty} \bm{q}(t) \psi^{*}\left(\frac{t-b}{a}\right) dt.
\end{equation}

In the above equation, $a$ and $b$ represent the scale factor and the translation factor, respectively, while $\psi(t)$ denotes the mother wavelet. $(\cdot)^*$ represents the complex conjugate operator.

A mother wavelet $\psi(t)$ is a function in $L^2(\mathbb{R})$ (the space of square-integrable functions), from which all other wavelet functions are generated through dilations and translations. A wavelet is a waveform of effectively limited duration with an average value of zero.

In practical applications, the Discrete Wavelet Transform (DWT) samples the CWT in a non-redundant manner, allowing for perfect reconstruction. The DWT of a signal $\bm{q}$ at a given scale level $j$ and translation $k$ is represented as:

\begin{equation}
DWT_{\bm{q}}[j,k] = \frac{1}{\sqrt{2^j}} \int_{-\infty}^{\infty} \bm{q}(t) \psi\left(\frac{t-2^j k}{2^j}\right) dt,
\end{equation}

In the above equation, $j$ and $k$ are integers representing the scale and the translation on the time axis, respectively. The 'db10' wavelet, also known as the Daubechies wavelet of order $10$, was chosen for this analysis. It has $10$ vanishing moments, expressed as:

\begin{equation}
\int t^k \psi(t) dt = 0, \quad \text{for } k=0,1,...,9.
\end{equation}

\begin{figure}
\centering
\includegraphics[width=0.8\textwidth]{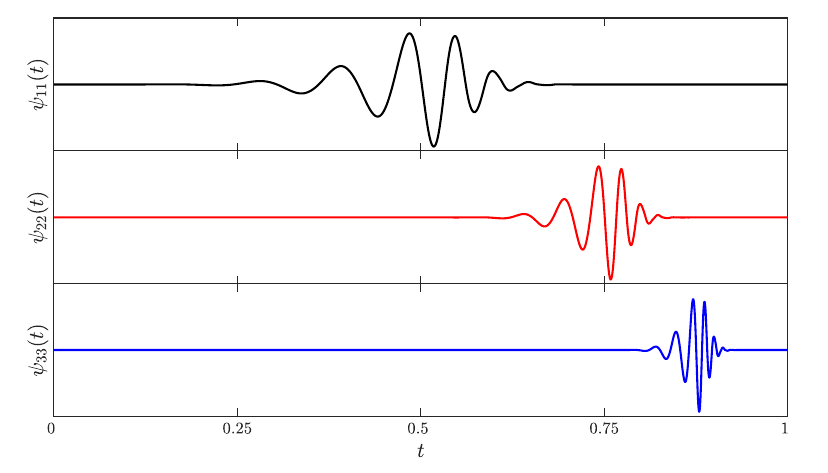}
\caption{The 'db10' wavelet and its decomposition of a time series signal.}
\label{fig:db10_wavelet}
\end{figure}

The Daubechies wavelet family, including the 'db10' wavelet, comprises orthogonal wavelets with varying vanishing moments. Figure~\ref{fig:db10_wavelet} showcases the 'db10' wavelets, illustrating how wavelet decomposition provides a multiresolution analysis that decomposes the signal into its constituent temporal scales.

The Inverse Continuous Wavelet Transform (ICWT) and Inverse Discrete Wavelet Transform (IDWT) are employed to reconstruct the signal from its continuous or discrete wavelet transform, respectively. In the continuous case, given $CWT_{\bm{q}}(a,b)$, the original signal can be retrieved by integrating across all scales (a) and translations (b):

\begin{equation}
\bm{q}(t) = \frac{1}{C_{\psi}} \int_{-\infty}^{\infty} \int_{-\infty}^{\infty} CWT_{\bm{q}}(a,b) \cdot \frac{1}{|a|} \psi^* \left(\frac{t-b}{a}\right) db da,
\end{equation}
where $C_{\psi}$ is a normalisation constant that depends on the wavelet function.

For the discrete wavelet transform, given $DWT_{\bm{q}}[j,k]$, the original signal can be reconstructed by summing over all scales (j) and translations (k):

\begin{equation}
\bm{q}(t) = \sum_{j=-\infty}^{\infty} \sum_{k=-\infty}^{\infty} DWT_{\bm{q}}[j,k] \cdot \psi\left(\frac{t-2^j k}{2^j}\right).
\end{equation}

The wavelet functions must be orthogonal for perfect reconstruction, as is the case for the Daubechies family of wavelets. Using the same wavelet function for both the wavelet transform and its inverse is also essential.

\section{Choice of the number of clusters $K$ for the Fluidic Pinball flow fields analysis.}\label{sec:RMSE}	

The construction of cluster based ROMs, regardless the specific variant, requires the phase space to be partitioned into $K$ clusters. Selecting an appropriate value for $K$ is crucial to balance model complexity and representational accuracy. In this work, unless otherwise specified, the number of clusters is determined ensuring that the variance ratio $>0.9$ (Section \ref{sec:flowfields}) and using the elbow method applied to the RMSE of the auto-correlation function between the training dataset and its reconstruction, as defined in Eq.~\eqref{eq:RMSE}.

\begin{figure}[ht]
	\centering
	\subfloat[$Re=120$]{%
		\includegraphics[trim=0cm 0cm 0cm 0cm, clip, width=0.48\columnwidth]{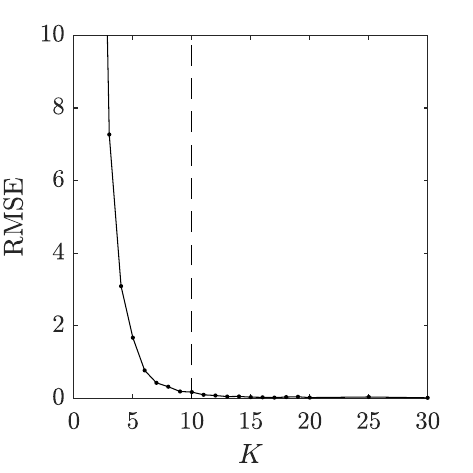}
	}
	\subfloat[$Re=160$]{%
		\includegraphics[trim=0cm 0cm 0cm 0cm, clip, width=0.48\columnwidth]{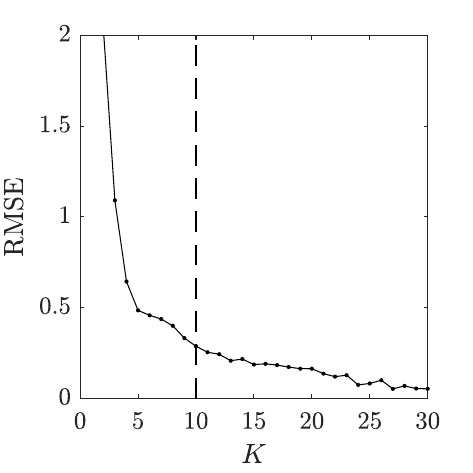}
	}
	\caption{RMSE of the autocorrelation function $R$ between the training dataset and its reconstruction. Panel (a): quasi-periodic regime ($Re = 120$). Panel (b): chaotic regime ($Re = 160$).}
	\label{fig:CLUchoice}
\end{figure}

Figure~\ref{fig:CLUchoice} shows the RMSE curves as a function of the number of clusters $K$ for the oCNM analysis, with Fourier basis, of the flow fields past the Fluidic Pinball. For both the quasi-periodic ($Re = 120$) and chaotic ($Re = 160$) regimes, we selected $K = 10$ as a good compromise between accuracy and model simplicity.

\section{Flow Field Comparison Between CNM and oCNM}\label{sec:ffcomparison}
\begin{figure}
	\centering
	\subfloat[$Re=120$]{%
		\begin{minipage}[c]{\columnwidth}
			\setlength{\tabcolsep}{0pt}
			\renewcommand{\arraystretch}{0.1} 
			\begin{tabular}{ccccc}
				\vspace{8pt}
				& \multicolumn{2}{c}{standard CNM} &  \multicolumn{2}{c}{ oCNM}  \\
				$t=5$	& & local error & & local error\\
				\includegraphics[trim= 1.7cm 0.6cm 2.6cm 0.6cm,clip,height=1.2cm]{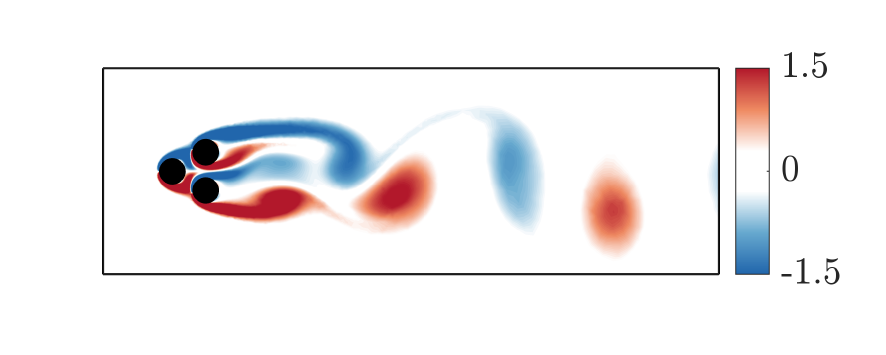}&
				\includegraphics[trim= 1.7cm 0.6cm 2.6cm 0.6cm,clip,height=1.2cm]{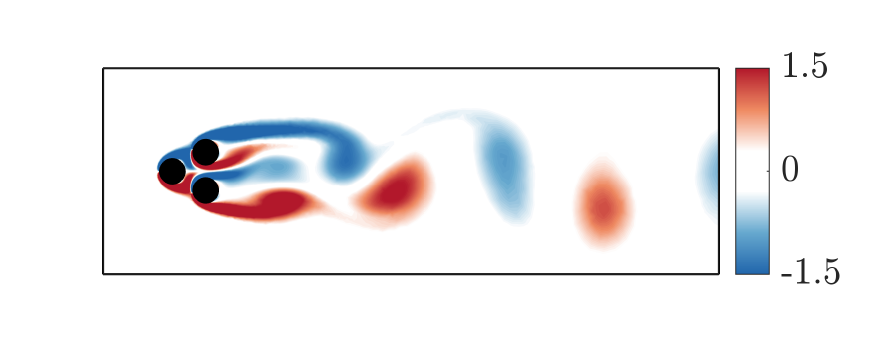}&
				\includegraphics[trim= 1.7cm 0.6cm 2.6cm 0.6cm,clip,height=1.2cm]{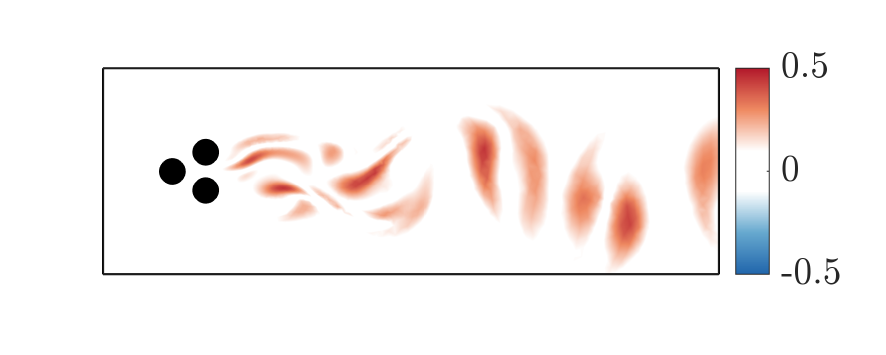}&
				\includegraphics[trim= 1.7cm 0.6cm 2.6cm 0.6cm,clip,height=1.2cm]{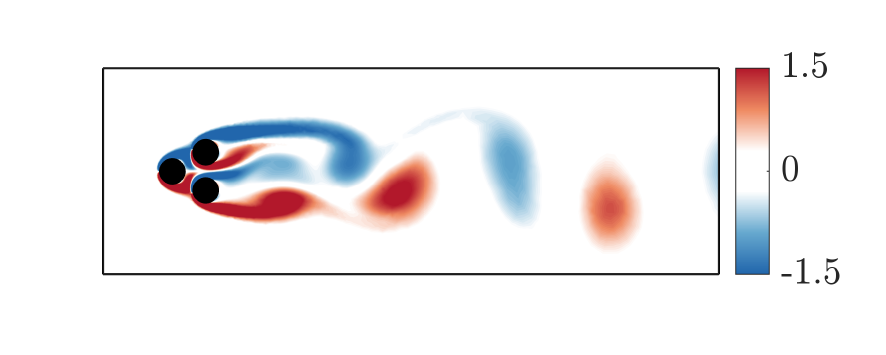}&
				\includegraphics[trim= 1.7cm 0.6cm 0.2cm 0.6cm,clip,height=1.2cm]{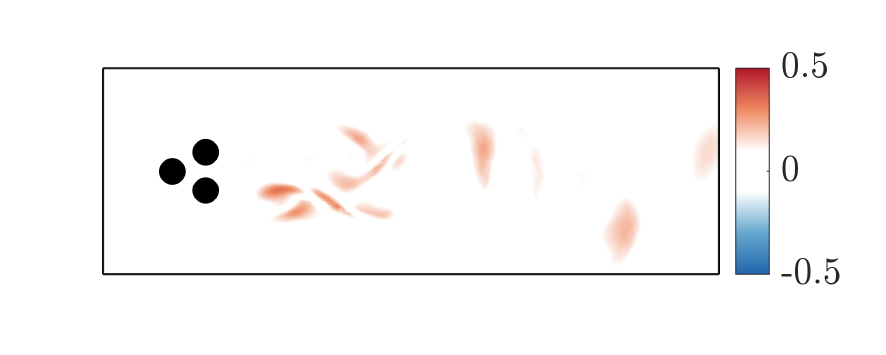}\\
				$t=10$	& &  & &\\
				\includegraphics[trim= 1.7cm 0.6cm 2.6cm 0.6cm,clip,height=1.2cm]{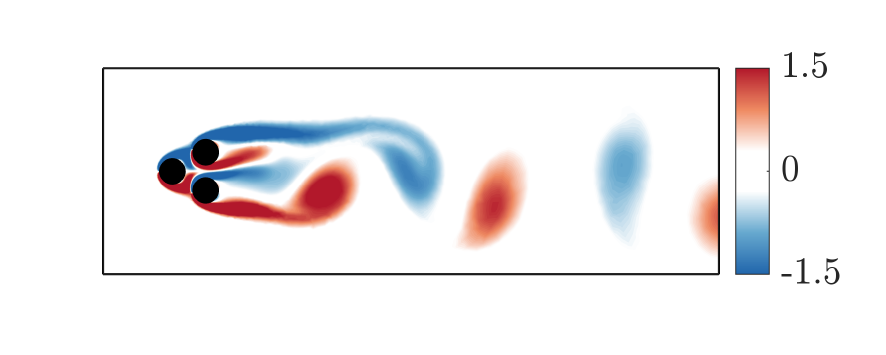}&
				\includegraphics[trim= 1.7cm 0.6cm 2.6cm 0.6cm,clip,height=1.2cm]{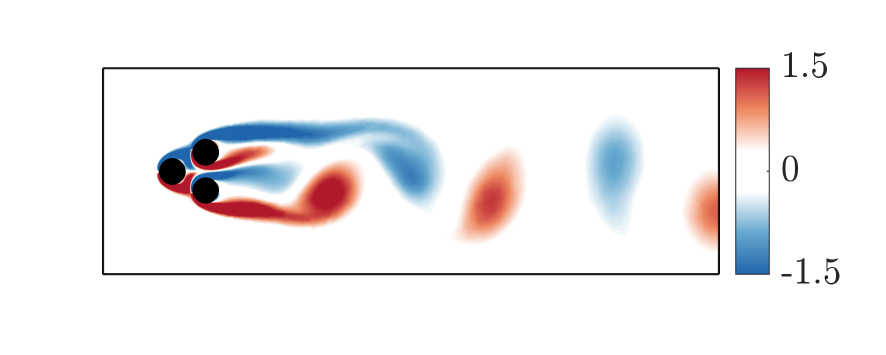}&
				\includegraphics[trim= 1.7cm 0.6cm 2.6cm 0.6cm,clip,height=1.2cm]{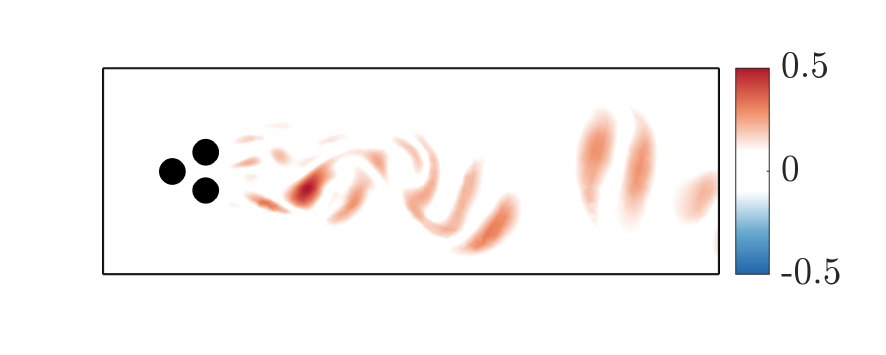}&
				\includegraphics[trim= 1.7cm 0.6cm 2.6cm 0.6cm,clip,height=1.2cm]{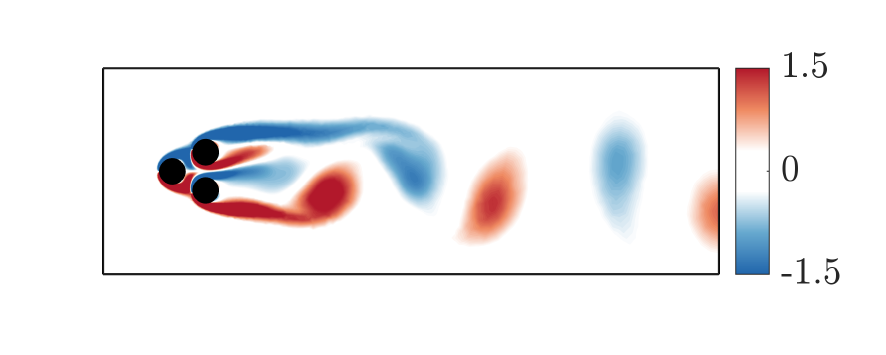}&
				\includegraphics[trim= 1.7cm 0.6cm 0.2cm 0.6cm,clip,height=1.2cm]{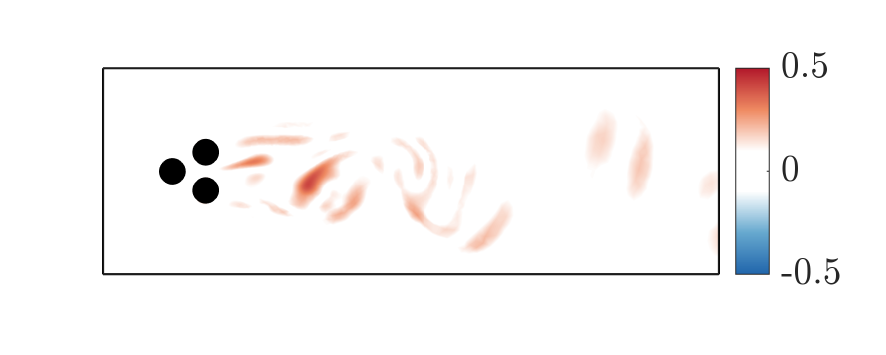}\\
				$t=100$	& &  & &\\
				\includegraphics[trim= 1.7cm 0.6cm 2.6cm 0.6cm,clip,height=1.2cm]{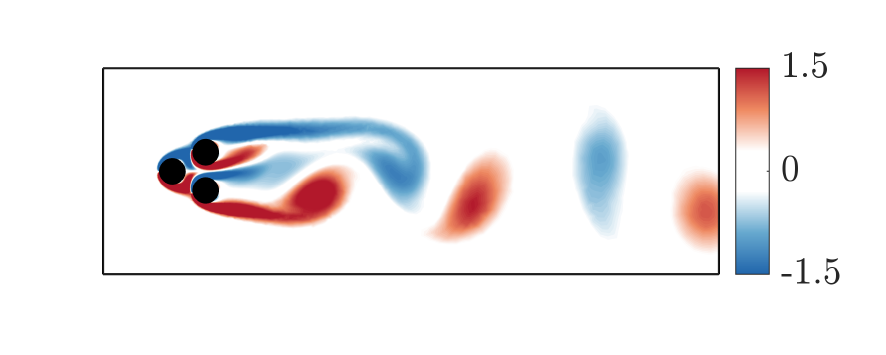}&
				\includegraphics[trim= 1.7cm 0.6cm 2.6cm 0.6cm,clip,height=1.2cm]{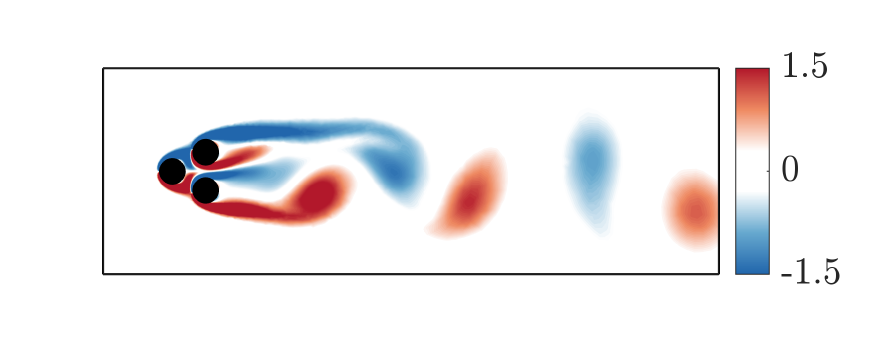}&
				\includegraphics[trim= 1.7cm 0.6cm 2.6cm 0.6cm,clip,height=1.2cm]{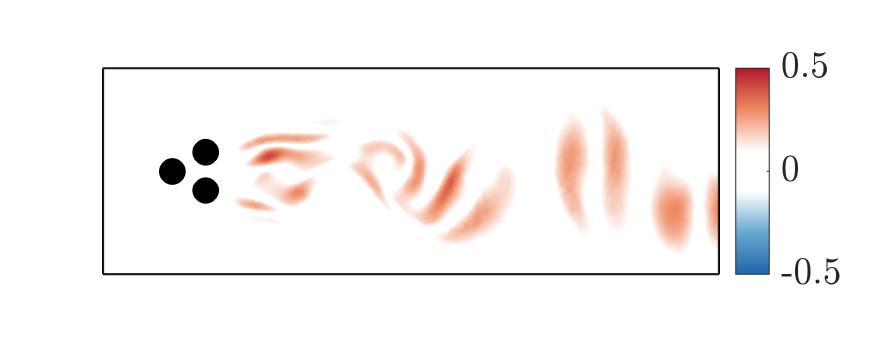}&
				\includegraphics[trim= 1.7cm 0.6cm 2.6cm 0.6cm,clip,height=1.2cm]{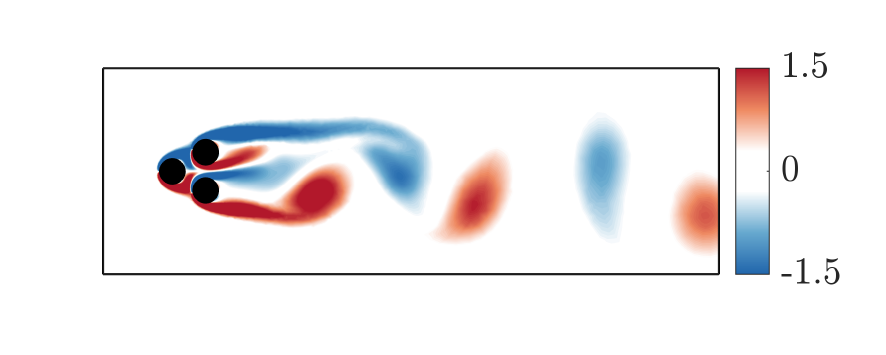}&
				\includegraphics[trim= 1.7cm 0.6cm 0.2cm 0.6cm,clip,height=1.2cm]{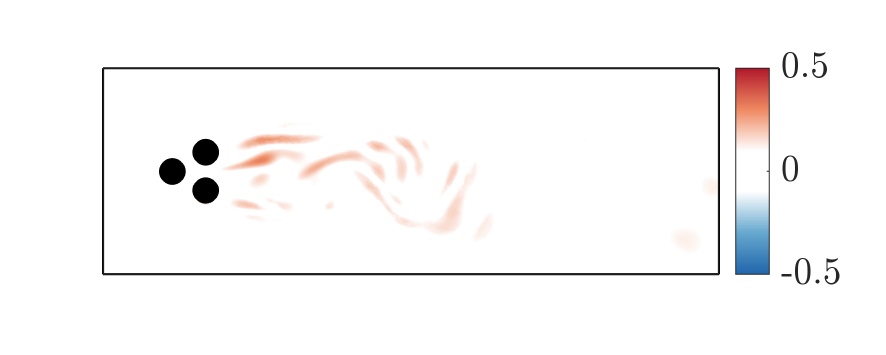}\\
				
			\end{tabular}
		\end{minipage}
	}\\
	\subfloat[$Re=160$]{%
		\begin{minipage}[c]{\columnwidth}
			\setlength{\tabcolsep}{0pt}
			\renewcommand{\arraystretch}{0.1} 
			\begin{tabular}{ccccc}
				$t=5$	& & local error & & local error\\
				\includegraphics[trim= 1.7cm 0.6cm 2.6cm 0.6cm,clip,height=1.2cm]{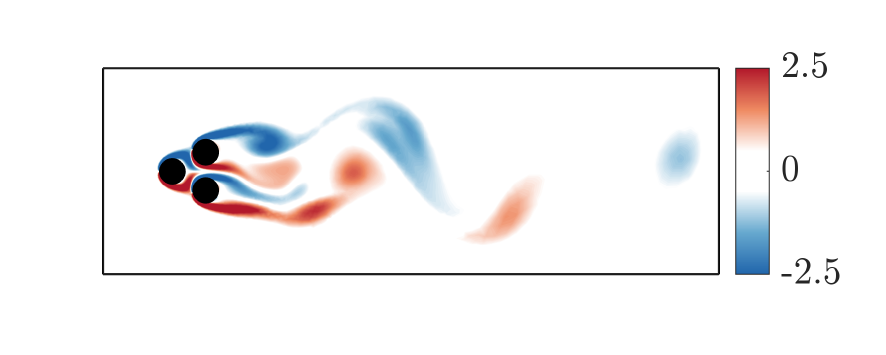}&
				\includegraphics[trim= 1.7cm 0.6cm 2.6cm 0.6cm,clip,height=1.2cm]{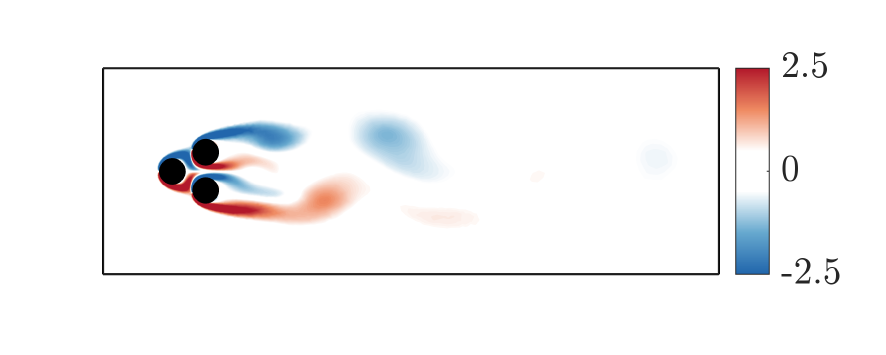}&
				\includegraphics[trim= 1.7cm 0.6cm 2.6cm 0.6cm,clip,height=1.2cm]{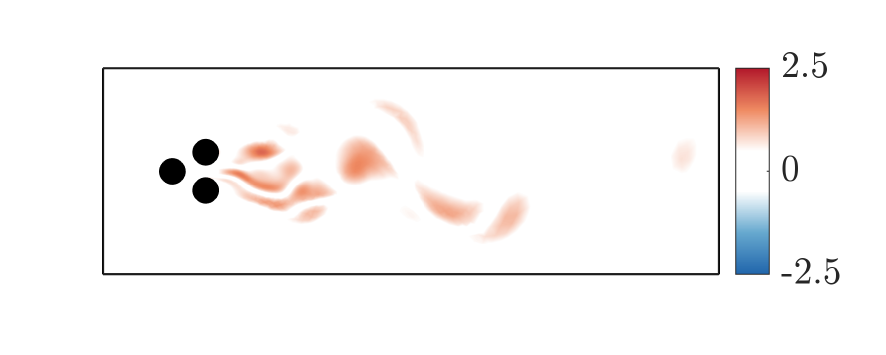}&
				\includegraphics[trim= 1.7cm 0.6cm 2.6cm 0.6cm,clip,height=1.2cm]{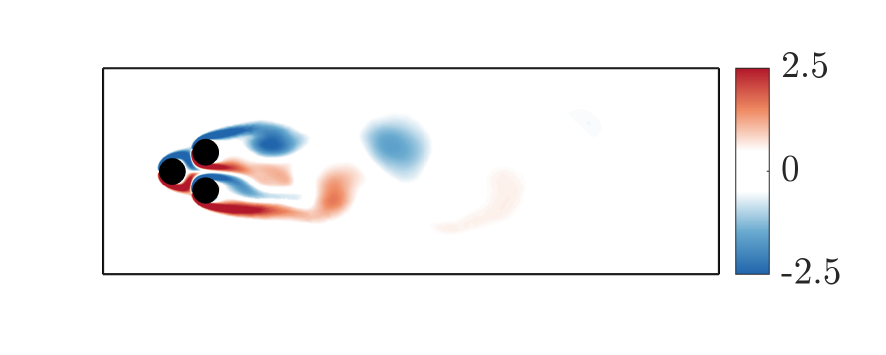}&
				\includegraphics[trim= 1.7cm 0.6cm 0.2cm 0.6cm,clip,height=1.2cm]{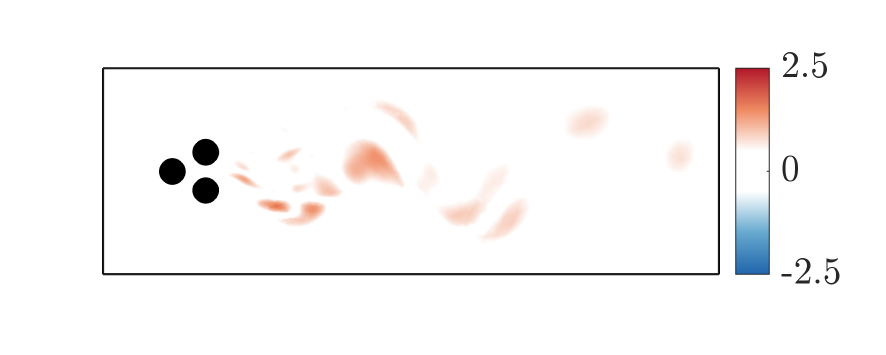}\\
				$t=10$	& &  & &\\
				\includegraphics[trim= 1.7cm 0.6cm 2.6cm 0.6cm,clip,height=1.2cm]{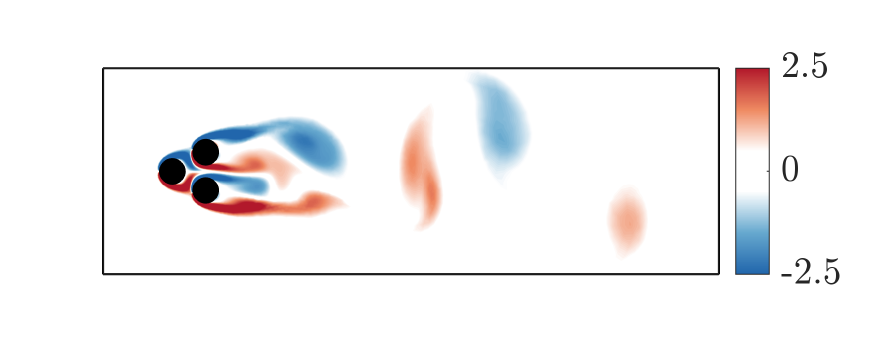}&
				\includegraphics[trim= 1.7cm 0.6cm 2.6cm 0.6cm,clip,height=1.2cm]{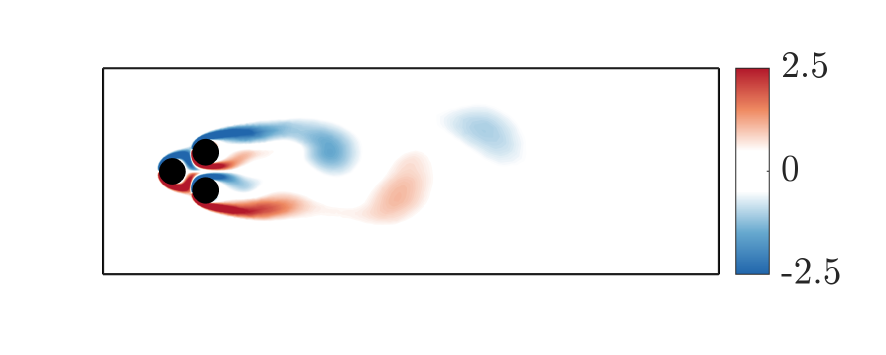}&
				\includegraphics[trim= 1.7cm 0.6cm 2.6cm 0.6cm,clip,height=1.2cm]{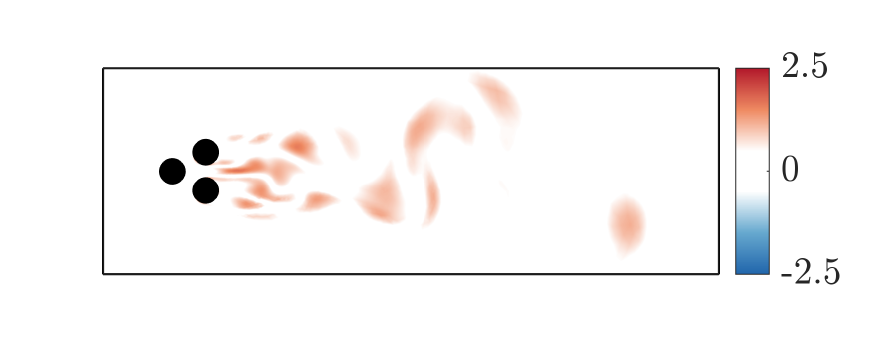}&
				\includegraphics[trim= 1.7cm 0.6cm 2.6cm 0.6cm,clip,height=1.2cm]{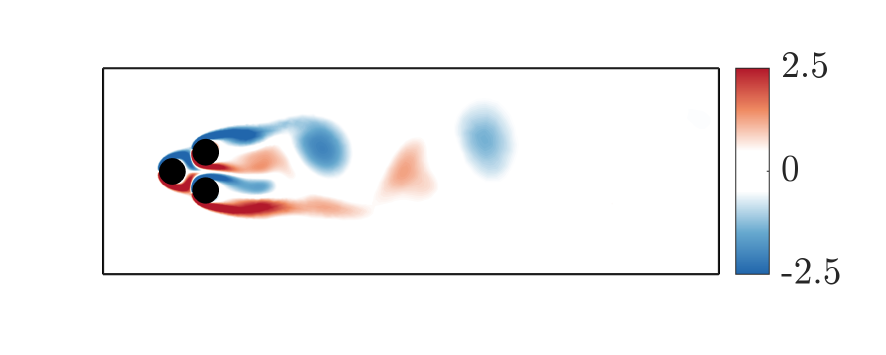}&
				\includegraphics[trim= 1.7cm 0.6cm 0.2cm 0.6cm,clip,height=1.2cm]{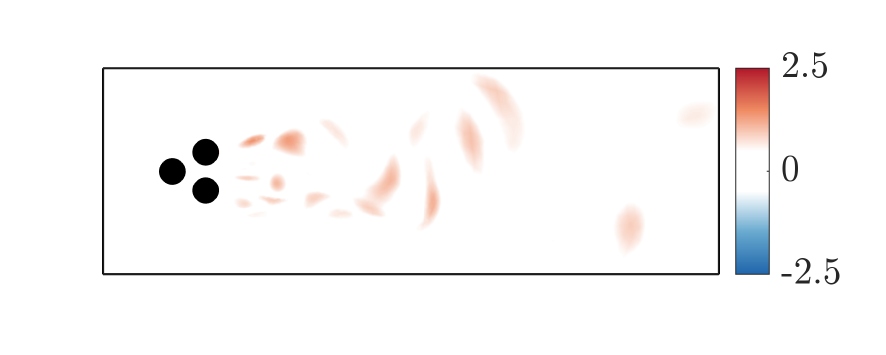}\\
				$t=100$	& &  & &\\
				\includegraphics[trim= 1.7cm 0.6cm 2.6cm 0.6cm,clip,height=1.2cm]{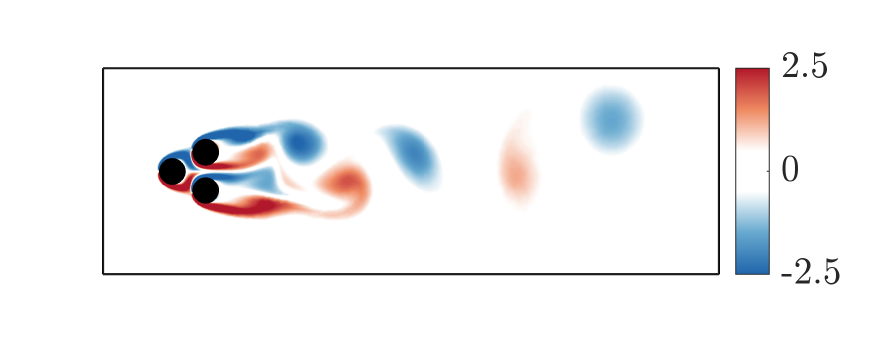}&
				\includegraphics[trim= 1.7cm 0.6cm 2.6cm 0.6cm,clip,height=1.2cm]{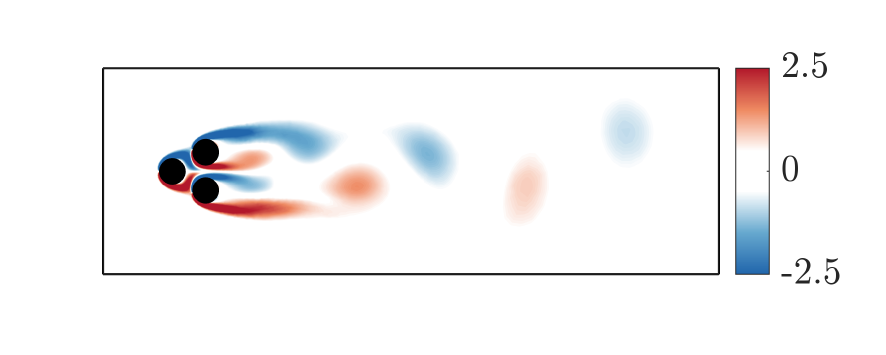}&
				\includegraphics[trim= 1.7cm 0.6cm 2.6cm 0.6cm,clip,height=1.2cm]{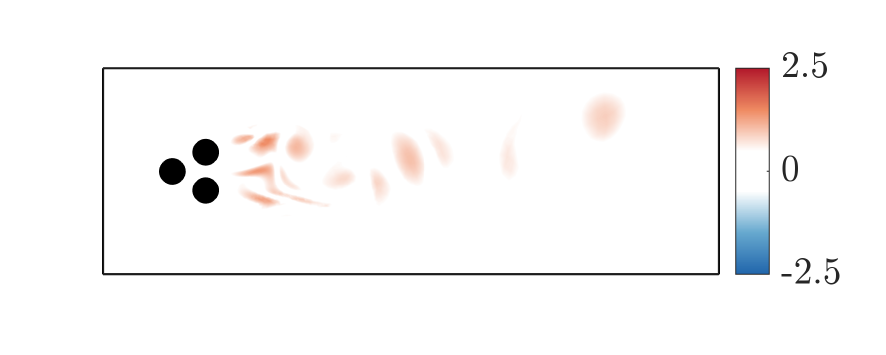}&
				\includegraphics[trim= 1.7cm 0.6cm 2.6cm 0.6cm,clip,height=1.2cm]{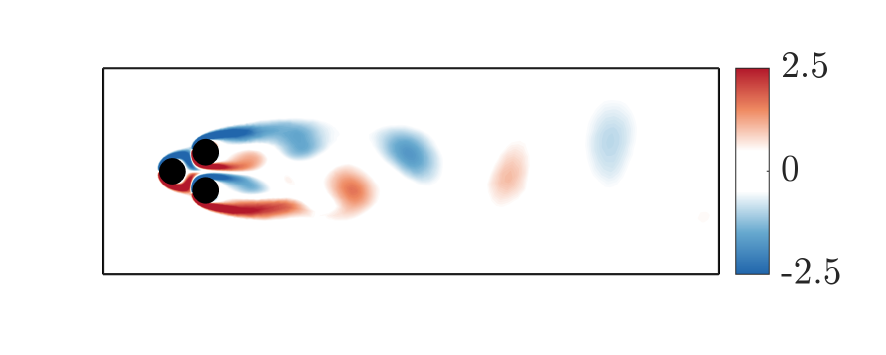}&
				\includegraphics[trim= 1.7cm 0.6cm 0.2cm 0.6cm,clip,height=1.2cm]{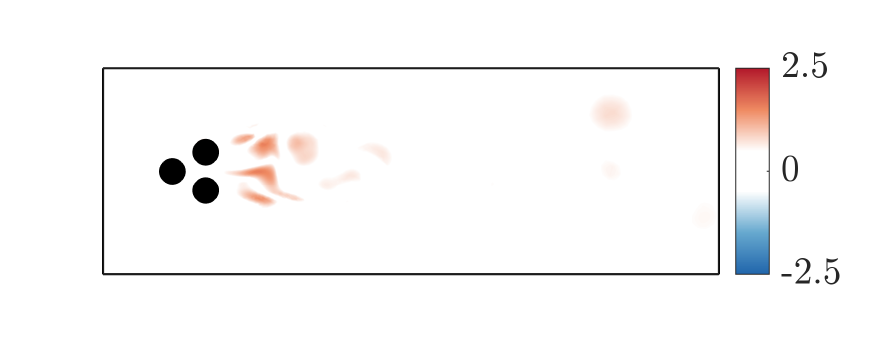}\\
				
			\end{tabular}
		\end{minipage}
	}
	
	\caption{Temporal evolution of vorticity snapshots at three different time instances for the test dataset (first column), compared with predictions from the standard CNM (second column) and the oCNM (fourth column). The absolute value of the local error is shown in the third and fifth columns. In all the panels abscissa and ordinate represent $x$ and $y$, respectively, in the domain $-4<x<20$ and $-4 < y < 4$. All variables in each row are normalized with respect to the maximum vorticity value observed across the three snapshots shown.}
	\label{fig:snapscomp}
\end{figure}

In this appendix, a comparison between the flow fields reconstructed by the standard CNM and the oCNM is provided. The comparison is performed at representative time instances in both the quasi-periodic and chaotic regimes. For further details on the CNM an oCNM parameters the reader is referred to section \ref{sec:flowfields}.
Figure~\ref{fig:snapscomp} shows snapshots of the predicted vorticity fields, along with the absolute error with respect to the test data. The results clearly demonstrate the improved reconstruction capabilities of the oCNM across the two flow regimes.
\section{Finite-Time Lyapunov Exponent}\label{sec:FTLE}

The Finite-Time Lyapunov Exponent (FTLE) is a Lagrangian metric used to identify coherent structures within complex fluid flows, as discussed by \citet{Haller2015}. It quantifies the divergence of pathlines within the flow, which can be computed forward or backward, each emphasising attracting or repelling structures within the flow, respectively. This study focuses on the backward/forward time FTLE, which highlights regions where fluid from different flow regions converges/diverges, revealing vortex structures within shear layers.

For FTLE calculation, a set of uniformly distributed spatial initial conditions, $\bm{x}_0$, is selected. These initial conditions' pathlines are determined using a fourth-order Runge-Kutta (RK4) solver. The backward/forward integration time, $\nu$, tailored to the specific study configurations, is set to match the characteristic period $1/f_c$. This approach maps the flow's initial state to its final points, represented by $\bm{\Phi}_\nu(\bm{x}_0)$. The spatial FTLE distribution which depends on the time instance $t^m$ and $\nu$, can be calculated as \citep{Haller2001}:

\begin{equation}
\lambda(\bm{x}_0, t_m, \nu) = \frac{1}{|\nu|} \ln \sqrt{\sigma_{\text{max}}},
\end{equation}
where $\sigma_{\mathrm{max}}$ is the maximum eigenvalue of the right Cauchy-Green deformation tensor $\bm{G}$, given by:

\begin{equation}
\bm{G} = (\bm{\nabla} \bm{\Phi}_\nu)^{\intercal} \bm{\nabla} \bm{\Phi}_\nu.
\end{equation}
Here, $\bm{\nabla}$ symbolises the gradient operator.

Within the framework of the orbital CNM, the flow map $\bm{\Phi}_\nu$ is constructed by considering the spatio-temporal clusters centroids $\bm{c}^{\bullet}_i(t)$ providing smooth temporal functions suitable for the backward-in-time RK4 solver, thereby enhancing our understanding of flow dynamics and coherent structures.

\bibliographystyle{elsarticle-harv} 
\bibliography{bibfile.bib}

\end{document}